\documentclass[final]{elsarticle}

\usepackage{url}
\usepackage{alltt, verbatim}
\usepackage{amsmath,amssymb}
\usepackage{latexsym}
\usepackage{stmaryrd}

\usepackage{alltt, verbatim}

\usepackage{tikz}
\usetikzlibrary{positioning,calc}

\usepackage[colorlinks]{hyperref}

\newcommand*\mtt[1]{\hbox{\tt#1}}
\newcommand*\mts[1]{\;\hbox{\tt#1}\;}
\newcommand*\syntaxSymb[1]{$\langle$\hbox{\rmfamily\itshape #1}$\,\rangle$}

\usepackage{hhline}

\newtheorem{example}{Example}

\long\def\comment#1{}

\newcounter{marginalnote}
\newcommand{\bigfracn}[3]{
\begin{array}[b]{c}
\displaystyle #1 \\\hline\displaystyle #2
\end{array}
\hbox to 0pt{\raisebox{0.7em}{{\tiny (#3)}}}
}

\renewcommand{\bigfracn}[3]{
  \begin{array}[b]{cc}
    \displaystyle #1  & {\makebox[0pt][l]{\raisebox{-1.33ex}[0mm][0mm]{\scriptsize{(#3)}}}}\\
    \hhline{-}\displaystyle #2
  \end{array}
}

\makeatletter
\newenvironment{prog}{\vspace{0.7ex}\par
\setlength{\parindent}{0cm}
\obeylines\@vobeyspaces\tt}{\vspace{0.0ex}\noindent
}
\makeatother
\newcommand{\startprog}{\begin{prog}}
\newcommand{\stopprog}{\end{prog}\noindent}

\makeatletter
\newenvironment{smallprog}{\vspace{0.7ex}\par
\setlength{\parindent}{0.7cm}
\obeylines\@vobeyspaces\tt\small}{\vspace{0.7ex}\noindent
}
\makeatother
\newcommand{\fstartprog}{\begin{smallprog}}
\newcommand{\fstopprog}{\end{smallprog}\noindent}

\makeatletter
\newenvironment{nismallprog}{\vspace{0.7ex}\par
\setlength{\parindent}{0.0cm}
\obeylines\@vobeyspaces\tt\small}{\vspace{0.7ex}\noindent
}
\makeatother
\newcommand{\fnistartprog}{\begin{nismallprog}}
\newcommand{\fnistopprog}{\end{nismallprog}\noindent}

\newcommand{\ul}[1]{\underline{#1}}

\def\defemb#1#2{\expandafter\def\csname #1\endcsname
{\relax\ifmmode #2\else\hbox{$#2$}\fi}}
\defemb{cA}{{\cal A}}
\defemb{cB}{{\cal B}}
\defemb{cC}{{\cal C}}
\defemb{cD}{{\cal D}}
\defemb{cE}{{\cal E}}
\defemb{cF}{{\cal F}}
\defemb{cG}{{\cal G}}
\defemb{cH}{{\cal H}}
\defemb{cI}{{\cal I}}
\defemb{cJ}{{\cal J}}
\defemb{cL}{{\cal L}}
\defemb{cM}{{\cal M}}
\defemb{cO}{{\cal O}}
\defemb{cP}{{\cal P}}
\defemb{cR}{{\cal R}}
\defemb{cS}{{\cal S}}
\defemb{cT}{{\cal T}}
\defemb{cU}{{\cal U}}
\defemb{cV}{{\cal V}}
\defemb{cX}{{\cal X}}
\defemb{cZ}{{\cal Z}}

\usepackage{listings}
\usepackage{xcolor}

\lstdefinelanguage{Maude}{
  breaklines=true,
  language=,
  numbers=left,
  numbersep=5pt,
  tabsize=4,
  basicstyle=\scriptsize\normalfont\ttfamily, %
  xleftmargin=2.5ex,
  keywordstyle=,
  commentstyle=\itshape, 
  identifierstyle=,
  numberstyle=\ttfamily\tiny,
  morekeywords={fmod, endfm, mod, omod, tmod, endm, endom, endtm, is, pr},
  morekeywords={protecting, crl, rl, op, ops, eq, ceq, sort, sorts, class},
  morekeywords={subsort, subsorts, subclass, subclasses, vars, var, if},
  morekeywords={Maude, result, cmb, tsearch, such, that},
  sensitive=true,
}

\newcommand{\norm}[1]{{\downarrow_ {#1}}}
\newcommand{\code}[1]{{\normalfont\ttfamily #1}}
\usepackage{fancyvrb}

\lstnewenvironment{maude}[1][]
    {\lstset{language=Maude, 
             numbers=none, 
             escapeinside={(*}{*)}, #1}
    }
    {}

\newcommand{\nt}[1]{\mbox{$\langle$\textit{{#1}}$\rangle$}}

\makeatletter
\DeclareTextCommandDefault\textcommaabove[1]{%
  \hmode@bgroup
  \ooalign{%
    \hidewidth
    \raise.7ex\hbox{%
      \check@mathfonts\fontsize\ssf@size\z@\math@fontsfalse\selectfont`%
    }%
   \hidewidth\crcr
   \null#1\crcr
  }%
  \egroup
}
\@ifundefined{textcommabelow}{%
  \DeclareTextCommandDefault\textcommabelow[1]
    {\hmode@bgroup\ooalign{\null#1\crcr\hidewidth\raise-.31ex
     \hbox{\check@mathfonts\fontsize\ssf@size\z@
     \math@fontsfalse\selectfont,}\hidewidth}\egroup}%
}{}
\makeatother

\journal{Journal of Logical and Algebraic Methods in Programming}
\myfooter[L]{\noindent\begin{tabular}{l}
	Accepted in Journal of Logical and Algebraic Methods in Programming. \hfill October 2019. \\[1ex]
	DOI: \href{https://doi.org/10.1016/j.jlamp.2019.100497}{10.1016/j.jlamp.2019.100497} \hfill \lower5pt\hbox{\includegraphics[scale=.5]{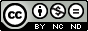}}
\end{tabular}}

\begin{document}

\begin{frontmatter}

\title{Programming and Symbolic Computation in Maude}


\address[malaga]{Universidad de M\'alaga, Spain.}
\address[sri]{SRI International, CA, USA.} 
\address[upv]{Universitat Polit\`ecnica de Val\`encia, Spain.}
\address[ucm]{Universidad Complutense de Madrid, Spain.} 
\address[illinois]{University of Illinois at Urbana-Champaign, IL, USA.}

\author[malaga]{Francisco Dur\'an}
\ead{duran@lcc.uma.es}
\author[sri]{Steven Eker}
\ead{eker@csl.sri.com}
\author[upv]{Santiago Escobar}
\ead{sescobar@upv.es}
\author[ucm]{Narciso Mart\'{\i}-Oliet}
\ead{narciso@ucm.es}
\author[illinois]{Jos\'e Meseguer}
\ead{meseguer@illinois.edu}
\author[ucm]{Rub\'en Rubio}
\ead{rubenrub@ucm.es}
\author[sri]{Carolyn Talcott}
\ead{clt@cs.stanford.edu}

\begin{abstract}
Rewriting logic is both a flexible \emph{semantic framework} within which
widely different concurrent systems can be naturally specified and
a \emph{logical framework} in which widely different logics can
be specified.  Maude 
programs are exactly rewrite theories.  Maude
has also a formal environment of verification tools.
\emph{Symbolic computation} is a powerful technique for
reasoning about the correctness of concurrent systems
and for increasing the power of
formal tools.  We present several new symbolic features 
of Maude that enhance formal reasoning about Maude programs and
the effectiveness of  formal tools.  They include:
(i) very general \emph{unification} modulo
user-definable equational theories, and (ii)
\emph{symbolic reachability analysis} of concurrent systems using narrowing.
The paper does not focus just on symbolic features: it also
describes several other new Maude features, including: (iii) Maude's 
\emph{strategy language} for controlling
rewriting, and (iv) \emph{external objects} that allow flexible interaction of
Maude object-based concurrent systems with the external world.
In particular, \emph{meta-interpreters} are external objects
encapsulating Maude interpreters that can interact with many other objects.
To make the paper self-contained and give a reasonably complete  language overview,
we also review  the
basic Maude features for equational rewriting and rewriting with
rules, Maude programming of concurrent object systems, and reflection.  Furthermore, we include
many examples illustrating all the Maude notions and features
described in the paper.
\end{abstract}


\end{frontmatter}

\tableofcontents

\section{Introduction}
\label{sec:intro}

\noindent {\bf What is Maude?}  The Maude book's title \cite{maude-book}
describes it as a \emph{High-Performance Logical Framework} and adds:
\emph{How to Specify, Program and Verify Systems in Rewriting Logic}.
Maude is indeed a declarative programming language based on rewriting
logic \cite{unified-tcs,bruni-meseguer-tcs,20-years}.  

So, what is rewriting logic?  It is a logic ideally suited to specify
\emph{and} execute computational systems in a simple and natural way.
Since nowadays most computational systems are concurrent, rewriting
logic is particularly well suited to specify concurrent systems
without making any a priori commitments about the 
model of concurrency in question, which can be synchronous or
asynchronous, and can vary widely in its shape and nature:  from a
Petri net \cite{stehr-meseguer-olveczky} to a process calculus 
\cite{Verdejo-Marti-Oliet00,stehr-cinni}, from an object-based system \cite{ooconc} to
asynchronous hardware \cite{DBLP:journals/jlp/KatelmanKM12}, from a
mobile ad hoc network protocol \cite{liu-etal-MANETS-JLAMP}   to a cloud-based 
storage system \cite{ACC-maude-techreport}, from a web browser 
\cite{IE-maude-analysis,DBLP:conf/facs2/SasseKMT12}
 to a programming language with
threads \cite{meseguer-rosu-tcs,DBLP:journals/iandc/MeseguerR13}, from
a distributed  control system \cite{DBLP:journals/tcs/MeseguerO12,multirate-pals-SCP}
to a model of mammalian
cell pathways \cite{pathways02,pathways04}, and so on.  And all \emph{without any encoding}:
what you see and get is a direct definition of the system itself, not some
crazy Turing machine or Petri net encoding of it.
All this means that rewriting logic is a flexible \emph{semantic framework}
to define and program computational systems.  But since in rewriting logic
\begin{quote}
  \emph{Computation} = \emph{Deduction}
\end{quote}
the exact same flexibility can be used to specify any \emph{logic} in
rewriting logic, used now as a \emph{logical framework}.  Indeed,
a logic's inference system can be naturally specified as a rewrite theory whose
(possibly conditional) rewrite rules are exactly
the logic's inference rules.  Again, logics
as different as linear logic, first-order logic, various modal logics, or all
the higher-order logics in Barendregt's lambda cube can be specified 
in rewriting logic (and mechanized in Maude) \emph{without any
  encoding} \cite{rwl-fwk,stehr-meseguer-ho,DBLP:conf/wrla/OlartePR18,20-years}.
 This explains the ``Logical Framework'' part in the
Maude book's title.

What about the ``High-Performance'' description?  You should not take our
word for it.  Instead, you may wish to take a look at the paper
\cite{garavel-etal-benchmarks}, where a thorough benchmarking 
by H. Garavel and his collaborators at INRIA Rh\^{o}ne-Alpes of
a wide range of functional and rule-based declarative languages based
on a large suite of benchmarks expressed in a language-independent manner
and mapped into each language 
is reported.  Although Maude is an interpreted language,
it ranks second in overall performance for that suite, closely after Haskell.

\vspace{1ex}

\noindent {\bf What are Maude Programs?}  Rewrite theories.
A rewrite theory is a triple $\mathcal{R}= (\Sigma,E,R)$,
where $(\Sigma,E)$ is an equational theory, with function symbols
$\Sigma$ and equations\footnote{As we explain in Section
  \ref{sec:functional}, the equational theory may also contain
  \emph{membership axioms} specifying the typing of some expressions.
  For the moment think of $E$ as containing both.} 
 $E$, specifying a concurrent
system's states as an algebraic data type,
and $R$ is a set of rewrite rules that specify the \emph{local
  concurrent transitions} that the concurrent system can perform.
In Maude this is declared as a \emph{system module} with syntax
\texttt{mod} $(\Sigma,E,R)$ \texttt{endm}.
The 
case when
$R = \emptyset$ gives rise to Maude's functional sublanguage of
\emph{functional modules}, which are declared with syntax
\texttt{fmod} $(\Sigma,E)$ \texttt{endfm} and specify
the algebraic data type defined by $E$ for the function symbols in
$\Sigma$.  Of course, this means
that when
writing and verifying Maude programs we \emph{never}  leave the realm of
mathematics.  This explains the  qualification:
``How to Specify, Program and Verify Systems in Rewriting Logic''
in Maude book's title.  Maude is not just a language: it has
a formal environment of \emph{verification tools}, some
internal to the language and others built as language extensions
(more on this in Sections \ref{sec:system}, \ref{sec:narrowing},
and \ref{sec:tools-apps}).

\vspace{1ex}

\noindent {\bf Why Another Paper on Maude?}  Maude is in her mid 20s.  The
first conference paper on Maude appeared in 1996
\cite{DBLP:journals/entcs/ClavelELM96}.  This was expanded into the 2002
journal paper \cite{maude-tcs}, which to this date remains the most cited
journal reference for the language.  Important new advances were
reported in the 2007 Maude book \cite{maude-book}, which is the most
highly cited reference on Maude to date.  But a lot has happened
since 2007.  From time to time we have reported on new advances in a piecemeal
way in a sequence of tool papers; but they are both quite brief
and scattered over numerous publications: no unified account of the
present state of Maude actually exists.  That is why we decided to
write this paper.

What is it like?  On the one hand \emph{repetition} of material already
available in previous publications should be avoided; but on the other
hand this paper should be a good entry point to learn about Maude as
it is in 2019 \emph{without assuming prior acquaintance with Maude}. 
Therefore, we have tried to strike a balance between: (i)
making the paper self-contained and providing a reasonably complete
\emph{overview} of the language; and (ii) making sure that all the
important \emph{new features} now available in Maude are explained and
illustrated.  The way this balance between generality and novelty is
attempted is reflected in the paper's organization. The most basic
introduction to the language is given in Section \ref{sec:functional}
on functional modules and Section \ref{sec:system} on system modules.
The first important new feature is Maude's \emph{strategy language},
treated in Section \ref{sec:strategies}.  Section
\ref{sec:oo} on \emph{object-based programming} is a mixture of old
and new: on the one hand we introduce new readers to the basic ideas
on how distributed object systems are programmed declaratively in
Maude.  On the other hand we explain several important new features
on how Maude objects can now interact with various \emph{external
  objects}.  Another mixture of old and new is
provided by Section~\ref{sec:reflection} on reflection and
meta-interpreters: reflection is a long-standing and crucial feature
of both rewriting logic and Maude; but meta-interpreters are an
entirely new feature.  A very
important additional theme with a host of new language features is reflected in the paper's title,
namely, Maude's
current support for \emph{symbolic computation}.  This theme
is developed along three sections: Section \ref{sec:unif+variants}
discusses unification, variants, and equational narrowing features; Section \ref{sec:narrowing}
discusses narrowing-based symbolic reachability analysis; and 
Section \ref{symb-comp-tools-and-apps}
discusses symbolic reasoning tools and applications.

\subsection{Features}
Let us summarize those Maude's features considered in this paper.\\

\noindent {\bf Strategies.} 
Most concurrent systems are intrinsically \emph{nondeterministic},
so that different transitions may lead the system into widely different
states.  The obvious consequence is that an expression $t$ in a system
module can be evaluated by its rules $R$ in many different ways.  For
example, the rewrite theory $(\Sigma,E,R)$ may describe the game of
chess, or the inference system of a theorem prover.  But many chess
moves may be stupid ones, and many inference steps may be useless.
In both cases we need a \emph{strategy} to apply the rules $R$ in a
way that achieves \emph{our} intended goals.  This is what Maude's
\emph{strategy language}, explained in Section \ref{sec:strategies},
makes possible.

\vspace{1ex}

\noindent {\bf Specification and Deployment of Concurrent Object
  Systems.}  Although, as already mentioned, Maude can naturally
express a wide variety of concurrent systems, many such systems
are best expressed as collections of concurrent
\emph{objects} which communicate with each other by message passing.
Section~\ref{sec:oo} explains how concurrent  object systems can
be programmed declaratively in Maude.  But this leaves open two issues:  (i) the
object-based view, by its very nature, should allow interactions with
\emph{any} kind of object, including the user seen as an ``object,''
and (ii) the Maude interpreter runs on a single machine, therefore
the concurrent system defined by the program can be \emph{simulated}
and analyzed in a Maude interpreter; but how can it be \emph{deployed} as a
distributed system?  Both issues are addressed by means of  several
kinds of \emph{external objects} with which standard Maude objects can
interact.  In particular, using socket external objects, Maude
programs can be deployed as distributed systems running on several
machines.

\vspace{1ex}

\noindent {\bf Reflection and Meta-Interpreters.}  Rewriting logic is
a \emph{reflective} logic.  This means that its \emph{meta-theory},
including notions such as theory and term, can 
be \emph{represented as data} at the so-called \emph{object level} of
  the logic in a \emph{universal theory}.  It also means that such
a universal theory, like in the case of universal Turing machines, can
\emph{simulate} any other theory, including itself.  This is
extremely powerful for (meta-)programming purposes and is efficiently
supported by Maude's \texttt{META-LEVEL} module.  Section \ref{sec:reflection}
explains reflection, and also illustrates how meta-programming can be
used to easily build advanced new tools such as an Eqlog \cite{eqlog-jlp}
functional-logic programming interpreter.  It also explains a very
powerful new reflective feature, namely, \emph{meta-interpreters},
which open the possibility of creating and interacting in a reflective manner with a hierarchy of 
Maude interpreters as \emph{external objects}.

\vspace{1ex}

\noindent {\bf Maude and Symbolic Computation.}  Because Maude is a
programming language \emph{and} a logical framework in which many
different logics and formal tools can be mechanized
\emph{and} has itself a formal environment of verification tools, 
support for \emph{symbolic reasoning} is very
important both for advanced formal reasoning about Maude programs and
to use Maude as a \emph{formal meta-tool} to build many other tools in
other logics.  From 2007 to the present, a sustained effort has
been made to endow Maude with powerful symbolic reasoning
capabilities.  At the \emph{equational logic} level, they focus around
the topic of Section \ref{sec:unif+variants}, namely,
 \emph{unification modulo an equational
  theory}, that is, solving systems of equations \emph{modulo} an
equational theory.
Maude's  unification features are extremely general in
\emph{three orthogonal  dimensions}, corresponding to three aspects of an
equational theory, which in Maude can have the form $(\Sigma,E \cup
B)$, where $\Sigma$ is an order-sorted signature
(more on this in Section \ref{sec:functional}),
$B$ are common
equational axioms such as associativity and/or commutativity and/or identity,
and $E$ are equations that are assumed convergent  (more on this in Section \ref{sec:functional})
modulo $B$.  The first dimension of generality is $\Sigma$: since
order-sorted signatures are strictly more general than many-sorted ones, which
are way more general than unsorted ones, order-sorted unification
algorithms are much more general than the usual unsorted ones.
The second dimension is unification modulo axioms $B$, which in Maude
can be \emph{any} combination of associativity and/or commutativity and/or
identity axioms.  The third dimension of generality is support for
order-sorted unification modulo \emph{any} theory
$E \cup B$, where the axioms $B$ are as explained and the equations $E$
are convergent modulo $B$.  Some very hard problems had
to be solved to make $E \cup B$-unification practical and to
characterize the cases when it terminates.  They were solved in \cite{variant-JLAP}
 thanks to the notion of \emph{variant}, as also explained in  Section
\ref{sec:unif+variants}.

At the \emph{rewriting logic level}, Section  \ref{sec:narrowing}
explains how 
the just-described support for
unification modulo $E \cup B$ becomes a key symbolic lever to support narrowing-based
\emph{symbolic reachability analysis}
for a rewrite theory (system module)
$\mathcal{R}=(\Sigma,E \cup B,R)$, where $E \cup B$ has the so-called \emph{finite
variant property} \cite{comon-delaune},
ensuring that $E \cup B$-unification terminates.  
Such reachability analysis provides a powerful form of \emph{symbolic
model checking} for $\mathcal{R}$, where possibly infinite sets of states are described by
symbolic expressions.  Using the new symbolic reasoning features
described in Sections \ref{sec:unif+variants}--\ref{sec:narrowing},
many symbolic reasoning tools can be developed covering many
applications.  Section \ref{symb-comp-tools-and-apps} focuses on
those tools and applications most directly related to Maude itself; but similar
formal tools can likewise be developed (and are developed)  for many
other logics \cite{20-years}.

\vspace{1ex}

\noindent {\bf Core Maude vs.\ Full Maude.}  Maude is also referred to
as \emph{Core Maude}.  This is done to distinguish it from \emph{Full
  Maude}.  But what is Full Maude?  What Maude is not yet but
\emph{will} be.  Most new features presented in this paper
---from the strategy language to variants, from unification algorithms
to symbolic model checking, from object-oriented features to parameterized modules---
first cut their teeth as features prototyped in Full Maude.  How does
Full Maude work?  By \emph{reflection}.  That is, Full Maude is a
\emph{reflective Maude program} extending Maude itself with new
language features \cite{maude-book}. As explained in \cite{DuranO09}, 
since many Maude verification tools need to manipulate Maude modules
reflectively and should be well integrated with Maude itself, they
can be built quite easily as \emph{extensions} of Full Maude.  Full Maude
is not directly discussed in this paper; but, as Alfred Hitchcock in
his movies, makes some interesting cameo appearances.  
A nice one takes place in Section \ref{sec:eqlog}, where we show
how the Eqlog \cite{eqlog-jlp}  functional-logic language can be easily implemented
using reflection and  Maude's narrowing-based symbolic reachability  and can be
given an execution environment as an extension of Full Maude.

\vspace{1ex}

\noindent {\bf Differences with the Conference Paper 
\cite{DBLP:conf/wrla/DuranEEMMT18}.}
This paper is a loose and very large extension of the conference paper 
\cite{DBLP:conf/wrla/DuranEEMMT18}.  Usually one says what has been
added, but that would take too long.  It is much shorter to say what has
been loosely imported from \cite{DBLP:conf/wrla/DuranEEMMT18}, namely,
some of the material in the ``symbolic'' Sections \ref{sec:unif+variants}--\ref{sec:narrowing}.
But even that needs a few grains of salt.  For example, since this
paper focuses on language features and what they are good for,
we have omitted the detailed description of Maude's
order-sorted unification algorithm modulo \emph{associativity} given
in \cite{DBLP:conf/wrla/DuranEEMMT18}.

\vspace{1ex}

\noindent {\bf Examples and Maude Executables}.  All the examples in the
paper run on Maude 3.0, which is available at \url{http://maude.cs.illinois.edu}.
The Maude code for all the examples in the paper can also be found at that site.
\label{urlmaude30}

\section{Functional Modules}\label{sec:functional}
Maude is a declarative language based on rewriting logic;
but rewriting logic has membership equational logic 
\cite{tarquinia,bouhoula-jouannaud-meseguer}
as its  functional sublogic. From the computational point of view the key
difference between rewriting logic and its membership equational
sublogic is that between: (i) the \emph{nondeterminism} of rewrite
theories, and (ii) the \emph{determinism} of equational ones.
That is, an equational program is a functional program in which a
functional expression (called a \emph{term}) is evaluated
using the equations in the
program as left-to-right rewrite rules, which
are assumed \emph{confluent} 
(see \cite{dershowitz-jouannaud} and Section \ref{rewriterel-sect}).
 If such an evaluation
terminates, it returns a \emph{unique} 
computed value (determinism), namely, its \emph{normal form}
after simplifying it with the (oriented) equations.
Instead, a rewrite theory usually
models a \emph{nondeterministic} and often \emph{concurrent} system,
which may never terminate and where the notion of a
computed value may be meaningless.

In this section we present 
Maude \emph{functional modules}, which are
conditional membership equational theories
of the form 
$(\Sigma,M \cup E \cup B)$ 
specifying functional programs, where:
(i) $\Sigma$ is the \emph{signature} specifying: the types, here
called \emph{sorts}, the subtype, i.e., \emph{subsort}, inclusions,
and the function symbols and constants used in the theory;
(ii) $E$ is a collection of (possibly conditional) equations
which are used as left-to-right rewrite rules to evaluate terms; (iii)
$B$ is a collection of equational axioms,
such as associativity and/or commutativity and/or identity 
satisfied by some of 
the function symbols in $\Sigma$; such axioms are viewed as
\emph{structural axioms}, so that rewriting with the equations $E$ is
performed \emph{modulo} the axioms $B$; and (iv) $M$ is a collection
of (possibly conditional)  \emph{memberships}, which can
lower the sort of a term if a membership's condition is satisfied
(more on this below).
In Maude, the functional module defined by $(\Sigma,M \cup E \cup B)$ 
is declared within keywords \code{fmod}  and \code{endfm},
and is also given a name, say, \code{FOO}, so that its declaration
has the form: \code{fmod} \code{FOO} \code{is} 
$(\Sigma,M \cup E \cup B)$ \code{endfm}.  

Maude's syntax for $\Sigma$, $E$, and $M$ is self-explanatory: it is
in essence the ASCII version of the standard textbook notation (see below).
The structural axioms $B$ are declared together with
the function symbols satisfying such axioms.  Furthermore,
the syntax for the signature $\Sigma$, i.e., the names and
syntactic form of the sorts, constants, and function symbols for
$\Sigma$, is completely \emph{user-definable}.
For example, if
\code{Nat} is the name we have chosen for the sort of natural
numbers, we may choose any syntax we wish to declare
a natural number addition function in $\Sigma$,  and, furthermore,
we may declare such a function as enjoying some
structural axioms $B$.  Suppose that \code{0} and \code{1}
have been declared as constants with the declaration:

\begin{maude}
  ops 0 1 : -> Nat [ctor] . 
\end{maude}

\noindent where the \code{ctor} declaration makes it
clear that the constants \code{0} and \code{1} are
\emph{data constructors} that will not be evaluated away to other values
by some equations. Then, we can choose any syntax we wish for the addition function.
The less imaginative choice is to adopt a \emph{prefix syntax},
such as \code{+(1,0)}, or \code{plus(1,0)}.  But we may
wish to use the more readable \emph{infix syntax}, so as to be able to
write the term \code{1 + 0}.  Suppose we decide to give addition such an
infix syntax and to declare it as enjoying
the \emph{associativity} axiom $(x + y)+z = x +(y+z)$,
the \emph{commutativity} axiom $x + y= y +x$,
and the \emph{identity} axioms $x + 0 = x=0+x$.  Then, we can
give this declaration in Maude as follows:

\begin{maude}
  op _+_ : Nat Nat -> Nat [ctor assoc comm id: 0] . 
\end{maude}

\noindent where the two underbar symbols indicate where the first and
second argument of the addition function must be placed before
and after the \code{+} character.   As before, the   \code{ctor} declaration makes it
clear that in this representation the addition symbol is
a \emph{data constructor}, which will not be evaluated away, except if
one of its arguments is \code{0}, so that the identity axiom can be
used.  That is, in this representation the natural numbers are:
$0$, $1$, $1 + 1,\ldots, 1 + \stackrel{n}{\ldots} + 1, \dots$.
Instead, had we chosen
the prefix syntax, say, \code{plus}, we would have given the alternative
declaration:

\begin{maude}
  op plus : Nat Nat -> Nat [ctor assoc comm id: 0] . 
\end{maude}

\emph{Order-sorted equational logic} \cite{osa1,tarquinia} is
a very useful sublogic of membership equational logic.
An order-sorted equational theory is a membership equational
theory $(\Sigma,M \cup E \cup B)$ such that $M = \emptyset$, i.e., it has
the form $(\Sigma,E \cup B)$, where 
$\Sigma = ((S,<),F)$ consists of a partially ordered set $(S,\leq)$ of
sorts, where $\leq$ denotes \emph{subsort inclusion}, and where $F$ is
a set of function symbols and constants \emph{typed} with sorts in 
$S$.  Function symbols in $F$ can be
\emph{subsort overloaded} (also called \emph{subtype polymorphic}).
For example,  we may introduce a sort \code{NzNat} of
non-zero natural numbers as a subsort of \code{Nat},
declare instead \code{1}  as a constant
of sort \code{NzNat}, and add the additional declaration:

\begin{maude}
  op _+_ : NzNat NzNat -> NzNat [ctor assoc comm id: 0] .
\end{maude}

\noindent The only requirement is that, as done above,
all subsort polymorphic function declarations must
satisfy the \emph{same} structural axioms.

\begin{example}\label{ex:TERM}
Consider 
the following order-sorted specification
of terms in prefix form, with an arbitrary number of constant and
function symbols, as elements of a sort \code{Term} having two
subsorts,  \code{Var} of variables, and \code{NvTerm} of
non-variable terms. Assuming that we import Maude's built-in modules
\code{NAT} of natural numbers, with main sort \code{Nat},
and \code{QID} of quoted identifiers, with main sort \code{Qid},
both modules in \emph{protecting} mode (i.e., the sorts in \code{NAT} and
\code{QID} are not modified, but they are protected, in such an importation
\cite{maude-book}) we can then define such a data type of terms as follows\footnote{%
For a discussion on attributes not explained here, such as \texttt{prec},
please see \cite{maude-manual}.}:

\begin{maude}
fmod TERM is
  protecting NAT + QID .

  sort Variable .
  op x{_} : Nat -> Variable [ctor] .

  sorts Term NvTerm .
  subsort Qid < NvTerm < Term .
  subsort Variable < Term .
  op _[_] : Qid NeTermList -> NvTerm [ctor prec 40] .

  sort NeTermList .
  subsort Term < NeTermList .
  op _,_ : NeTermList NeTermList -> NeTermList [ctor assoc] .
endfm
\end{maude}

\noindent where, since no equations have been declared (only the
associativity structural axiom for non-empty lists of terms), all
operators are \emph{data constructors}.
Given a term $t$,  
the \emph{least sort}\footnote{Under a simple syntactic condition on $\Sigma$
checked by Maude, called \emph{preregularity} \cite{osa1}
 (more generally, preregularity
\emph{modulo} the structural axioms $B$ \cite{maude-book}),
any $\Sigma$-term $t$ always has a smallest possible typing with a sort
called its \emph{least sort}.} of $t$ is denoted $\mathit{ls}(t)$.
For example, assuming a countable set of variables, say,
$x_{1}, x_{2},\ldots, x_{n}\ldots$ and arbitrary names for 
constants and function symbols,
the term $f(g(x_{3},b,x_{1}),k(x_{2}))$ is here represented
as the term: \verb+'f['g[x{3},'b,x{1}],'k[x{2}]]+
of least sort  
\code{NvTerm}.
The term 
\code{'b} has least sort
\code{Qid}, and \verb+x{3}+ has least sort \code{Variable}.  But 
all these terms share the common supersort \code{Term}.
\end{example}

Note that any finite poset, and in particular the poset of sorts 
$(S,\leq)$, can be viewed as the reflexive-transitive closure of a directed acyclic
graph (DAG), and that the set of nodes of  such a DAG breaks into a set of \emph{connected
components}.  For example, in the signature $\Sigma = ((S,\leq),F)$
of the \code{TERM} module there are \emph{three} connected
components: (i) one involving the sort \code{Bool}, since the
Booleans are imported by \code{NAT}, (ii) another involving the sort
\code{Nat} and its subsort \code{NzNat}, and (iii) yet another
involving the sorts \code{Qid}, \code{Var}, \code{NvTerm}, and
\code{Term}. Maude automatically adds a new so-called \emph{kind}
supersort  at the top of each connected component in the poset $(S,\leq)$
declared by the user, where kinds are indicated with a bracket
notation.  For this example, Maude
will add kinds \code{[Bool]}, \code{[Nat]} and \code{[Term]}
at the top of each of these three components.  Furthermore, for each
function symbol, say $f: s_{1} \ldots s_{n} \rightarrow s$, in $\Sigma$
a new subsort-overloaded symbol
$f: [s_{1}] \ldots [s_{n}] \rightarrow [s]$ is also added by Maude at the kind
level.  Intuitively, terms whose least sort is a kind are viewed as
\emph{error terms}.  For example, the least sort of the term
\verb+`f[`a][`g[`c,x{2}],'b]+  is the kind \code{[Term]}.
Since \verb+`f[`a]+ is \emph{not} a quoted identifier (sort \code{Qid}),
and therefore cannot be used as a function symbol,
the operator declaration 
\verb+op _[_] : [Qid] [NeTermList] -> [NvTerm]+
is used to construct the term.
Note that the kind of sort \code{Term}, denoted by \code{[Term]},
coincides with the kind of, e.g., sort \code{TermList}, denoted by \code{[TermList]},
since \code{Term} is a subsort of \code{TermList}.
 This
is very useful to \emph{give functional expressions the benefit of the
doubt}, because at parse time only partial type information
may be available, but as a computation progresses some
typing problems may go away.  For example, in a data type
\code{RAT} of rational numbers an expression like
\verb+3 / (2 - 7)+ can only be parsed with least sort 
\code{[Rat]}, but will happily evaluate to \verb+- 3 / 5+
with least sort \code{NzRat}.  Instead, the
evaluation of the term \verb?3 / (7 - (4 + 3))?
will yield the error term
\verb+3 / 0+, whose least sort is \code{[Rat]}.

\subsection{Predicate Subtyping with Membership Predicates}

The full generality of Maude functional modules as
membership equational theories can be illustrated by means of
the following module specifying an algebraic data type of
finite partial functions on the natural numbers, that is,
\emph{arrays} whose values are natural numbers.
A finite partial function is just a finite relation
that is \emph{single-valued} on its
domain of definition.  For example, $\{(0,2),(3,0),(4,1)\}$ is a
finite partial function, but
$\{(0,2),(3,0),(3,1),(4,1)\}$ is \emph{not} a finite partial function, because
$3$ is mapped to both $0$ and $1$.
Using order-sorted algebra alone, we cannot specify
an algebraic data type having a subsort  \verb+PFun < Rel+  exactly
characterizing those finite relations that are finite partial functions; but
we can do so using membership equational logic.

\begin{example}\label{ex:pfun}
We define in Figure~\ref{fig:PFUN1} a functional module \code{PFUN} of finite 
partial functions on the natural numbers.

\begin{figure}[htb]
\begin{maude}
fmod PFUN is
  protecting NAT .
  sorts Pair Magma PFun Rel Nat? .
  subsorts Pair < Magma .
  subsorts PFun < Rel .
  subsort Nat < Nat? .
  op undef : -> Nat? [ctor] .

  vars I J K : Nat .
  var M : Magma .
  var F : PFun .
  var R : Rel .

  op [_,_] : Nat Nat -> Pair [ctor] .
  op null : -> Magma [ctor] .          *** empty magma
  op _,_ : Magma Magma -> Magma [ctor assoc comm id: null] .
  op {_} : Magma -> Rel [ctor] .
  eq [I,K], [I,K] = [I,K] .            *** idempotency
  mb {null} : PFun .
 cmb {[I, K], M} : PFun if def(I, {M}) = false /\ {M} : PFun .

  op def : Nat Rel -> Bool .           *** is key defined in relation?
  eq def(I, {null}) = false .
  eq def(I, {[J, K], M}) = if I == J then true else def(I, {M}) fi .

  op _[_] : PFun Nat -> Nat? .         *** partial function application
  eq {null}[K] = undef .
 ceq {[I, K], M}[J] = if I == J then K else {M}[J] fi if {[I, K], M} : PFun .
endfm
\end{maude}
\caption{\label{fig:PFUN1}\code{PFUN} module}
\end{figure}

%
%
%
%
%
%
\end{example}

A few things are worth mentioning about this example. 
First of all, an if-then-else operator\footnote{The \texttt{if\_then\_else\_fi} operator should not be confused with the condition of conditional equations, introduced by \texttt{if}.} with ``mix-fix'' syntax
\verb+if_then_else_fi+ and with the obvious equational definition
is added automatically by Maude to
any module importing the \code{BOOL} module, which
is always imported by default unless the user indicates otherwise
\cite{maude-book}.  Second, a built-in equality predicate
\verb+_==_+ is also automatically added by Maude for each connected
component.  However, neither of these built-in operators are really needed:
the user can easily define his/her own if-then-else,
as well as an equality predicate for natural numbers, or
for many other data types (see \cite{DBLP:journals/scp/GutierrezMR15}).
These two built-in operators have been used in the
definitions of the \code{def} predicate and of partial function application.
Third, because of the idempotency equation, all terms of sort 
\code{PFun} in normal form are finite \emph{sets} of pairs,
and therefore partial functions in the mathematical sense.

The key new feature used here is the use of \emph{memberships}
(introduced with keywords \code{mb} and \code{cmb} for, respectively, 
unconditional and conditional membership axioms) defining the sort
\code{PFun} of finite partial functions as a subsort of  the sort of
finite relations \texttt{Rel}.  We follow exactly a typewriter version
of the set-theoretic description of finite relations.  To do so,
we distinguish between the finite \emph{relation} itself, e.g.,
$\{(0,2),(3,0),(3,1),(4,1)\}$  and its underlying \emph{multiset} of pairs
$(0,2),(3,0),(3,1),(4,1)$, which we call a \emph{magma}.  Union of
magmas is defined by the
associative-commutative operator \verb+_,_+ with identity the empty
magma \texttt{null}.
The subsort \verb+PFun < Rel+ 
is defined by memberships distinguishing two cases.  A partial function \code{F} is either:
(i)~the empty relation \verb+{null}+, or 
(ii)~a relation \code{\{M\}} that is a partial function and to which a 
new pair \code{[I,K]} has been added, provided \code{\{M\}} is undefined 
for the input \code{I}.  Of course, many other auxiliary functions,
such as, for example, array updating, relation union and intersection,
relation and partial function composition,
(multi-valued) relation application, and so on, could easily be added
to this module.  Rather than doing so
ourselves, we encourage the reader, particularly if not yet acquainted
with Maude, to do it him/herself in order to have some fun playing
with Maude and get a taste for how one can specify finitary
set theory operations in Maude just as one would like to do it out of a textbook.
Also, for a taste of how one can specify finitary set theory
(the so-called hereditarily finite sets) in
a single Maude functional module we refer the reader to \cite{duran-meseguer-rocha-JLAMP}.

Membership predicates are unary predicates in
postfix notation $\_\!:\!s$, where $s \in S$ is a sort.
Applied to a term $t$ whose least sort belongs to the connected component
of $s$ (and could even be its kind sort), the predicate states that $t$
has sort $s$.   A single such membership predicate
is called an \emph{unconditional membership} and is introduced
with the keyword \code{mb}.  In general, both equations and
memberships can be \emph{conditional}, can be labeled, and have, respectively, the general form:
\[\texttt{ceq}\ \texttt{[}l\texttt{]:}\ t =t' \;\; \mathtt{if} \;\; u_{1} = v_{1}\, \wedge \, \ldots \, \wedge \, u_{n} = v_{n} \wedge w_1:s_1 \wedge \ldots \wedge w_j : s_j \ . \]
\[\texttt{cmb}\ \texttt{[}l\texttt{]:}\ t\! :\! s \;\; \mathtt{if} \;\; u_{1} = v_{1}\, \wedge \, \ldots \, \wedge \, u_{n} = v_{n} \wedge w_1:s_1 \wedge \ldots \wedge w_j : s_j \ . \]
That is, both equations and memberships may appear in their
conditions.  Using ASCII symbols, the conjunction symbol $\wedge$ is rendered
in Maude as \verb+/\+.  Since an equation $t=t'$ will be applied as
a left-to-right rewrite rule $t \rightarrow t'$ to simplify terms,
sort information should \emph{increase} as such simplification
proceeds.  This is captured by the requirement that all
equations $t=t'$ (conditional or not) in a functional module 
 should be \emph{sort-decreasing}.  That is, for any substitution
 $\theta$ we should have $\mathit{ls}(t \theta) \geq
 \mathit{ls}(t' \theta)$, where $t \theta$ and $t' \theta$
denote the respective instantiations of $t$ and $t'$ by $\theta$.
In the most common cases, all the variables appearing
in such formulas also appear in the term $t$ at the left of 
the equation $ t =t'$ or the membership $ t\! : \!s$.  Furthermore, the
equations and memberships in a condition can appear in different orders.
However,  for greater expressiveness Maude allows conditional equations
and memberships whose conditions \emph{can have extra variables}
that are \emph{incrementally instantiated} by matching, provided they
obey the syntactic requirements explained in
\cite{maude-manual,maude-book}.  We explain the incremental
evaluation of conditions by means of
Example \ref{ex:LP-system}
in Section \ref{sec:system}.

In any confluent and operationally terminating \cite{DLMMU-hosc}
functional module (more on this in Section \ref{rewriterel-sect}), 
any term can be evaluated to its unique \emph{normal form}
having a least possible sort by applying to it both the module's
equations as left-to-right rewrite rules, and the memberships to lower
its sort, where equations and memberships are applied \emph{modulo}
the axioms $B$.  For example, the idempotency equation 
\code{[N,M],[N,M] = [N,M]} can be applied
modulo the associativity-commutativity axioms for \verb+_,_+
 to simplify the term \verb+[1,2],[3,7],[1,2]+
to the term \verb+[1,2],[3,7]+,
even though
the two instances of \verb+[1,2]+ are not contiguous.
This evaluation to normal form is performed with the \code{reduce} command
(which can be abbreviated to \code{red}). For example, 
in the above functional module 
in Example~\ref{ex:pfun} above
we can perform the following evaluations:

\begin{maude}
reduce in PFUN : {[1,2],[1,2],[3,7],[5,17],[3,7]} .
result PFun: {[1,2],[3,7],[5,17]}

reduce in PFUN : {[1,2],[3,7],[5,17]}[3] .
result NzNat: 7
\end{maude}

\subsection{Equational Simplification Modulo Axioms} 
\label{rewriterel-sect}

To further explain how Maude's \texttt{reduce} command, illustrated in
the \texttt{PFUN} module above, is performed in general, we 
restrict ourselves to the simpler case of an \emph{order-sorted
  unconditional} functional module, i.e., a module of the form
\texttt{fmod FOO is} $(\Sigma,E \cup B)$ \texttt{endfm}, where the
equations $E$ are unconditional and there are no memberships $M$.
Full treatments for the general case where the equations $E$ may be
conditional or the set $M$ of memberships may be non-empty can be
found in
\cite{lucas-meseguer-normal-th-JLAMP,bouhoula-jouannaud-meseguer}.
Consider, for example, the following functional module
\texttt{AC-NAT}, which defines addition and multiplication of natural
numbers with constants \texttt{0} and \texttt{1} and \texttt{+}
declared as a data constructor modulo associativity and commutativity:

\begin{maude}
fmod AC-NAT is
  sorts NzNat Nat .
  subsorts NzNat < Nat .
  op 0 : -> Nat [ctor] .
  op 1 : -> NzNat [ctor] .
  op _+_ : Nat Nat -> Nat [assoc comm] . 
  op _+_ : NzNat NzNat -> NzNat [ctor assoc comm] .
  op _*_ : Nat Nat -> Nat [assoc comm] . 
  op _*_ : NzNat NzNat -> NzNat [assoc comm] .

  vars N M K : Nat .

  eq N + 0 = N .
  eq N * 0 = 0 .
  eq N * 1 = N .
  eq N * (M + K) = (N * M) + (N * K) .
endfm
\end{maude}

\noindent Note the somewhat subtle fact that \texttt{+}
is only declared to be a data constructor for non-zero
natural numbers of sort \texttt{NzNat}.  This is
because, thanks to the equation \texttt{N + 0 = N},
any ground (i.e., without variables) natural number expression
fully simplified by the above equations is either \texttt{0} or \texttt{1},
which are both constructor constants, or a number of the
form \texttt{1 +} $\cdots$ \texttt{+ 1}, which has sort 
\texttt{NzNat}.

Intuitively, equational simplification of an arithmetic expression,
say $t$, with the above four equations means that we
\emph{simplify} $t$ as much as possible by applying the
equations as simplification rules from left to right, but
\emph{modulo} the declared axioms $B$.  In this example,
such axioms $B$ are the associativity and commutativity of
both \texttt{+} and \texttt{*}, which we abbreviate to just
AC.  Of course, we intuitively expect that,
after applying the \texttt{reduce} command, any ground
arithmetic expression $t$ will be fully simplified by the
above equations to a \emph{data} expression
(i.e., an expression involving only constructor symbols
declared with the \texttt{ctor} keyword)
of the form \texttt{0}, or \texttt{1}, or \texttt{1 +} $\cdots$ \texttt{+ 1},
called the \emph{normal form} of $t$, and that
such a normal form will be \emph{unique} up to AC
equality.

This subsection (which can be safely skipped by readers
familiar with term rewriting modulo axioms)
makes all these intuitions precise, yet in an informal style; 
for a fully formal presentation see \cite{dershowitz-jouannaud}.
Recall that, given an equation
$t=t'$, its orientation as a left-to-right simplification
rule, denoted $t \rightarrow t'$, is called
a \emph{rewrite rule}. Therefore, 
given a set $E$ of equations, we get the
set of rewrite rules
$\vec{E} = \{ t \rightarrow t' \mid (t=t') \in E\}$.
For executability purposes in any rule $t \rightarrow t'$
\emph{all variables in $t'$ should  also appear in $t$}.
As already mentioned, for conditional equations this requirement can
be relaxed under suitable conditions.
We need to make precise the intuitive idea of 
``applying the simplification rules'' $\vec{E}$
to an expression
(also called a  \emph{term})
 $t$ modulo axioms $B$.  To do this, we first need
to explain a few simple notions, which are
best understood when a term $t$ is represented 
as its parse \emph{tree}.   In such a tree, a
path from its root to some subexpression (i.e., subtree) 
can be uniquely
described by a sequence of natural
numbers.  Consider, for example, the expression
$(1 + 1)+((1 + 0)+1)$, and the subexpression $1+0$.
The path from the root of the tree to the subtree
$1+0$ can be uniquely
described by the sequence $2 \, 1$.  This is
because $2$ instructs us to select the \emph{second argument} 
$(1 + 0)+1$ of the overall expression, and then $1$
instructs us to select the \emph{first argument} $1+0$
of $(1 + 0)+1$.  Of course, the empty string
$\epsilon$ describes the empty path from the root to
the entire expression $(1 + 1)+((1 + 0)+1)$.
A symbol $f$ in a general signature $\Sigma$
can have 0 arguments (a constant), 1 argument (unary symbol),
2 (binary symbol), or an arbitrary number $n$ of arguments
($n$-ary symbol).  Therefore, a path $p$ in a $\Sigma$-expression
$t$ will in general be a sequence of natural numbers.
We call such a path $p$ a \emph{position} in $t$, since
it uniquely identifies a position in the tree, and
therefore the subexpression (subtree) at that position.
We then denote the subexpression of $t$ at position
$p$ by $t|_{p}$.
For example, $((1 + 1)+((1 + 0)+1))|_{2 \, 1} = 1+0$.
A position $p$ also allows us to ``perform surgery'' on
a term/tree $t$ by \emph{replacing} the subterm/subtree
$t|_{p}$ by another subterm/subtree $v$ at the exact same
position $p$.  We then denote by $t[v]_{p}$ the
term obtained after replacing $t|_{p}$ by $v$
at position $p$.  For example,
if $t =(1 + 1)+((1 + 0)+1)$, then
$t[0]_{2 \, 1}=(1 + 1)+(0+1)$.

We next need to explain the notion of a substitution $\sigma$,
and its application $t \sigma$ to a term $t$ (which need not be
ground, i.e., may have variables).  A \emph{substitution} $\sigma$
is a finite mapping, i.e., a finite set of pairs of the form
$\sigma = \{ (x_{1},u_{1}), \ldots
(x_{n},u_{n})\}$ where the
$x_{1},\ldots,x_{n}$ are different variables, and
the $u_{1},\ldots,u_{n}$ are $\Sigma$-terms in
some signature $\Sigma$.  Furthermore,
$\sigma$ must be \emph{sort-preserving},
i.e., if $x$ has sort $s$, then the least sort of
$\sigma(x)$ must be less that or equal to $s$.
Then, given a $\Sigma$-term $t$, the \emph{application}
$t \sigma$ of substitution $\sigma$ to $t$
is the term obtained by replacing each variable $x$
appearing in both $t$ and in the domain of $\sigma$
by $\sigma(x)$ at all positions where $x$ occurs in $t$.
For example, if $t = x + (x+y)$
and $\sigma = \{(x,1 +z),(y,z+z')\}$,
then $t \sigma = (1+z)+((1 + z)+(z + z'))$.

Let $R$ be a set of rewrite rules between $\Sigma$-terms,
and $B$ a set of  equational axioms between some operations in
$\Sigma$.  By definition, the relation $u \rightarrow_{R,B} v$
holds between two $\Sigma$-terms $u$ and $v$ if and
only if there is a position $p$ in $u$, a rule $l \rightarrow r$
in $R$, and a substitution $\sigma$ such that:
(i) $u|_{p} =_{B} l \sigma$, and (ii) $v = u[r \sigma]_{p}$,
where $=_{B}$ denotes provable equality with the axioms $B$.
This allows us to precisely capture the notion
of equational simplification modulo axioms $B$.
For example, for $\vec{E}$ the rules associated
to the functional module \texttt{AC-NAT}, 
the term $1 + (0 + 1)$ can be simplified to the term $1+1$
modulo AC,
because $1 + (0 + 1) \rightarrow_{\vec{E},\mathit{AC}} 1+1$
in \emph{two} different ways with the rule $N+0 \rightarrow N$,
namely: (i) at position 2 with substitution $\sigma_{1}=\{(N,1)\}$,
since  $(N+0) \sigma_{1} = 1+0 =_{\mathit{AC}} 0 +1$,
and $1 + (0 + 1)[N \sigma_{1}]_{2}=1 + (0 + 1)[1]_{2}=1 +1$;
and (ii) at position $\epsilon$
with substitution $\sigma_{2}=\{(N,1+1)\}$,
since  $(N+0) \sigma_{2} = (1 + 1)+0 =_{\mathit{AC}} 1 + (0 + 1)$,
and 
$1 + (0 + 1)[N \sigma_{2}]_{\epsilon}=1 + (0 + 1)[1 + 1]_{\epsilon}=1 +
1$.

We can now  explain  the \texttt{reduce}
command.  The simplification steps performed
by \texttt{reduce} can be made explicit by giving to
Maude the command:

\begin{maude}
set trace on .
\end{maude}

With tracing on, the equation, substitution, and the lefthand and
righthand subterms are displayed for each rewrite step.
Thus, for the reduction of the term $(1 + (0 + 1))+(0 * 1)$
in \texttt{AC-NAT} we obtain the following
trace:

\begin{maude}
Maude> reduce (1 + (0 + 1)) + (0 * 1) .
reduce in AC-NAT : (1 + 0 + 1) + 0 * 1 .
*********** equation
eq 0 + N = N .
N --> 1
0 + 1
--->
1
*********** equation
eq 0 * N = 0 .
N --> 1
0 * 1
--->
0
*********** equation
eq 0 + N = N .
N --> 1 + 1
0 + 1 + 1
--->
1 + 1
rewrites: 3 in 0ms cpu (0ms real) (12500 rewrites/second)
result NzNat: 1 + 1
\end{maude}

This trace describes the following sequence of rewrites
modulo AC:
\[ (1 + (0 + 1)) + (0 * 1)
\rightarrow_{\vec{E},\mathit{AC}}
(1 + 1) + (0 * 1)
\rightarrow_{\vec{E},\mathit{AC}}
(1 + 1) + 0
\rightarrow_{\vec{E},\mathit{AC}}
1 + 1
\]
where the first rewrite happens at position $1 \, 2$,
the second at position 2, and the third at position $\epsilon$.

As already pointed out
when discussing the \texttt{PFUN} module,
to have good executability properties
any functional module
\texttt{fmod FOO is} $(\Sigma,E \cup B)$ \texttt{endfm}
should be such that its rules
$\vec{E}$ are: (i) \emph{sort-decreasing},
(ii) \emph{operationally terminating} modulo $B$,
and (iii) \emph{confluent} modulo $B$.
Conditions (i)--(iii) can be abbreviated by
a single notion: the rules $\vec{E}$ are then called
\emph{convergent}\footnote{Properly speaking, convergence requires
a fourth condition: (iv) \emph{strict coherence}
\cite{meseguer-strict-coherence}:
if $u\rightarrow_{\vec{E},B}v$ and $u=_{B} u'$, there exists $v'=_{B} v$
with $u'\rightarrow_{\vec{E},B}v'$.  But strict coherence is
automatically guaranteed by Maude by adding ``extension rules''
  (see Section 4.8 in \cite{maude-book}).}
 modulo $B$.
Let us make this more precise.
Sort decreasingness (condition (i)) 
 has already been explained when discussing \texttt{PFUN}.
In the
order-sorted unconditional case, 
condition (ii) just means that the rules $\vec{E}$ are
\emph{terminating modulo AC}, i.e.,
no infinite rewrite sequences
\[t_{0}\rightarrow_{\vec{E},\mathit{AC}} t_{1} \rightarrow_{\vec{E},\mathit{AC}}
\cdots \rightarrow_{\vec{E},\mathit{AC}} t_{n}\rightarrow_{\vec{E},\mathit{AC}} t_{n+1} \rightarrow_{\vec{E},\mathit{AC}} \cdots
\]
are possible. In the context of (ii),  condition
 (iii)  of \emph{confluent modulo AC}
just means
that for any term $t$ all terms $t'$ that are
\emph{normal forms} of $t$ by rewriting with $\vec{E}$
modulo AC, i.e., such that
$t\rightarrow^{*}_{\vec{E},\mathit{AC}} t'$
(where $\rightarrow^{*}_{\vec{E},\mathit{AC}}$ denotes 
the reflexive-transitive
closure of $\rightarrow_{\vec{E},\mathit{AC}}$)
and $t'$ cannot be further rewritten with $\vec{E}$
modulo AC, are AC-equal to each other,
i.e., are \emph{unique} up to AC-equality.
This ensures that term simplification always yields
a \emph{unique result}.  The \texttt{reduce}
command simplifies each term $t$ to its
unique normal form.

The module  \texttt{AC-NAT} is convergent; it also satisfies the
property of being \emph{sufficiently complete}, which
means that the normal form of any ground term $t$
is a constructor term according to the \texttt{ctor} declarations
in the module, i.e., it is
either \texttt{0}, \texttt{1}, or a term of the 
form \texttt{1 +} $\cdots$ \texttt{+ 1}.  Both convergence
and sufficient completeness can be checked by
tools in Maude's Formal Environment 
\cite{tool-interoperability-mfe}.

\subsection{Initial Algebra Semantics}

What is the
\emph{mathematical} meaning of a Maude functional module, say,
\code{fmod} \code{FOO} \code{is} 
$(\Sigma,M \cup E \cup B)$ \code{endfm}?  That is, what does
such a module declaration \emph{denote}?  The answer is simple:
Maude has an \emph{initial algebra semantics} for such
modules, so that what \code{FOO} denotes is the initial
algebra \cite{tarquinia}
$T_{\Sigma/M \cup E \cup B}$ of the theory $(\Sigma,M \cup E \cup B)$.
There are two possible descriptions of $T_{\Sigma/M \cup E \cup B}$,
one more abstract, and another very concrete.  In the
abstract description an element
$[t]\in T_{\Sigma/M \cup E \cup B}$ is the $=_{M \cup E \cup B}$-equivalence class of  a
ground $\Sigma$-term $t$ (i.e., $t$ has no variables), where
$=_{M \cup E \cup B}$ is the \emph{provable equality} equivalence relation in 
the theory $(\Sigma,M \cup E \cup B)$ \cite{tarquinia}.  
Under the already-mentioned
executability condition that the rules $\vec{E}$ 
and the memberships $M$ 
are \emph{convergent} modulo $B$,
the more concrete and informative
description is given by the isomorphic algebra
$C_{\Sigma/M \cup E, B} \cong T_{\Sigma/M \cup E \cup B}$,
called the \emph{canonical term algebra} of $(\Sigma,M \cup E \cup
B)$, whose elements
$[u] \in C_{\Sigma/M \cup E, B}$ are $=_{B}$-equivalence classes
of ground terms $u$ that are in normal form by the equations $E$
and the memberships $M$ modulo $B$. 
 $C_{\Sigma/M \cup E, B}$ provides
the most concrete possible semantics for
\code{FOO}, since it is just the semantics of Maude's
\code{reduce} command in the following sense: 
a ground term $t$ that is evaluated  to a term $u$ by Maude's 
\code{reduce} command has as its \emph{value} the
$B$-equivalence class $[u] \in C_{\Sigma/M \cup E, B}$.  
Furthermore, thanks to the
Church-Rosser Theorem for membership equational logic
\cite{bouhoula-jouannaud-meseguer}, what the isomorphism
$C_{\Sigma/M \cup E, B} \cong T_{\Sigma/M \cup E \cup B}$ ensures is the \emph{full
agreement} between the 
\emph{mathematical semantics} provided by $T_{\Sigma/M \cup E \cup B}$ 
and the rewriting-based \emph{operational semantics}
(for details see \cite{bouhoula-jouannaud-meseguer}, and for the
conditional order-sorted subcase modulo $B$ see \cite{lucas-meseguer-normal-th-JLAMP}), whose algebra
of normal forms is precisely $C_{\Sigma/M \cup E, B}$.
For example, for $(\Sigma,M \cup E \cup B)$ our specification of 
\texttt{PFUN}, $C_{\Sigma/M \cup E, B}$ is just the algebraic data
type giving to finite relations and finite partial functions on the
natural numbers \emph{the exact same mathematical meaning} 
as in set theory.
Likewise, 
 the elements of $C_{\Sigma/E, B}$ for the
 \texttt{AC-NAT} module
are the AC-equivalence classes
of constructor terms of the form 
\texttt{0}, or \texttt{1}, or 
\texttt{1 +} $\cdots$ \texttt{+ 1}; and
addition and multiplication are interpreted in $C_{\Sigma/E, B}$
as natural number addition and multiplication 
in this representation of the natural numbers.

\subsection{Theories, Views and Parameterized Functional Modules}

Maude,
like its OBJ3 predecessor \cite{intro-obj}, 
supports a very expressive form of \emph{parametric polymorphism}
\cite{strachey-concepts-hosc} by means of its parameterized modules.
The extra expressiveness has to do with the fact that
\emph{parameters} are \emph{not} just parametric \emph{types},
but are instead specified by \emph{parameter theories}.
That is, not only types (sorts) can be parametric: constants and function symbols
can also be parametric, and, furthermore, parameter theories can
impose \emph{semantic requirements}, in the form of logical axioms,
that must be satisfied by any instantiation of a parameter  theory with
actual parameters to be correct.  Roughly speaking,\footnote{\label{init-ctr-th}In fact,
parameter theories may also contain \emph{initiality constraints} 
in the sense of, e.g., \cite{institutions,DBLP:journals/tcs/DuranM03},
which can impose the requirement that some sorts and functions
must be interpreted as the initial model of an imported subtheory.
For example, a theory $T$ may import the theory \code{NAT}
of natural numbers in \code{protecting} mode, so that only models
where \code{NAT} is interpreted as the natural numbers are accepted.}
a \emph{parameter theory} called, say, \code{FOO},
is a membership equational 
theory $(\Sigma,M \cup E \cup B)$,
which is declared in Maude with syntax:
\code{fth} \code{FOO} \code{is} 
$(\Sigma,M \cup E \cup B)$ \code{endfth}.  

What is the mathematical semantics of such a functional theory \code{FOO}?
Unlike the case of a  functional module, whose semantics is
the initial algebra $T_{\Sigma/M \cup E \cup B}$, the semantics
of a functional theory is the \emph{class} of all $(\Sigma,M \cup E
\cup B)$-algebras, denoted ${\bf Alg}_{(\Sigma,M \cup E\cup B)}$.
That is, functional theories have a ``loose semantics'' that specifies
all the possible instantiations of the parameter theory
$(\Sigma,M \cup E \cup B)$ by an algebra
$A \in {\bf Alg}_{(\Sigma,M \cup E\cup B)}$ as an actual parameter.

Let us illustrate the extra power of parameterized theories, as
opposed to just parameterized sorts, by describing
some examples at a high level (further details can be found in \cite{maude-book}). 
The case of having just a parameterized sort is handled by the
trivial parameter theory \code{TRIV}:

\begin{maude}
fth TRIV is
  sort Elt .
endfth
\end{maude}

\noindent Note that the class of algebras of this theory,
${\bf Alg}_{\texttt{TRIV}} = {\bf Set}$,  
is precisely \emph{the class} ${\bf Set}$ \emph{of all sets}.
Therefore, \code{TRIV} is exactly the theory of a \emph{parametric
  type} in the standard sense.  For example, the parameterized
functional module of lists \verb+LIST{X :: TRIV}+ can be
instantiated by any set, say $A$, as actual parameter to obtain the
data type of lists with elements in $A$.
Let us consider two other examples where the parameter theories
are nontrivial.  The functional module
\verb+SORTING{X :: TOSET}+ provides a parameterized functional
module to sort lists of elements for any totally ordered set 
$(A,\leq)$, that is, for any
$(A,\leq) \in {\bf Alg}_{\texttt{TOSET}}$, where \code{TOSET}
is the functional theory of totally ordered sets.\footnote{The fact
  that, as explained in Footnote \ref{init-ctr-th}, a functional theory can
  include initiality constraints is useful in this case, since
  \code{TOSET} can be easily defined by importing the functional
  module \code{BOOL} in \code{protecting} mode (see \cite{maude-book}).}
Yet a third example is the functional module 
\verb+POLY{R :: RING, X :: TRIV}+ of \emph{polynomials},
which has \emph{two} parameter theories.  The first is the
theory \code{RING} of commutative rings, so that its actual
parameters are commutative rings $(R,-,+,*,0,1) \in {\bf Alg}_{\texttt{RING}}$
providing the ring of coefficients used in the polynomials.
The actual parameters for the second theory \code{TRIV}
are precisely sets $X \in {\bf Set}$ providing the set of
\emph{variables} used in the polynomial expressions.
Of course, for parameter instantiations to be  \emph{correct},
all the \emph{axioms} in theories such as \code{TOSET}
or \code{RING} must be \emph{satisfied} by their actual parameters,
$(A,\leq)$ or $(R,-,+,*,0,1)$.
Maude does not check the semantic correctness of
instantiations.  However, tools like the Maude's Inductive Theorem Prover
(ITP) \cite{itp-manual} can be used for this purpose.

But how are parameter theories \emph{instantiated} in Maude?
By \emph{theory interpretations}. Suppose that we 
want to instantiate the parameterized module 
 \verb+POLY{R :: RING, X :: TRIV}+ 
to polynomials with rational coefficients and with  quoted
identifiers as variables.  We can, for example, use Maude's modules 
\code{RAT} of rational numbers and \code{QID} of quoted identifiers
 in Maude's standard prelude as actual parameters.  But any functional
 module \emph{is} a theory, namely, a membership equational theory
\emph{with the initiality constraint} that a model belongs to its
class of models iff it is an initial algebra for the theory.
This means that not only the axioms explicitly mentioned in the
functional module, but also all its \emph{inductive consequences}
are true in such models and therefore valid under the initiality
constraint.  So we just need two theory interpretations:
$\texttt{ring2RAT} : \texttt{RING} \rightarrow \texttt{RAT}$
to get the actual ring of coefficients, and
$\texttt{Qid} : \texttt{TRIV} \rightarrow \texttt{QID}$
to select the sort \code{Qid} in module \code{QID} as the set of
variables. What is a theory interpretation?
Given two membership equational theories, say
$(\Sigma,M \cup E \cup B)$ and $(\Sigma',M' \cup E' \cup
B')$,  a \emph{theory
  interpretation} (called a \emph{view} in Maude)
$V: (\Sigma,M \cup E \cup B) \rightarrow (\Sigma',M' \cup E' \cup
B')$ is a \emph{signature map} $V: \Sigma \rightarrow \Sigma'$
that preserves all the axioms $M \cup E \cup B$, in the sense that the
translated axioms $V(M) \cup V(E) \cup V(B)$ 
 are \emph{logical consequences} of the theory
$(\Sigma',M' \cup E' \cup
B')$.  What is a signature map $V: \Sigma \rightarrow
\Sigma'$?  It is a mapping of sorts and function symbols such that:
(i) If $(S,\leq)$ and $(S',\leq')$ are the posets of sorts for
$\Sigma$ and $\Sigma'$, then $V$ is a \emph{monotonic function} on
sorts, and (ii) each constant $a$ in $\Sigma$
of sort $s$ in $S$ is mapped to a ground
$\Sigma'$-term $V(a)$ 
with $ls(V(a)) \leq V(s)$, and each function symbol $f: s_{1} \ldots s_{n}
\rightarrow s$ in $\Sigma$ is mapped to a $\Sigma'$-\emph{term}
$V(f) = t'$ with  $\mathit{ls}(t') \leq V(s')$ and with variables among the
$x_{1}\!:\!V(s_{1}) , \ldots , x_{n}\!:\!V(s_{n})$ in such a way that
$V$ preserves subtype polymorphism.  In Maude such theory
interpretations are defined with syntax of the form (see
\cite{maude-book} for more details):

\begin{maude}
view V from T to T' is 
  sort S1 to S'1 . 
  ... 
  op f(X1:S1,...,Xn:Sn) to term t'(X1:V(S1),...,Xn:V(Sn)) . 
  ... 
endv 
\end{maude}

A parameterized functional module 
\verb+M{X1 :: T1, ... , Xm :: Tm}+
can be \emph{instantiated} by \emph{replacing} its
formal parameter theories \code{T1}, $\ldots,$ \code{Tm}
by corresponding views \code{V1}, $\ldots,$ \code{Vm}
from \code{T1}, $\ldots,$ \code{Tm} to \code{T'1}, $\ldots,$
\code{T'm}, where the \code{T'1}, $\ldots,$ \code{T'm} need not
be all different. In this way, we get the instance
\verb+M{V1, ... ,Vm}+.
 For example, polynomials with rational coefficients
and quoted identifiers as variables are defined as follows:

\begin{maude}
fmod RAT-POLY is protecting POLY{Ring2RAT,Qid} . endfm
\end{maude}

We say that a parameterized module 
\verb+M{X1 :: T1, ... , Xm :: Tm}+ is
\emph{fully instantiated} by the views \code{V1}, $\ldots,$
\code{Vm} if their target theories \code{T'1}, $\ldots,$
\code{T'm}  are all (unparameterized) functional modules.
But this is not the only possibility: a module may be instantiated in
an \emph{incremental} way. For example, we can define
a view:

\begin{maude}
view triv2TOSET from TRIV to TOSET is 
  sort Elt to Elt . 
endv 
\end{maude}

\noindent to instantiate the list module \verb+LIST{X :: TRIV}+
to the module \verb+LIST{triv2TOSET}+ which is still parameterized,
but now by the \code{TOSET} theory; and we can then use
\verb+LIST{triv2TOSET}+ as part of the definition of a 
\verb+SORTING{X :: TOSET}+ module.

Let us see an example  illustrating all
the ideas discussed so far. 

\begin{example}\label{ex:pfunXY}
A parameterized module for finite partial functions generalizing 
the \code{PFUN} module of Example~\ref{ex:pfun} can be found in Figure~\ref{fig:PFUN2}.
This 
module is such a straightforward
\emph{generalization} of the \code{PFUN} module 
of Example~\ref{ex:pfun}
that not much needs to
be said about it, except, perhaps, for some syntax details.  First of
all, note that \verb+PFUN{X :: TRIV, Y :: TRIV}+ has \emph{two}
parameters, both with parameter theory \code{TRIV}, but of course
these two occurrences of \code{TRIV} are different and can be instantiated
quite differently.  This means that two different \emph{copies} of
\code{TRIV} must be used to avoid a confusion of sorts.  In the
first copy, the sort \code{Elt} in \code{TRIV}
is automatically renamed to \verb+X$Elt+,
and in the second copy to \verb+Y$Elt+.
Furthermore, the sorts \verb+Pair{X,Y}+, \verb+Magma{X,Y}+, \verb+PFun{X,Y}+  
and \verb+Rel{X,Y}+ are now \emph{parametric} on both \code{X} and \code{Y}.
Finally, the role 
formerly played by the supersort \code{Nat < Nat?}, where the
\code{undef} constant was added in \code{PFUN}, is now played
by the supersort \verb+Y$Elt < ?{Y}+, which is of course parametric
on \code{Y}.

\begin{figure}[htb]
\begin{maude}
fmod PFUN{X :: TRIV, Y :: TRIV} is
  sorts Pair{X,Y} Magma{X,Y} PFun{X,Y} Rel{X,Y} ?{Y} .
  subsorts Pair{X,Y} < Magma{X,Y} .
  subsorts PFun{X,Y} < Rel{X,Y} .
  subsort Y$Elt < ?{Y} .
  op undef : -> ?{Y} [ctor] .
  vars I J : X$Elt .
  var  K : Y$Elt .
  var M : Magma{X,Y} .
  var F : PFun{X,Y} .
  var R : Rel{X,Y} .
  op [_,_] : X$Elt Y$Elt -> Pair{X,Y} [ctor] .
  op null : -> Magma{X,Y} [ctor] .          *** empty Magma{X,Y}
  op _,_ : Magma{X,Y} Magma{X,Y} -> Magma{X,Y} [ctor assoc comm id: null] .
  op {_} : Magma{X,Y} -> Rel{X,Y} [ctor] .
  eq [I,K], [I,K] = [I,K] .                 *** idempotency
  mb {null} : PFun{X,Y} .
 cmb {[I, K], M} : PFun{X,Y} 
    if def(I, {M}) = false /\ {M} : PFun{X,Y} .

  op def : X$Elt Rel{X,Y} -> Bool .         *** is key defined in relation?
  eq def(I, {null}) = false .
  eq def(I, {[J, K], M}) = if I == J then true else def(I, {M}) fi .

  op _[_] : PFun{X,Y} X$Elt -> ?{Y} .       *** partial function application
  eq {null}[I] = undef .
 ceq {[I, K], M}[J] = if I == J then K else {M}[J] fi 
    if {[I, K], M} : PFun{X,Y} .
endfm
\end{maude}
\caption{\label{fig:PFUN2}\code{PFUN} module}
\end{figure}
\end{example}

Since a view from \code{TRIV} into any theory \code{T} is fully determined 
by the name \code{Foo} of the sort in \code{T} to which the
sort \code{Elt} is mapped, Maude has a collection of such views
already predefined in its standard prelude.  Therefore, to
define a module of partial functions from the natural numbers to
the rationals we can just write:

\begin{maude}
fmod Nat2RaT-PFUN is
  protecting PFUN{Nat,Rat} .
endfm
\end{maude}

\noindent and we can then evaluate some expressions in this module as
follows:

\begin{maude}
reduce in Nat2RaT-PFUN : {[1,1/2],[1,1/2],[3,1/7],[5,1/17],[3,1/7]} .
result PFun{Nat,Rat}: {[1,1/2],[3,1/7],[5,1/17]}

reduce in Nat2RaT-PFUN : {[1,1/2],[3,1/7],[5,1/17]}[3] .
result PosRat: 1/7
\end{maude}

\noindent {\bf Further Reading}.  Besides \cite{maude-book}, further
details on the executability conditions for a (possibly conditional)
functional module can be found in: (i) for operational
termination \cite{DLMMU-hosc}; for confluence and sort-decreasingness
\cite{bouhoula-jouannaud-meseguer,crc-alp};
for rewriting modulo axioms $B$, the canonical term algebra 
$C_{\Sigma/M\cup E, B}$, and
the agreement between mathematical and operational semantics
\cite{lucas-meseguer-normal-th-JLAMP,meseguer-strict-coherence}.  
For the semantics of
parameterized functional modules see 
\cite{institutions,DBLP:journals/tcs/DuranM03,os-param-ind}.

\section{System Modules}\label{sec:system}

Maude system modules model concurrent systems as 
conditional rewrite theories \cite{unified-tcs,bruni-meseguer-tcs}
 of the form 
$\mathcal{R}= (\Sigma, M \cup E\cup B, R, \phi)$, 
where: (i) $(\Sigma,M \cup E\cup B)$  is
a membership equational theory satisfying the 
executability conditions of a functional
module (i.e., convergence), (ii) $R$ is a set of (possibly
conditional) rewrite rules specifying the system's
concurrent transitions,
and (iii) $\phi$ is a frozenness map (more on this below).
 In Maude, a system module
for the above theory named \code{FOO}  is declared with syntax:
\code{mod} \code{FOO is} $(\Sigma, M \cup E\cup B, R, \phi)$ \code{endm}.

What is the concurrent system defined by $\mathcal{R}$?
The membership equational theory $(\Sigma,M \cup E\cup B)$ 
defines the \emph{states} of such a system as the elements
of the algebraic data type $C_{\Sigma/M \cup E, B}$.
We can call this aspect the \emph{static} part of the
specification $\mathcal{R}$.  Instead, its \emph{dynamics}, i.e.,
how states \emph{evolve}, is described by the rewrite rules $R$,
which specify  the possible \emph{local} concurrent transitions
of the system thus specified.  The system's concurrency is
naturally modeled by the fact that in a given state
$[u] \in C_{\Sigma/M \cup E, B}$ several rewrite rules in $R$ may be
applied concurrently to different subterms of $u$, 
producing several concurrent 
local state changes, and that rewriting logic itself
models those concurrent transitions as logical deductions
(see \cite{unified-tcs,bruni-meseguer-tcs} and the later discussion on semantics).
The only restrictions imposed when applying rules in $R$
are specified  by the \emph{frozenness map} 
$\phi : \Sigma \longrightarrow \mathcal{P}(\mathbb{N})$, which
 assigns to each operator $f\; :\; k_1 \ldots k_n \rightarrow k$ in
 $\Sigma$ the subset $\phi(f) \subseteq \{1,\ldots,n\}$ of its
 \emph{frozen arguments}, that is, those argument positions under
 which rewriting with rules in $R$ is forbidden.

The rules in $R$ can be unconditional rewrite rules
of the form $t \rightarrow t'$, where $t,t'$ are $\Sigma$-terms
of the same kind.  They are then specified in Maude with 
syntax \code{rl} $t$ \code{=>} $t'$ \code{.} but,
by making them conditional,
rules can become considerably more expressive.
Conditional rules in $R$ have the general form:

{\small
\begin{align*}
\texttt{crl}\ \texttt{[}l\texttt{]:}\ t \rightarrow t' \;\; \mathit{if} \;
& u_{1} = v_{1} \wedge  \ldots  \wedge  u_{n} = v_{n}\ \wedge \\
& w_1:s_1 \wedge \ldots \wedge w_m : s_m \ \wedge \\
& l_{1} \rightarrow r_{1} \wedge  \ldots  \wedge  l_{k} \rightarrow r_{k} 
 \ .
\end{align*}

 }

\noindent
where in Maude the $\rightarrow$ symbol
is rendered in ASCII as \code{=>} and the
conjunction $\wedge$ as \verb+/\+, and
where $t$ and $t'$ are $\Sigma$-terms of
the same kind,
$ u_{i} = v_{i}, 1 \leq i \leq n$, are $\Sigma$-equations,
$ w_{i}: s_i, 1 \leq i \leq m$, are memberships,
and
$l_{i} \rightarrow r_{i}, 1\leq i\leq k$, are 
rewrite conditions understood as \emph{reachability predicates},
that is, the arrow in them (rendered in ASCII as \code{=>})
should be implicitly
understood in a reflexive-transitive closure sense as
$l_{i} \rightarrow^{*} r_{i}$.   Of course, in their full generality,
so that, for example, new variables may appear in a rule's condition
in an arbitrary manner, a conditional rule may not be executable
in Maude.  However, Maude allows
conditional rules 
to have
extra variables in their conditions provided they
appear in a disciplined manner 
that allows such extra variables to be \emph{incrementally instantiated}
by incrementally evaluating the conditions in the rule from left to
right;
see \cite{maude-book} for further details.  
We postpone a more detailed explanation of conditional rule
evaluation until after Example \ref{ex:LP-system}.
Rule labels are optional, but may be useful when using formal 
tools or controlling the execution using strategies 
(see Section~\ref{sec:strategies}).

Besides the syntactic requirements for a
conditional rule in $R$ having extra variables in its condition
to be executable in Maude spelled out in \cite{maude-book},
some further \emph{executability requirements} are
needed: (i) first of all, the equational
part $(\Sigma,M \cup E\cup B)$ must be executable as a 
functional module, i.e., the oriented equations
$\vec{E}$ and the memberships $M$ should be \emph{convergent}
modulo $B$; and  (ii) the
rules $R$ should ``commute'' with the equations
$E$ modulo $B$ in the precise sense of having
the \emph{ground coherence property}.
This exactly
means that if $t$ is a ground term having a normal
form $[u] \in C_{\Sigma/M \cup E, B}$,
 and we can perform a rewrite $t \rightarrow_{R,B} t'$ with a rule
in $R$ modulo $B$, then we can also perform
a rewrite $u \rightarrow_{R,B} t''$ 
so that $t'$ and $t''$ have the \emph{same} normal
form modulo $B$, say, $[w] \in C_{\Sigma/M \cup E, B}$.
Ground coherence can be checked by Maude's ChC tool 
\cite{crc-alp}.   This  allows Maude to always normalize terms
with the equations $E$ modulo $B$ \emph{before}
performing a transition with $R$, under the assurance
that no state transitions will ever be missed by following
this strategy.

Let us explain
how terms are evaluated in a system
module.  As pointed out at the beginning of Section
\ref{sec:functional}, the key
difference between an equational program (functional module)
and a rewriting logic one (system module) is that
evaluation to normal form of a term $t$ in a functional module
by means of the \code{reduce} command yields
a \emph{unique} result (determinism) under the
confluence and operational termination assumptions.
Instead, rewrite theories
are intrinsically \emph{nondeterministic}.
What should Maude do to evaluate a term $t$ in a
system module?  $t$ can be rewritten in many
different ways to many different terms, and the process may never
terminate.  Maude offers the following options:

\begin{itemize}
\item 
A \emph{rule fair} sequence of rewrite steps starting
from a term $t$ can be obtained by giving the
command:  \code{rewrite} $t$ \code{.} but since
such a rewrite sequence may not terminate,
a bound limiting the number of rewrite
steps can be specified (see \cite{maude-book}).
Rule fairness means that if more than one rule is applicable to the terms of a rewrite sequence,
different rules are applied, avoiding the repetition of the very same rule for each term in a rewrite sequence.

\item A \emph{rule and position fair} sequence of 
rewrite steps starting
from a term $t$ can be obtained by giving the
command:  \code{frewrite} $t$ \code{.} 
and a bound limiting the number of rewrite 
steps can likewise be specified (see \cite{maude-book}).
Position fairness is similar to rule fairness but refers to the positions of a term where a rule is applicable.

\item The entire, possibly infinite, state space of terms 
reachable from a term $t$ by a sequence of rewrite
steps can be explored with Maude's \code{search}
command, which searches such a state space in a breadth-first manner. 
Moreover, a search graph, instead of a search tree, is constructed by 
storing previously seen states; this allows the search command to terminate
when there is a finite graph associated to the state space.
 The general form of the command is:
\code{search} $t$ \code{=>}$\diamond$ $t'$ \code{s.t.} $C$ 
\code{.} where $t'$ is a term pattern, so that we are looking
for terms reachable from $t$ that are instances of $t'$
by a substitution $\theta$,
and $C$ is an equational condition such 
that only reachable terms of the form
$t'\theta$ such that $C \theta$ holds are selected.
The  $\diamond$ symbol is a place holder for the 
options: $\diamond=$ \code{1} (exactly one rewrite step), $\diamond=$
\code{+} (one or more steps), $\diamond=$
\code{*} (zero or more steps), and $\diamond=$
\code{!} (terminating states).
Since the search may either never terminate
and/or find an infinite number of solutions, two
\emph{bounds}\label{bounds-search} can be added to a \code{search}
command: one bounding the number of solutions
requested, and another bounding the depth of the rewrite
steps from the initial term $t$ (see the examples in
Example \ref{ex:relatives-system} and
\cite{maude-book} for details).
Note that Maude's \code{search} command 
provides a quite expressive form of model checking by
\emph{reachability analysis}, in addition to its LTL-based model checking.
\end{itemize}


\begin{example}
\label{hanoi-smod}
Let us consider a simple example of a system module. The \code{HANOI} module in 
Figure~\ref{fig:HANOI} specifies the Tower of Hanoi puzzle, invented by the French 
mathematician \'Eduard Lucas in 1883. His story tells that in an Asian temple there 
are three diamond posts; the first is surrounded by sixty-four golden disks of 
increasing size from the bottom to the top. The monks are committed to move them from 
one post to another respecting two rules: only a disk can be moved at a time, and 
they can only be laid either on a bigger disk or on the floor. Their objective is to 
move all of them to the third post, and then the world will end.

In the \code{HANOI} module, the golden disks are modeled as natural numbers 
describing their size, and the posts are described as lists of disks in bottom-up 
order. In general, we have terms describing the states of the game and rules that 
model the moves allowed by the game. Rewriting with these rules allows going from a 
given initial state to other states, hopefully including the desired final state. In 
\cite[Chapter 7]{maude-book} there is a collection of game examples that follow this 
general pattern. 

\begin{figure}
\begin{maude}
mod HANOI is
  protecting NAT-LIST .
  sorts Tower Hanoi .
  subsort Tower < Hanoi .
  op (_)[_] : Nat NatList -> Tower [ctor] .
  op empty : -> Hanoi [ctor] .
  op __ : Hanoi Hanoi -> Hanoi [ctor assoc comm id: empty] .
  vars S T D1 D2 : Nat .   vars L1 L2 : NatList .

  crl [move] : (S) [L1 D1]  (T) [L2 D2]
   =>          (S) [L1]     (T) [L2 D2 D1]
   if D2 > D1 .
   rl [move] : (S) [L1 D1]  (T) [nil]
   =>          (S) [L1]     (T) [D1] .
endm
\end{maude}
\caption{\label{fig:HANOI}\code{HANOI} module}
\end{figure}

If we try to rewrite the initial puzzle setting

\begin{maude}
Maude> rewrite in HANOI : (0)[3 2 1] (1)[nil] (2)[nil] .
\end{maude}

\noindent 
the command does not terminate because the disks are being moved in a loop. We can 
instead rewrite with a bound on the number of rewrites, like 23 in the following 
command example

\begin{maude}
Maude> rewrite [23] in HANOI : (0)[3 2 1] (1)[nil] (2)[nil] .
result Hanoi: (0)[3 2] (1)[1] (2)[nil]
\end{maude}

\noindent 
Even if the example has non-terminating rewrite sequences, as shown above with the rewrite command, 
the number of reachable configurations is 
finite and the following search command terminates.

\begin{maude}
Maude> search in HANOI : (0)[3 2 1] (1)[nil] (2)[nil] =>* H .
.... 27 solutions
\end{maude}

\noindent
Indeed, the solution to the initial configuration 
\code{(0)[3 2 1] (1)[nil] (2)[nil]}
is 
\code{(0)[nil] (1)[nil] (2)[3 2 1]} 
and is found at state $26$.
\end{example}

In Section~\ref{sec:strategies} we will consider again this example with the help of strategies. 


\subsection{Logic Programming Running Example}\label{sec:prolog-system}

One of the key strengths of rewriting logic, inherited by Maude,
is that a wide range of concurrent and nondeterministic
systems can be naturally specified as rewrite theories
and executed and analyzed as system modules in Maude.
Such systems include: (i) a very wide range of concurrency
models \cite{concur96,20-years}, (ii) the executable semantic
definition of concurrent programming languages~
\cite{meseguer-rosu-tcs,serbanuta-rosu-meseguer-ic,DBLP:journals/iandc/MeseguerR13},
and (iii) a very wide range of logics, specified as rewrite
theories using rewriting logic as a \emph{logical framework}
\cite{rwl-fwk, 20-years}.  Section \ref{sec:oo} will
give examples of how concurrent object systems
can be naturally specified in Maude.  Here we
give an example that straddles cases (ii)--(iii) above,
namely, a computational logic (Horn Logic \cite{horn_1951}) that can
at the same time be used as a programming language.
Since this logic programming
example uses symbolic computation in an essential
manner, we will present several variants of
it at various places in the paper to illustrate various Maude
symbolic computation features.

\begin{example}[LP-Syntax]\label{ex:LP-functional-syntax}
To define the semantics of logic programs as a system module\footnote{As explained at the end of Section~\ref{urlmaude30}, the Maude code is available at
\url{http://maude.cs.illinois.edu}.
}
we first specify an \code{LP-SYNTAX} functional module that
 imports the \code{TERM} functional module with sort \code{Term}
 in Example~\ref{ex:TERM}.
An atomic predicate
is defined as 
a \code{Qid} symbol applied to a non-empty list of terms in parentheses:

{\footnotesize
\begin{maude}
  sort Predicate .
  op _`(_`) : Qid NeTermList -> Predicate [ctor] .
\end{maude}
}

\noindent
Since for Horn clauses we need both empty and non-empty lists of 
predicates,  we define them in Maude as 
sorts \texttt{PredicateList} and \texttt{NePredicateList} 
(the view for sort \texttt{Predicate} is not included):

\begin{maude}
  protecting (LIST * (sort List{X} to PredicateList,
              sort NeList{X} to NePredicateList,
              op __ to _`,_ [prec 50])) {Predicate} .
\end{maude}

\noindent
We define a Horn clause using the standard symbol \verb!:-! for $\Leftarrow$. 
We express an axiom, e.g. \code{'father('john,'peter)}, as a Horn clause with an empty body,
e.g. \code{'father('john,'peter) :- nil}.
\begin{maude}
  sort Clause .
  op _:-_ : Predicate PredicateList -> Clause [ctor prec 60] .
\end{maude}
Finally, a logic program is given by a colon-separated list of Horn clauses, of sort \texttt{Program}:
\begin{maude}
  protecting (LIST * (sort List{X} to Program,
              sort NeList{X} to NeProgram,
              op __ to _;_ [prec 70])) {Clause} .
\end{maude}
\end{example}

Then, we should provide some basic 
notion of substitution. 

\begin{example}[LP-Substitution Module]\label{ex:LP-substitution}
Using the syntax for predicates and Horn clauses
given in Example~\ref{ex:LP-functional-syntax},
we  define a substitution as a partial function according to Example~\ref{ex:pfunXY}.
We need to define views 
from the sort \code{Elt} to sorts \code{Variable} and \code{Term}.

\begin{maude}
view Variable from TRIV to TERM is
  sort Elt to Variable .
endv
\end{maude}

\begin{maude}
view Term from TRIV to TERM is
  sort Elt to Term .
endv
\end{maude}

\noindent
And we import the parametric module \code{PFUN} instantiated to the views \code{Variable} and \code{Term}
but rename 
sort \code{PFun\{Variable,Term\}} 
into \code{Substitution},
operator \verb!_,_! for combining bindings into \verb!_;_!,
and
operator \verb![_,_]! into the more standard syntax 
\verb!_->_! for substitution bindings.

\begin{maude}
fmod LP-SUBSTITUTION is
  protecting (PFUN * (op _,_ to _;_,
                      op [_,_] to _->_)) {Variable,Term}
                * (sort PFun{Variable,Term} to Substitution,
                   sort Pair{Variable,Term} to Binding) .
endfm
\end{maude}
\end{example}

Now, we are able to provide some basic syntactic unification functionality.

\begin{example}[LP-Unification Module]\label{ex:LP-functional-unification}
Continuing Example~\ref{ex:LP-substitution},
module \linebreak\code{LP-UNIFICATION} in Figure~\ref{fig:LP-UNIFICATION} defines a syntactic 
unification procedure, without the occurs check, for both predicates and terms 
that will be used by our Horn logic interpreter in Example~\ref{ex:LP-system} below.
Note that the failure to unify is represented by an expression at the kind 
\code{[Substitution]}.

\begin{figure}
\begin{maude}
fmod LP-UNIFICATION is
  protecting LP-SYNTAX .
  protecting LP-SUBSTITUTION .

  op unify : Predicate Predicate Substitution -> [Substitution] .
  eq unify(F(NeTL1), F(NeTL2), S)
   = unify(NeTL1, NeTL2,S) .

  vars C F : Qid .
  var V : Variable .
  vars NeTL1 NeTL2 : NeTermList .
  var NVT : NvTerm .
  var S : Substitution .
  vars T T1 T2 : Term .

  op unify : NeTermList NeTermList Substitution -> [Substitution] .
  eq unify(C, C, S) = S .
  eq unify(V, T1, (V -> T2) ; S) = unify(T1, T2, (V -> T2) ; S) .
 ceq unify(T1, V, (V -> T2) ; S) = unify(T1, T2, (V -> T2) ; S) if not T1 :: Variable .
 ceq unify(V, T, S) = (V -> T) ; S if not def(V, S) .
 ceq unify(NVT, V, S) = (V -> NVT) ; S if not def(V, S) .
  eq unify(F[NeTL1], F[NeTL2], S) = unify(NeTL1, NeTL2, S) .
 ceq unify((T1,NeTL1), (T2, NeTL2), S)
   = unify(NeTL1, NeTL2, unify(T1, T2, S))
   if unify(T1, T2, S) :: Substitution .
endfm
\end{maude}
\caption{\label{fig:LP-UNIFICATION}\code{LP-UNIFICATION} module}
\end{figure}

%
%
%
%
%
%
%
\end{example}

\begin{example}[LP System Module]\label{ex:LP-system}
Using the LP syntax defined in
Example~\ref{ex:LP-functional-syntax}
and the unification algorithm
in Example~\ref{ex:LP-functional-unification},
the system module 
	\linebreak
\code{LP-SEMANTICS} in Figure~\ref{fig:LP-SEMANTICS} defines
the semantics of logic programs
and provides an interpreter with a breadth-first strategy.
We first define logic programming configurations to hold the execution state, e.g.
\code{PL \$ S} where \code{PL} is a predicate list and \code{S} is a substitution, 
 that represents the pending objectives and
the bindings carried from already executed clauses.
The execution of a query predicate w.r.t. a logic program will be defined by means of transition rules
transforming an expression of the form
\verb!< N | PL $ S | PG >! where 
\code{N} is a natural number used for renaming clauses before unification,
\code{PL} and \code{S} are as explained above, 
and \code{PG} is the logic program.

\begin{figure}
\begin{maude}
mod LP-SEMANTICS is
  protecting LP-UNIFICATION .
  sort Configuration .
  op <_|_$_|_> : Nat PredicateList Substitution Program -> Configuration .

 crl [clause] :
   < N1 | P1, PL1 $ S1 | Pr1 ; P2 :- PL2 ; Pr2 >
   => < N2 | PL3, PL1 $ S2 | Pr1 ; P2 :- PL2 ; Pr2 >
   if P3 :- PL3 := rename(P2 :- PL2, N1)
   /\ S2 := unify(P1, P3, S1)
   /\ N2 := max(N1, last$(P3 :- PL3)) .

  ...
  
  op <_|_> : PredicateList Program -> Configuration .
  eq < PL | Pr > = < last$(PL) | PL $ null | Pr > .
endm
\end{maude}
\caption{\label{fig:LP-SEMANTICS}\code{LP-SEMANTICS} module}
\end{figure}
\noindent
%


\noindent
Its semantics is defined by a single conditional
rule that invokes a variable renaming function \code{rename}
not shown here.
Note that a program is defined as a list, instead of a set, of clauses of the form \verb!P2 :- PL2!
making the interpreter closer to how logic programming languages work.
%
Finally, we add an initialization function to evaluate a predicate list
\code{PL} in \code{Pr}.
\end{example}

Since the condition in the above conditional rule only
involves equations, its incremental evaluation is exactly
the same as if it were the
condition of a conditional equation or membership in
a functional module.  Therefore, we can use this example
to explain the incremental evaluation of conditions
in all cases.  After this explanation we will briefly
summarize the more general case of a conditional
rule that also has rewrite conditions.
To syntactically indicate the fact that extra variables appear in a condition,
the equality sign  \code{:=}
is used instead of the usual sign \code{=}.
At the left of the \code{:=} sign a \emph{pattern with
new variables} is placed.  Here three such
patterns are given, one for each condition,
namely, the term \code{P3 :- PL3}
and the variables \code{S2} and \code{N2}.
Operationally, the substitution
instantiating the new variables in these three
patterns is obtained by \emph{incrementally} (one condition at a time, from
left to right): (i) reducing to normal
form with the module's \emph{equations and memberships}
the substitution instance of the condition's  righthand side.
But instance under \emph{which} substitution?  For the first condition
---here the condition \verb+P3 :- PL3 := rename(P2 :- PL2,N1)+---
its righthand side variables ---here \code{P2}, \code{PL2} and \code{N1}---
must always appear in the conditional rule's lefthand side.
Therefore, the righthand side \code{rename(P2 :- PL2,N1)} is
instantiated by the \emph{matching substitution} $\theta_{0}$ with which
we have instantiated the rule's lefthand side to attempt a rewrite.
Suppose, for example, that in $\theta_{0}$ \code{P2} was instantiated to 
\verb+`p(x{1})+, \code{PL2}  to \verb+nil+, and \code{N1} to $5$.
Then we \emph{reduce} to normal form with the module's equations the
instance by $\theta_{0}$
of the condition's righthand side \code{rename(P2 :- PL2,N1)},
that is, the term \verb+rename(`p(x{1}) :- nil,5)+.
Since the rename function just renames the variables to fresh ones
above the given index, we will get the result
\verb+`p(x{6}) :- nil+.  Then, (ii) we now incrementally extend $\theta_{0}$
by instantiating the new variables in the condition's lefthand side
\verb+P3 :- PL3+ by \emph{matching} \verb+`p(x{6}) :- nil+
against it.  That is, \code{P3} is instantiated to \verb+`p(x{6})+
and \code{PL3} is instantiated to \code{nil}, thus getting
an extended substitution $\theta_{0} \uplus \theta_{1}$.
But this extended substitution can now instantiate all
the variables in the righthand side of the \emph{second} condition
\verb+S2 := unify(P1,P3,S1)+, so that we can again reduce the instance
by $\theta_{0} \uplus \theta_{1}$ of its righthand side to normal form and then
instantiate \code{S2} to \emph{that} normal form to obtain
a new extended substitution $\theta_{0} \uplus \theta_{1} \uplus \theta_{2}$.
We then proceed in the same manner to evaluate the third
condition using $\theta_{0} \uplus \theta_{1} \uplus \theta_{2}$
and in that way we finally get an extended substitution
$\theta_{0} \uplus \theta_{1} \uplus \theta_{2} \uplus \theta_{3}$
with which we can now instantiate the conditional rule's righthand
side, thus ending the conditional rule's evaluation.
Note that the second and third conditions exactly correspond to \code{where}
clauses in some functional languages.

Since rewrite conditions are not used in this example (but are used in Figure~\ref{fig:LP-EXTRA-NEGATION} below),
we briefly explain their use and execution.
Further details can be found in \cite{maude-book}.
 In general, a reachability condition 
$l \rightarrow r$ may have extra variables in its 
righthand side $r$.  It succeeds for a substitution  instance $l \theta$
(where, given the incremental way conditions are evaluated,
$\theta$ will extend the original substitution $\theta_{0}$
obtained by matching the rule's lefthand side)
iff $l \theta$ can be rewritten in 0, 1 or more steps with the
module's rules $R$
to a term whose normal form by the module's equations $E$
and memberships $M$ is a substitution
instance of $r$ up to $B$-equality.
In this way, if $l \rightarrow r$ was the $k$th condition,
we obtain an extended substitution $\theta \uplus \theta_{k}$
that we then use to evaluate condition $k+1$ or, if no more
conditions are left, to instantiate the rule's righthand side.
Note that, in general, the equalities,
memberships, and reachability conditions in a rule's condition need
not appear in any particular order, provided that the 
variables appearing in either the righthand side of an equational condition $u := v$
or the lefthand side of a rewrite condition $l \rightarrow r$ have
already appeared in \emph{previous} conditions and/or in the
rule's lefthand side.

We can now evaluate some logic programs
using our semantic definition as a breadth-first
logic programming interpreter.  An interesting
question is what pattern $t'$ to use in a search from
an initial state 
\code{< PL | Pr >}, where \code{PL} is the
list of predicates that we wish to find a solution for,
and \code{Pr} is the given Horn logic program.
The answer is that we should look for a
pattern of the form:
\code{< N | nil \$ S1 | Pr >},
indicating that
 the list of objectives has become empty
and therefore the substitution \code{S1}
is a solution.  Therefore,
calls to our interpreter will have the
general form:
\verb+ search < PL | Pr > =>* < N | nil $ S1 | Pr > .+

\begin{example}[Search LP-evaluation]\label{ex:relatives-system}\label{ex:relatives}
Consider the following logic program defining several family relations
between  Jane, Mike, Sally, John, and Tom.

{\footnotesize
\begin{maude}
 mother(jane, mike) .
 mother(sally, john) .
 father(tom, sally) .
 father(tom, erica) .
 father(mike, john) .
 sibling(X1, X2) :- parent(X3,X1), parent(X3,X2) .
 parent(X1, X2) :- father(X1,X2) .
 parent(X1, X2) :- mother(X1,X2) . 
 relative(X1, X2) :- parent(X1,X3), parent(X3,X2) .
 relative(X1, X2) :- sibling(X1,X3), relative(X3,X2) .
\end{maude}
}

\noindent
This logic program is expressed using the syntax of Example~\ref{ex:LP-functional-syntax}
as follows:

{\footnotesize
\begin{maude}
 'mother('jane, 'mike) :- nil ;
 'mother('sally, 'john) :- nil ;
 'father('tom, 'sally) :- nil ;
 'father('tom, 'erica) :- nil ;
 'father('mike, 'john) :- nil ;
 'sibling(x{1}, x{2}) :- 'parent(x{3}, x{1}), 'parent(x{3}, x{2}) ;
 'parent(x{1}, x{2}) :- 'father(x{1}, x{2}) ;
 'parent(x{1}, x{2}) :- 'mother(x{1}, x{2}) ;
 'relative(x{1}, x{2}) :- 'parent(x{1}, x{3}),'parent(x{3}, x{2}) ;
 'relative(x{1}, x{2}) :- 'sibling(x{1}, x{3}),'relative(x{3}, x{2})
\end{maude}
}

We can now evaluate different initial calls for this program
by specifying specific lists of atoms that we seek a solution for.
We can do so by searching for a set of configurations
reachable from the given initial call that
contains a solution.  As already explained,
a solution will correspond to a configuration of the form \verb!nil $ Sub!. 
Depth and solution bounds are automatically provided by  Maude's
\code{search} command (see page~\pageref{bounds-search}). 
In order to simplify 
the presentation,
we have abbreviated our program to \code{P}
and we do not show all the bindings returned by the search command, 
just the binding associated to the logic programming computed substitution.
%

First, we can ask whether Sally and Erica are sisters; the associated reachability graph is finite and no bound is needed.

\begin{maude}
Maude> search < 'sibling('sally, 'erica)  | P > =>* < N | nil $ S1 | Pr > .
Solution 1 (state 7)
S1 --> (x{1} -> 'sally) ; (x{2} -> 'erica) ; (x{3} -> x{4}) ; 
       (x{4} -> 'tom) ; (x{5} -> 'sally) ; (x{6} -> x{4}) ; x{7} -> 'erica
\end{maude}

\noindent
Who are the siblings of Erica? Sally and herself.

\begin{maude}
Maude> search < 'sibling(x{1},'erica) | P > =>* < N | nil $ S1 | Pr > .
Solution 1 (state 19)
S1 --> (x{1} -> x{2}) ; (x{2} -> x{6}) ; (x{3} -> 'erica) ; (x{4} -> x{5}) ; 
       (x{5} -> 'tom) ; (x{6} -> 'sally) ; (x{7} -> x{5}) ; x{8} -> 'erica

Solution 2 (state 20)
S1 --> (x{1} -> x{2}) ; (x{2} -> x{6}) ; (x{3} -> 'erica) ; (x{4} -> x{5}) ; 
       (x{5} -> 'tom) ; (x{6} -> 'erica) ; (x{7} -> x{5}) ; x{8} -> 'erica
\end{maude}

\noindent
How many possible siblings are there? Sally and Sally, Sally and Erica, Erica and Sally, Erica and Erica, John and John, and Mike and Mike.

{\footnotesize
\begin{maude}
Maude> search < 'sibling(x{1},x{2}) | P > =>* < N | nil $ S1 | Pr > .
Solution 1 (state 19)
S1 --> (x{1} -> x{3}) ; (x{2} -> x{4}) ; (x{3} -> x{7}) ; (x{4} -> x{9}) ; 
       (x{5} -> x{6}) ; (x{6} -> 'tom) ; (x{7} -> 'sally) ; (x{8} -> x{6}) ; 
       x{9} -> 'sally
...

Solution 7 (state 25)
S1 --> (x{1} -> x{3}) ; (x{2} -> x{4}) ; (x{3} -> x{7}) ; (x{4} -> x{9}) ; 
       (x{5} -> x{6}) ; (x{6} -> 'sally) ; (x{7} -> 'john) ; (x{8} -> x{6}) ; 
       x{9} -> 'john
\end{maude}
}

\noindent
Seven solutions are given. Are Jane and John relatives? Yes

{\footnotesize
\begin{maude}
Maude> search < 'relative('jane,'john) | P > =>* < N | nil $ S1 | Pr > .
Solution 1 (state 11)
S1 --> (x{1} -> 'jane) ; (x{2} -> 'john) ; (x{3} -> x{5}) ; (x{4} -> 'jane) ; 
       (x{5} -> 'mike) ; (x{6} -> x{5}) ; x{7} -> 'john
\end{maude}
}

\noindent
Who are the relatives of John? Tom and Jane.

{\footnotesize
\begin{maude}
Maude> search [2] < 'relative(x{1},'john) | P > =>* < N | nil $ S1 | Pr > .
Solution 1 (state 28)
S1 --> (x{1} -> x{2}) ; (x{2} -> x{5}) ; (x{3} -> 'john) ; (x{4} -> x{6}) ; 
       (x{5} -> 'tom) ; (x{6} -> 'sally) ; (x{7} -> x{6}) ; x{8} -> 'john

Solution 2 (state 29)
S1 --> (x{1} -> x{2}) ; (x{2} -> x{5}) ; (x{3} -> 'john) ; (x{4} -> x{6}) ; 
       (x{5} -> 'jane) ; (x{6} -> 'mike) ; (x{7} -> x{6}) ; x{8} -> 'john
\end{maude}
}

\noindent
This last call produces an infinite search
and we must restrict the search, by asking for two solutions only.
\end{example}

\subsection{Initial Model Semantics and Parameterization}

What is the mathematical meaning of
a system module
\code{mod} \code{FOO is} $(\Sigma, M \cup E\cup B, R, \phi)$
\code{endm}, that is, of a rewrite theory 
$\mathcal{R}=(\Sigma, M \cup E\cup B, R, \phi)$?
It is its \emph{initial model} $\mathcal{T}_{\mathcal{R}}$ in the class
(indeed, category) of all models of $\mathcal{R}$
\cite{unified-tcs,bruni-meseguer-tcs}.  What does 
$\mathcal{T}_{\mathcal{R}}$ look like?  Why, of course, it models a
concurrent system!  Its \emph{states}, as already pointed out,
are the normal forms $[w] \in C_{\Sigma/M \cup E, B}$
in the canonical term algebra of $(\Sigma, M \cup E\cup B)$.
What about its transitions?  $\mathcal{T}_{\mathcal{R}}$
provides a \emph{true concurrency semantics}.  That is,
not only are one-step transitions modeled: concurrent
\emph{computations} are also modeled.  Furthermore, $\mathcal{T}_{\mathcal{R}}$
provides a notion of \emph{equivalence} between two different
descriptions of the \emph{same} concurrent computation.  
Mathematically, what all this means
is that for each kind $[s]$ in $\Sigma$ the concurrent computations
of $\mathcal{R}$ form a \emph{category}
\cite{unified-tcs,bruni-meseguer-tcs},
whose objects/states are precisely the normal forms in 
$C_{\Sigma/M \cup E, B,[s]}$.  But how are concurrent computations
modeled? They coincide with rewriting logic  \emph{proofs} 
in $\mathcal{R}$, with the category structure providing a natural
notion of \emph{proof equivalence}.  This is what one should
expect, since any declarative programming language worth its salt
should satisfy the equivalence:
\begin{quote}
\emph{Computation = Deduction}
\end{quote}
and of course this is exactly what also happens at the equational logic
level of functional modules, where
the initial algebra $T_{\Sigma/M \cup E \cup B}$ is
built up out of the proof theory of membership equational logic \cite{tarquinia}.

What about \emph{parameterized} system modules?
They are completely analogous to parameterized functional ones.
That is, a \emph{parameterized system module} 
\verb+M{X1 :: T1, ... , Xm :: Tm}+ is a rewrite theory with
\code{T1}, $\ldots,$ \code{Tm} its parameter theories
 and is instantiated by
views \code{V1}, $\ldots,$ \code{Vm}
from \code{T1}, $\ldots,$ \code{Tm} to \code{T'1}, $\ldots,$
\code{T'm} to the instance \verb+M{V1, ... , Vm}+.
The only new feature is that, although often the
parameter theories are functional, some of the theories among the
\code{T1}, $\ldots,$ \code{Tm} may be \emph{system theories},
i.e., theories of the form \code{th} \code{FOO is} $(\Sigma, M \cup E\cup B, R, \phi)$
\code{endth}, where, as done for functional theories, the rewrite theory 
$\mathcal{R}=(\Sigma, M \cup E\cup B, R, \phi)$ is given a ``loose
semantics,'' so that actual parameter instances of this formal
parameter range over the class of all models of $\mathcal{R}$.
Of course, if the parameter theory \code{Ti} is a rewrite theory,
then the target theory \code{T'i} should also be a rewrite theory, and
the view \code{Vi} :  \code{Ti} $\rightarrow$ \code{T'i} is a
\emph{theory interpretation} between rewrite theories.

\vspace{1ex}

\noindent {\bf Further Reading}.  Besides \cite{maude-book}, the
following references may be helpful: (i) for the semantics of
rewrite theories \cite{unified-tcs,bruni-meseguer-tcs};
for the modeling of concurrent systems, programming languages, and logical systems
in rewriting logic
\cite{20-years,concur96,meseguer-rosu-tcs,serbanuta-rosu-meseguer-ic,DBLP:journals/iandc/MeseguerR13,rwl-fwk};
(iii) for the ground coherence property and how to check it
\cite{crc-alp}; (iv) for automatically
transforming a rewrite theory into a semantically equivalent one that
is ground coherent \cite{DBLP:conf/wrla/Meseguer18};
and for an even more general notion of rewrite theory well suited
for symbolic computation \cite{DBLP:conf/wrla/Meseguer18}.

\section{The Maude Strategy Language}
\label{sec:strategies}


%
%



Rewriting with rules is a highly
nondeterministic process, since at every step many rules could be applied at various positions, as described in Section~\ref{sec:system}. Sometimes, a finer control on how rules are applied may be desirable. This responsibility has traditionally been given to the Maude reflective features, or has been achieved by means of specific data and rule representations that force rewriting to happen in some desired ways. However, such data and rule representation usually have to be changed if a different execution is desired, and make specifications harder to understand. To avoid this, a strategy language has been proposed \cite{towardsStrategy,strategies06,rewSemantics} as a specification layer above those of equations and rules. This provides a cleaner way to control the rewriting process, respecting a \emph{separation of
concerns} principle, and allowing the same system module to be controlled by different strategy specifications. The design of this language was influenced, among others, by ELAN~\cite{elan} and Stratego~\cite{stratego}.

Strategies are seen as recipes to control rewriting, but they are usually described as transformations from an initial term to a set of terms. These are the results of the strategy, which can be multiple because of its nondeterminism. The usual Maude command for rewriting with strategies is:
\begin{maude}[escapechar=^,basicstyle=\normalfont\small\ttfamily]
srewrite [^\syntaxSymb{Bound}^] in ^\syntaxSymb{ModuleName}^ : ^\syntaxSymb{Term}^ by ^\syntaxSymb{StrategyExpr}^ .
\end{maude}
\noindent
It rewrites the given term according to the given strategy, and prints all the results. Like in the standard rewriting commands, we can optionally specify the module where to rewrite after \texttt{in}, and a bound on the number of solutions to be shown with \texttt{[$N$]} just after the command keyword, which can be shortened to \texttt{srew}. For example, going back to the Tower of Hanoi example introduced in Section~\ref{sec:system} (page~\pageref{hanoi-smod}), we consider a basic strategy which is just rule application, invoked by mentioning the rule label as follows:

\begin{maude}
Maude> srew [3] in HANOI : (0)[3 2 1] (1)[nil] (2)[nil] using move .

Solution 1
rewrites: 1
result Hanoi: (0)[3 2] (1)[1] (2)[nil]

Solution 2
rewrites: 2
result Hanoi: (0)[3 2] (1)[nil] (2)[1]

No more solutions.
rewrites: 2
\end{maude}
The two results of applying the \code{move} rule to the initial term are shown, and the interpreter tells us that there are no more solutions, because we have requested three, exceeding the possible one-step moves. The order in which solutions appear is implementation dependent. Notice that in this way we have a standard approach to model a game: terms represent states, rules represent allowed moves, and the strategy language can be used to model a (hopefully winning) strategy for solving or playing the game.

\begin{table}\centering
	\newcommand*\sem[2][\theta, t]{\ensuremath{\llbracket #2\rrbracket(#1)}}
\begin{tabular}{c|c}
	Strategy $\zeta$ 		& Results $\sem\zeta$ 							\\ \hline
	\multicolumn2{l}{}											\\[-1.33em] \hline
	\code{idle}			& $\{ t \}$								\\ \hline
	\code{fail}			& $\emptyset$								\\ \hline
	$\mathit{rlabel}$		& $\{\; t' \in T_{\Sigma} \mid t \to_{R,B}^{\mathit{rlabel}} t' \;\}$	\\ \hline
	\lower1.5pt\hbox{$\alpha \mtt; \beta$}
					& $\bigcup_{t' \in \strut\sem\alpha} \sem[\theta, t']{\beta}$ 		\\ \hline
	$\alpha \mtt| \beta$		& $\sem\alpha \cup \sem\beta$						\\ \hline
	$\alpha \mtt*$			& $\bigcup_{n=0}^\infty \llbracket\alpha\rrbracket^n(\theta, t)$ 	\\ \hline
	\ttfamily match $P$ s.t.\ $C$	& $\begin{cases}
		\{t\} 		& \text{if } \mathrm{matches}(P, t, C, \theta) \neq \emptyset \\
		\emptyset 	& \text{otherwise}
	\end{cases}$ 												\\ \hline
	$\alpha \;\texttt?\; \beta \;\texttt:\; \gamma$	& $\begin{cases}
		\sem{\alpha \mtt; \beta} & \text{if } \sem\alpha \neq \emptyset \\
		\sem\gamma		 & \text{if } \sem\alpha = \emptyset \\
	\end{cases}$ 												\\ \hline
	\ttfamily \begin{tabular}{l}
		matchrew $P$ s.t.\ $C$ by \\
		\quad $X_1$ using $\alpha_1$, $\ldots$, \\
		\quad $X_n$ using $\alpha_n$
	\end{tabular} & $\begin{array}{l}
		\bigcup_{\sigma \in \mathrm{matches}(P, t, C, \theta)} \left( \; \bigcup_{t_1 \in \sem[\sigma, \sigma(X_1)]{\alpha_1}} \; \cdots \right. \\[3pt]
		\quad \left. \bigcup_{t_n \in \sem[\sigma, \sigma(X_n)]{\alpha_n}} \; \sigma[x_1/t_1, \ldots, x_n/t_n](P) \right)
	\end{array}$ \\ \hline
	$sl(t_1, \ldots, t_n)$		& $\bigcup_{(\mathit{lhs}, \delta, C) \in \mathrm{Defs}} \bigcup_{\sigma \in \mathrm{matches}(\zeta, \mathit{lhs}, C, \mathrm{id})} \; \sem[\sigma, t]{\delta}$

\end{tabular}

\caption{Main strategy combinators and their informal semantics} \label{tab:strategy}
\end{table}

 In order to use more elaborate strategies we will need to introduce the complete strategy language, whose constructors are summarized in Table~\ref{tab:strategy}. As we have seen, rule application is its basic building block. Besides the rule label, further restrictions can be imposed; its most general syntax has the form:
\[
label \verb@ [@ X_1  \verb@ <- @ t_1, \ldots, X_n \verb@ <- @ t_n \verb@] {@ \alpha_1, \ldots, \alpha_m \verb@}@
\]
When this strategy is invoked, all rules with the given label and exactly $m$ rewriting conditions are applied nondeterministically at any position of the subject term. Rewriting inside those conditions is controlled by the strategies $\alpha_1, \ldots, \alpha_m$ between curly brackets, which must be omitted when $m = 0$. Between square brackets, we can optionally specify an initial ground substitution to be applied to both sides of the rule. Moreover, restricting the application of the rule to the top position is possible by surrounding it by the \texttt{top($\alpha$)} modifier.

The other basic element of the language are the tests, used for checking conditions on the subject term. Their syntax has the form
\code{match} $P$ \code{s.t.}\ $C$
where $P$ is a pattern and $C$ is an equational condition. On a successful match and condition check, the result is the initial term; otherwise, the test does not provide any solution. The initial keyword of the test indicates where to match, either on top (\texttt{match}), anywhere (\texttt{amatch}), or on the flattened top modulo axioms (\texttt{xmatch}). For example, we can test whether the Tower of Hanoi puzzle is solved:

\begin{maude}
Maude> srew (0)[nil] (1)[nil] (2)[3 2 1] using
            match (N)[3 2 1] H s.t. N =/= 0 .

Solution 1
result Hanoi: (0)[nil] (1)[nil] (2)[3 2 1]

No more solutions.
\end{maude}

We present now various combinators that build more complex strategies out of rule applications and tests. 
The concatenation $\alpha \mtt; \beta$ executes the strategy $\alpha$ and then the strategy $\beta$ on each $\alpha$ result. 
The disjunction or alternative $\alpha \mtt| \beta$ executes $\alpha$ or $\beta$; in other words, the results of $\alpha \mtt| \beta$ are both those of $\alpha$ and those of $\beta$. The iteration \texttt{$\alpha$*} runs $\alpha$ zero or more times consecutively. These combinators resemble  similar constructors for regular expressions. The empty word and empty language constants are here represented by the \code{idle} and \code{fail} operators; the result of applying \code{idle} is always the initial term, while \code{fail} generates no solution.

We say that a strategy \emph{fails} when no solution is obtained. Remember that failures can happen in less explicit situations: when a rule cannot be applied to the term, when a test fails, etc. Thus, strategies will explore rewriting paths that can later be discarded.

A conditional strategy is written $\alpha \;\mtt?\; \beta \;\mtt:\; \gamma$. It executes $\alpha$ and then $\beta$ on its results, but if $\alpha$ does not produce any, it executes $\gamma$ on the initial term. That is, $\alpha$ is the \emph{condition}; $\beta$ the \emph{positive branch}, which applies to the results of $\alpha$; and $\gamma$ the \emph{negative branch}, which is applied only if $\alpha$ fails.
Some common patterns are defined as derived operators with their own names such as:
\begin{itemize}
  \item The \code{or-else} combinator is defined by $\alpha \mts{or-else} \beta \equiv \alpha \mts? \mtt{idle} \mts: \beta$. It executes $\beta$ only if $\alpha$ has failed.
  \item The \emph{negation} is defined as $\mtt{not(}\alpha\mtt) \equiv \alpha \mts? \mtt{fail} \mts: \mtt{idle}$. It fails when $\alpha$ succeeds, and succeeds as an \code{idle} when $\alpha$ fails.
  \item The \emph{normalization operator} $\alpha\mtt! \equiv \alpha\mtt{* ; not(} \alpha \mtt)$ applies $\alpha$ until it cannot be further applied.
\end{itemize}

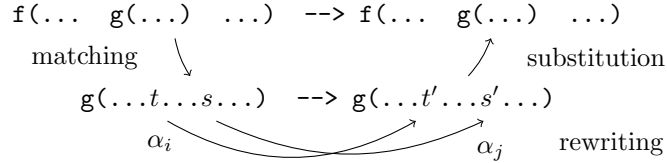
\begin{figure}\centering
\begin{tikzpicture}
	\node (TP) {\ttfamily f(... g(...) ...) --> f(... g(...) ...)};
	\node[below=1.5em of TP] (T) {\ttfamily g(...$t$...$s$...) --> g(...$t'$...$s'$...)};

	\draw (-2, -1.4) edge[bend right,->] (1.3, -1.4);
	\draw (-1.35, -1.4) edge[bend right=25,->] (2.2, -1.4);
	\draw (-1.9, -0.3) edge[bend right=10,->] (-1.7, -0.8);
	\draw (2, -0.8) edge[bend right=10,->] (2.3, -0.3);

	\node at (-3.1, -0.55) {matching};
	\node at (3.7, -0.55) {substitution};
	\node at (3.9, -1.7) {rewriting};
	\node at (-2.1, -1.7) {$\alpha_i$};
	\node at (2.3, -1.8) {$\alpha_j$};
\end{tikzpicture}
\caption{Behavior of \code{amatchrew}} \label{fig:matchrew}
\end{figure}

The \emph{match and rewrite} operator \code{matchrew} restricts the application of a strategy to a specific subterm of the subject term. Moreover, we can use it to obtain information about the term, by means of pattern matching or equational calculations, and bind it to new variables which can then be used to parameterize its substrategies. Its syntax is
%
\begin{maude}[mathescape]
matchrew $P(X_1, \ldots, X_n)$ s.t. $C$ by $X_1$ using $\alpha_1$, $\ldots$, $X_n$ using $\alpha_n$
\end{maude}
\noindent
where $P$ is a pattern with variables $X_1, \ldots, X_n$ among others, and $C$ is an optional equational condition. The \code{using} clauses associate variables in the pattern, which are matched by subterms of the matched term, with strategies that will be used to rewrite them. These variables must be distinct and must appear in the pattern. 
The semantics of the \texttt{amatchrew} operator is illustrated in Figure~\ref{fig:matchrew}. All matches of the pattern in the subject term that satisfy the (maybe empty) condition are considered. If there is none, the strategy fails. Otherwise, for each match, the subterms bound to each $X_i$ are extracted, rewritten according to $\alpha_i$ in \emph{parallel}\footnote{The various subterms are rewritten independently, and their progresses are interleaved in a fair way. However, the current implementation does not use hardware parallelism.}, and their solutions are put in place of the original subterms. Hence, all the results from all the subterms are combined to generate the reassembled solutions.

Strategies can be given a name and are defined in strategy
modules. Named strategies facilitate the definition of complex strategies, and extend the expressiveness of the
language by means of recursive and mutually recursive strategies. As
for functional and system modules, a \emph{strategy module} is
declared in Maude 
using the keywords
\begin{maude}[escapechar=^,basicstyle=\small\normalfont\ttfamily]
smod ^\syntaxSymb{ModuleName}^ is ^\syntaxSymb{DeclarationsAndStatements}^ endsm
\end{maude}
A typical strategy module imports the system module it will control, declares some strategies, and specifies their definitions.
\begin{maude}[escapechar=^, basicstyle=\small\normalfont\ttfamily]
strat ^\syntaxSymb{StratName}^ ^\normalfont[^ : ^\syntaxSymb{InputType}*^ ^\normalfont]^ @ ^\syntaxSymb{SubjectType}^ .

sd ^\syntaxSymb{StratCall}^ := ^\syntaxSymb{StrategyExpr}^ .
csd ^\syntaxSymb{StratCall}^ := ^\syntaxSymb{StrategyExpr}^ if ^\syntaxSymb{EqCondition}^ .
\end{maude}

	Let us come back to the Tower of Hanoi puzzle, to show an example of strategy module. First, we define
the auxiliary operator \code{third} which calculates the third of two posts in a
functional module \code{HANOI-AUX} (Figure~\ref{fig:HANOI-AUX}). Then, we define a recursive strategy \code{moveAll} to solve the puzzle given three arguments: the source and target posts, and the number of disks to be moved. This is specified in the strategy module \code{HANOI-SOLVE} in
Figure~\ref{fig:HANOI-SOLVE}. Now, we can execute it:

\begin{figure}[b]
\begin{maude}
fmod HANOI-AUX is
  protecting SET{Nat} .
  op third : Nat Nat ~> Nat .
  var N M K : Nat .
  ceq third(N, M) = K if N, M, K := 0, 1, 2 .
endfm
\end{maude}
\caption{\label{fig:HANOI-AUX}\code{HANOI-AUX} module}
\end{figure}

\begin{figure}
\begin{maude}
smod HANOI-SOLVE is
  protecting HANOI-RULES .
  protecting HANOI-AUX .

  strat moveAll : Nat Nat Nat @ Hanoi .

  var S T C : Nat .

  sd moveAll(S, S, C) := idle .
  sd moveAll(S, T, 0) := idle .
  sd moveAll(S, T, s(C)) := moveAll(S, third(S, T), C) ;
                            move[S <- S, T <- T] ;
                            moveAll(third(S, T), T, C) .
endsm
\end{maude}
\caption{\label{fig:HANOI-SOLVE}\code{HANOI-SOLVE} module}
\end{figure}

\begin{maude}
Maude> srew (0)[3 2 1] (1)[nil] (2)[nil] using moveAll(0, 2, 3) .

Solution 1
result Hanoi: (0)[nil] (1)[nil] (2)[3 2 1]

No more solutions.
\end{maude}

Although strategies control and restrict rewriting, they do not make the process deterministic. A strategy may allow multiple rewriting paths, produced by alternative local decisions that may appear during its execution. The \code{rewrite} command solves this nondeterminism by choosing a single alternative using a fixed criterion. On the contrary, the \code{search} command explores all the possible rule applications looking for a term matching a given goal. The \code{srewrite} command also performs an exhaustive exploration of the alternatives, but in this case, looking for strategy solutions. Hence, \code{srewrite} can be seen as a search, not in the complete rewriting tree of \code{search}, but in a subtree pruned by the effect of the strategy. How this tree is explored has implications for the command output and performance.

The \code{srewrite} command explores the rewriting graph following a fair policy which ensures that all solutions that are reachable in a finite number of steps are eventually found, unless the interpreter runs out of memory. Without being a breadth-first search, multiple alternative paths are explored in parallel. An alternative rewriting command \code{dsrewrite} (\emph{depth strategic rewrite}) explores the strategy rewriting graph in depth. Its syntax coincides with \code{srewrite} except for the starting keyword, which can be abbreviated to \code{dsrew}.
\begin{maude}[escapechar=^,basicstyle=\small\normalfont\ttfamily]
dsrewrite [^\syntaxSymb{Bound}^] in ^\syntaxSymb{ModuleName}^ : ^\syntaxSymb{Term}^ by ^\syntaxSymb{StrategyExpr}^ .
\end{maude}
The disadvantage of the depth-first exploration is its incompleteness,
since it can get lost in an infinite branch before finding some reachable solutions; see the next section 
for an example.
The advantages are that \code{dsrewrite} can be faster and requires less memory than \code{srewrite}.

\subsection{Logic Programming Running Example} \label{strat:prolog}

In addition to the specification of strategies to solve games, like in the previous Tower of Hanoi example, another typical application of strategies is the execution of operational semantics for programming languages, which are specified by means of rules that require in many cases to be executed in a specific way. Again, language expressions become terms in its Maude representation, operational semantics rules become rewrite rules, and strategies are used to control rewriting in the appropriate way. In this sense, the Maude strategy language has already been used to define the operational semantics of the parallel functional programming language Eden \cite{eden,MOPV07}, and also in the definition of modular structural operational semantics \cite{operational}.
In this section, the strategy language is used to define the semantics of a logic programming language similar to Prolog~\cite{prolog}, continuing with Examples~\ref{ex:LP-functional-syntax} and~\ref{ex:LP-system} in Section~\ref{sec:prolog-system}. Strategies will be used to discard failed proofs, to enforce the Prolog search strategy, and to implement advanced features like negation. Although it is also possible to implement the Prolog cut in this way, this feature will not be shown here (see \cite{maude-manual}).

The \code{rewrite} command is not useful as a logic programming
interpreter because it simply explores a single rewriting path, thus a
single proof path. This is clearly not enough to show multiple
solutions, but it may also be insufficient to find even a single
one. An admissible logic programming interpreter must consider all
possible proof paths and be able to resume them when the execution
arrives to a dead end. In Example~\ref{ex:LP-system} we used the
\code{search} command as a possible solution. Here, strategies will be
used instead.

First, we define an auxiliary predicate \code{isSolution} in module
\code{LP-EXTRA} in Figure~\ref{fig:LP-EXTRA} to decide whether a given
configuration is a solution. We then define, in the strategy module
\code{PROLOG} in Figure~\ref{fig:PROLOG}, the recursive strategy
\code{solve} that applies the \code{clause} rule in
Example~\ref{ex:LP-system} until a solution is found, and implicitly rejects any
rewriting path that does not end in one.

\begin{figure}[htb]
\begin{maude}
mod LP-EXTRA is
  protecting LP-SEMANTICS .
  op isSolution : Configuration -> Bool .
  var N : Nat .          var S : Substitution .
  var Pr : Program .     var Conf : Configuration .
  eq isSolution(< N | nil $ S | Pr >) = true .
  eq isSolution(Conf) = false [owise] .
endm
\end{maude}
\label{fig:LP-EXTRA}\caption{\code{LP-EXTRA} module\protect\footnotemark}
\end{figure}

\footnotetext{The equation attribute {\tt owise} denotes the word {\em otherwise} and allows for a simple equational definition of this predicate just by 
searching for the only interesting pattern.}

\begin{figure}[htb]
\begin{maude}
smod PROLOG is
  protecting LP-EXTRA .
  strat solve @ Configuration .
  var Conf : Configuration .
  sd solve := match Conf s.t. isSolution(Conf)
                       ? idle : (clause ; solve) .
endsm
\end{maude}
\caption{\label{fig:PROLOG}\code{PROLOG} module}
\end{figure}

Now, the \code{solve} strategy can be applied to the previous examples, where \texttt{family} is the kinship predicates program of Example~\ref{ex:relatives}. The exhaustive search of the \code{srewrite} command shows all reachable solutions for the initial predicate.

\begin{maude}[escapechar=^]
Maude> srew < 'parent('tom, x{1}) | family > using solve .

Solution 1
rewrites: 453
result Configuration: 
  < 3 | nil $ x{1} -> x{3} ; x{2} -> 'tom ; x{3} -> 'sally
      | ^\textit{(omitted)}^ >

Solution 2
rewrites: 489
result Configuration: 
  < 3 | nil $ x{1} -> x{3} ; x{2} -> 'tom ; x{3} -> 'erica
      | ^\textit{(omitted)}^ >

No more solutions.
rewrites: 605
\end{maude}

As the example above shows, the resulting configurations are not
easily readable, because they are overloaded with intermediate data like mappings on variables
that do not occur in the initial predicate, and the
full program, which has been omitted here. To display solutions in a
more readable form, we will use a wrapper strategy. It will use the
\texttt{solution} rule defined in \code{LP-SIMPLIFIER-BASE} (Figure~\ref{fig:LP-SIMPLIFIER-BASE}) that
restricts the substitution to the variables in a given set, after resolving them by transitivity. Some auxiliary
functions are required for this task.

\begin{figure}[hb]
\begin{maude}
mod LP-SIMPLIFIER-BASE is
  extending LP-SEMANTICS .
  sort VarSet .
  subsort Variable < VarSet .
  op empty : -> VarSet .
  op _;_ : VarSet VarSet -> VarSet [ctor assoc comm id: empty] .

  op occurs : Configuration -> VarSet .
  op simplify : Substitution VarSet -> Substitution .
  op solution : Substitution -> Configuration [ctor format (g! o)] .
  *** [...] the functions above are defined by equations
  var N : Nat .         var S : Substitution .
  var Pr : Program .    var VS : VarSet .

  rl [solution] : < N | nil $ S | Pr > 
               => solution(simplify(S, VS)) [nonexec] .
endm
\end{maude}
\caption{\label{fig:LP-SIMPLIFIER-BASE}\code{LP-SIMPLIFIER-BASE} module}
\end{figure}

The strategy \texttt{wsolve} in the strategy  module \code{PROLOG-SIMPLIFIER} in
Figure~\ref{fig:PROLOG-SIMPLIFIER} records in \texttt{VS} the variables that occur in the initial 
configuration predicate, then executes the previous \texttt{solve} strategy, and 
finally applies the \texttt{solution} rule with the initial variable set, thus 
restricting the substitution to those variables.

\begin{figure}
\begin{maude}
smod PROLOG-SIMPLIFIER is
  protecting LP-SIMPLIFIER-BASE .
  extending PROLOG .
  extending LP-EXTRA .
  extending LP-SEMANTICS .

  strat wsolve @ Configuration .

  var Conf : Configuration .   var VS : VarSet .
  
  sd wsolve := matchrew Conf s.t. VS := occurs(Conf)
                 by Conf using (solve ; solution[VS <- VS]) .
endsm
\end{maude}
\label{fig:PROLOG-SIMPLIFIER} \caption{\code{PROLOG-SIMPLIFIER} module\protect\footnotemark}
\end{figure}

Now, the previous example can be rerun with \texttt{wsolve}, obtaining more concise and clearer answers.

\begin{maude}
Maude> srew < 'parent('tom, x{1}) | family > using wsolve .

Solution 1
rewrites: 511
result Configuration: solution(x{1} -> 'sally)

Solution 2
rewrites: 569
result Configuration: solution(x{1} -> 'erica)

No more solutions.
rewrites: 641
\end{maude}

\footnotetext{Although the system modules \texttt{LP-EXTRA} and \texttt{LP-SEMANTICS} are already included via \texttt{PROLOG}, explicit \emph{extending} inclusions are written in Figure~\ref{fig:PROLOG-SIMPLIFIER}, since \texttt{PROLOG-SIMPLIFIER} is not a protecting extension of these. By the Maude convention, looser inclusions do not degrade the mode of stricter ones held by transitivity (see~\cite{maude-book}).}

We can also observe that the order in which solutions appear depends on the way the 
rewriting tree is explored. With the \texttt{dsrewrite} command the results will 
appear in the same order as in Prolog, because both explore the derivation tree in 
depth. However, the \texttt{srewrite} command will often obtain shallower solutions 
first.
\begin{maude}
Maude> dsrew < 'p(x{1}) | 'p(x{1}) :- 'q(x{1}) ; 'p('a) :- nil ;
                            'q('b) :- nil > using wsolve .

Solution 1
rewrites: 82
result Configuration: solution(x{1} -> 'b)

Solution 2
rewrites: 111
result Configuration: solution(x{1} -> 'a)

No more solutions.
rewrites: 117

Maude> srew < 'p(x{1}) | 'p(x{1}) :- 'q(x{1}) ; 'p('a) :- nil ;
                           'q('b) :- nil > using wsolve .

Solution 1
rewrites: 105
result Configuration: solution(x{1} -> 'a)

Solution 2
rewrites: 117
result Configuration: solution(x{1} -> 'b)

No more solutions.
rewrites: 95
\end{maude}
As we discussed before, the benefit of using \texttt{srewrite} is that all reachable solutions are shown. In Prolog and with \texttt{dsrewrite} some of them may be missed by going down a non-terminating branch.

\begin{maude}
Maude> dsrew < 'p(x{1}) | 'p(x{1}) :- 'p(x{1}) ; 'p('a) :- nil >
         using wsolve .

Debug(1)> abort .  *** non-terminating

Maude> srew < 'p(x{1}) | 'p(x{1}) :- 'p(x{1}) ; 'p('a) :- nil >
         using wsolve .

Solution 1
rewrites: 109
result Configuration: solution(x{1} -> 'a)

Debug(1)> abort .  *** non-terminating
\end{maude}

We consider now the negation feature.  In logic programming, the
concept of negation is 
complicated: facts and predicates express positive knowledge, so we
could either explicitly assert what is false or assume that any
predicate that cannot be derived from the facts is considered
false. The latter approach is known as \emph{negation as failure}: the
negation of a predicate holds if the predicate cannot be proved, no
matter the values its variables take. This cannot be expressed with
Horn clauses but it can be easily implemented using strategies and an
extra rewriting rule \texttt{negation} added in the extension \code{LP-EXTRA-NEGATION} (Figure~\ref{fig:LP-EXTRA-NEGATION}) of \texttt{LP-EXTRA}.  Like in ISO Prolog, in the system module
 in Figure~\ref{fig:LP-EXTRA-NEGATION}
negation is represented as a normal predicate\footnote{Hence, its
  argument is written as a term, i.e.\ brackets should be used instead
  of parentheses.} named \verb|\+|, which can be seen as the ASCII
representation of the \emph{not provable} symbol $\nvdash$.

\begin{figure}[h]
\begin{maude}
mod LP-EXTRA-NEGATION is
  including LP-EXTRA .

  var Q : Qid .                 var PL : PredicateList .
  var Conf : Configuration .    var NeTL : NeTermList .

  crl [negation] : 
    < N | '\+(Q[NeTL]), PL $ S | Pr > 
    => < N | PL $ S | Pr >
    if < N | Q(NeTL) $ S | Pr > => Conf .
endm
\end{maude}
\caption{\label{fig:LP-EXTRA-NEGATION}\code{LP-EXTRA-NEGATION} module}
\end{figure}

\begin{figure}[h]
\begin{maude}
smod PROLOG-NEGATION is
  protecting LP-EXTRA-NEGATION .

  strat solve-neg @ Configuration .

  var Conf : Configuration .

  sd solve-neg := match Conf s.t. isSolution(Conf) ? idle :
    ((clause | negation{not(solve-neg)}) ; solve-neg) .
endsm
\end{maude}
\caption{\label{fig:PROLOG-NEGATION}\code{PROLOG-NEGATION} module}
\end{figure}

The \texttt{negation} rule only removes the negation predicate from
the objectives list if its rewriting condition holds. By its own
semantics, negation never binds variables, so the substitution remains
unchanged. The initial term of the rewriting condition contains the
negated predicate as its only objective. Whether this term can be
rewritten to a solution configuration determines whether the negated
predicate can be satisfied. Hence, we need to control the condition
with a strategy that fails whenever that happens. We do so in the
strategy module
\code{PROLOG-NEGATION} in Figure~\ref{fig:PROLOG-NEGATION}.
The strategy is similar to the original \texttt{solve} strategy, but the \texttt{negation} rule can be applied when a negated predicate is on top of the objective list. The strategy \verb|not(solve-neg)| fails if \texttt{solve-neg} finds a solution for the negated predicate. Otherwise, it behaves like an \texttt{idle}, triggering the rule application. Thus, it is a suitable strategy for the rewriting condition.

We can illustrate the negation feature using the family tree example. Again, to obtain simplified results, we use the strategy \texttt{wsolve-neg}, defined from \texttt{solve-neg} as \texttt{wsolve} was defined from \texttt{solve} in \texttt{PROLOG-SIMPLIFIER} (a generic implementation is possible using parameterized strategy modules, see~\cite{maude-manual}). A predicate \texttt{'no-children} claims that someone does not have descendants:
\begin{maude}
Maude> srew < 'no-children('erica) | family ;
  'no-children(x{1}) :- '\+('parent[x{1}, x{2}]) > using wsolve-neg .

Solution 1
rewrites: 887
result Configuration: solution(empty)

No more solutions.
rewrites: 887

Maude> srew < 'no-children('mike) | family ;
  'no-children(x{1}) :- '\+('parent[x{1}, x{2}]) > using solve-neg .

No more solutions.
rewrites: 894
\end{maude}
The second predicate does not succeed since Mike is the father of John.

As mentioned at the beginning of this section, the Prolog cut has also
been implemented using strategies in this way (details will appear in
\cite{maude-manual}). These examples show that our framework can be
used to fully realize Kowalski's motto ``Algorithm = Logic + Control''
\cite{Kowalski79}, putting into practice the separation of concerns
allowed by our strategy language. The logic of a system (be it a game
or a language operational semantics or whatever) is declaratively
specified by means of equations and rules. The concrete, controlled
way of executing such rules, when desired or when necessary, is
written as a strategy on top of them. The separation between logic and
control allows us to have different controls for the same logic, like,
for example, having a logic programming interpreter which is complete
because it uses breadth-first search instead of the standard
depth-first search used by Prolog.

\vspace{1ex}

\noindent {\bf Further Reading}. 
The Maude strategy language was introduced in a series of conference papers \cite{towardsStrategy,strategies06,rewSemantics} and applied in different areas, including operational semantics (see \cite{eden,MOPV07,operational}, among others). More current work includes the extension of the language to include parameterized strategies \cite{pssm}, and the development of model-checking techniques for systems controlled by strategies \cite{modelCheckingStrat}. More examples and further details can be found in~\cite{maude-manual}.

\section{Object-Based Programming} \label{sec:oo}

In the design of distributed systems, the motto \emph{think globally, act 
locally} expresses the essential philosophy. Each object in a distributed system 
has only a quite limited partial view of the global state and can only 
\emph{act locally}, typically by communicating with other objects and changing 
its local state, to achieve some \emph{global} system goals. A well-designed 
distributed system uses such local actions to achieve a desired global behavior.  

Rewriting logic \cite{unified-tcs} is precisely a logic to express local 
actions in a concurrent system by means of rewrite rules.  As
explained in Section~\ref{sec:intro}, the concurrent systems that can be
specified in rewriting logic, and therefore in Maude, can be widely
different.  In this sense, 
rewriting logic and Maude are completely ecumenical, since they do not 
prescribe any particular style of concurrent, synchronous or asynchronous, 
interaction at all: any such style can be supported.  Nevertheless, the 
overwhelming majority of distributed systems and communication protocols can 
be most naturally expressed as made up of \emph{concurrent objects} having 
their own \emph{local states} that communicate with each other by 
\emph{message passing}.  Given the great importance of distributed 
object-based systems, Maude provides special support for such systems in 
the following ways: (i) a special notation is supported both in Maude and in 
its Full Maude extension; (ii) the \code{frewrite} command, when applied 
to object-based systems, provides an object and message fair rewriting 
strategy for simulation purposes; (iii) several kinds of 
\emph{external objects} allow regular Maude objects to interact with the 
external world; and (iv) using such external objects, a Maude object-based 
distributed system design can be seamlessly transformed (within Maude) into an 
actual distributed system \emph{implementation}. We discuss all these aspects 
in this section.

\subsection{Modeling Concurrent Object Systems in Maude}

To begin with, we explain below Maude's syntax support for concurrent
objects, and illustrate how a concurrent object system design can be
expressed in Maude using such a syntax.  As a running example we consider
the goal of designing a communication protocol that can achieve in-order, 
fault-tolerant communication in an asynchronous medium where messages can 
arrive out-of-order and can furthermore be lost. The first order of business 
is to specify the \emph{distributed states} of such a system that we will call 
\emph{configurations}. After this is done, we can then specify its 
\emph{concurrent behavior} by means of rewrite rules that define the 
\emph{local actions} that each object in such a system can perform to achieve 
in-order fault-tolerant communication. 

A system's distributed state or 
configuration can be naturally understood as a ``soup'' or ``ether'' medium 
in which both objects and messages ``float.''  In such a fluid medium, 
objects and messages can come together and participate in concurrent 
\emph{actions}. We can model such a fluid medium mathematically by means of 
\emph{structural axioms} of associativity and commutativity.  That is, we 
can think of a configuration as a \emph{multiset} of objects and messages.  
Since each object should have a unique identifier, the objects in the system 
should form a \emph{set}.  However, there can be several copies of a message 
floating around in the system.  

Given sorts \code{Object}, \code{Msg}, and \code{Configuration}, objects 
and messages can be declared as wished, and configurations of them can also 
be defined as needed. Indeed, what make them special, whatever the operators 
used to define them are the \code{object}, \code{msg} (or \code{message}) 
and \code{config} (or \code{configuration}) attributes. 
However, to simplify their use, several definitions are included in the 
predefined module \code{CONFIGURATION} in Figure~\ref{fig:CONFIGURATION}. 
With these declarations, objects of a class $C$ are record-like structures 
of the form 
\mbox{\code{$<$\ $O$\ :\ $C$\ $|$\ $a_1$\ :\ $v_1$,\ \ldots,\ $a_n$\ :\ $v_n$\ $>$}}, 
where $O$ is the name of the object and $v_i$ are the current values of its 
attributes. 

\begin{figure}[htb]
\begin{maude}
mod CONFIGURATION is
  sorts Attribute AttributeSet .
  subsort Attribute < AttributeSet .
  op none : -> AttributeSet [ctor] .
  op _,_ : AttributeSet AttributeSet -> AttributeSet 
    [ctor assoc comm id: none] .

  sorts Oid Cid Object Msg Portal Configuration .
  subsort Object Msg Portal < Configuration .
  op <_:_|_> : Oid Cid AttributeSet -> Object [ctor object] .
  op none : -> Configuration [ctor] .
  op __ : Configuration Configuration -> Configuration 
    [ctor config assoc comm id: none] .
  op <> : -> Portal [ctor] .
endm
\end{maude}
\caption{\label{fig:CONFIGURATION}\code{CONFIGURATION} module}
\end{figure}

The essential facts about concurrent object configurations are all stated 
in the \texttt{CONFIGURATION} module. 
They are multisets of objects and messages belonging, respectively,  
to the subsorts \code{Object} and \code{Msg}. These multisets are built with 
the ``empty syntax'' (juxtaposition) associative-commutative union operator 
\verb#__#, having \code{none} (empty configuration) as its identity element.
Objects themselves are constructed with the operator \code{<\_:\_|\_>}, which 
takes a \emph{name} or object identifier of sort \code{Oid}, belonging to an 
\emph{object class} whose name has sort \code{Cid} of class identifiers, and 
having an associative-commutative \emph{set} of attribute-value pairs of sort 
\code{AttributeSet} built with the associative-commutative set union operator 
\verb#_,_# with empty set \code{none} as its identity. Each such attribute-value 
pair has sort \code{Attribute} and can have any syntax we like. Likewise, we 
can use any syntax we like to define different kinds of messages. However, 
a message operator should be of sort \code{Msg} or a subsort of it, should have
the attribute \code{msg}, and the \emph{first argument} of any message operator 
should be the \code{Oid} of the message's \emph{addressee}.  

All these ideas can be illustrated by defining configurations of objects and 
messages for our fault-tolerant communication protocol in the  
functional module \code{FT-COMM-CONF} in Figure~\ref{fig:FT-COMM-CONF}.
Objects belong to classes \code{Sender} or \code{Receiver}.  
In addition to importing the \code{CONFIGURATION} module, the
\code{FT-COMM-CONF} module imports \code{QID-LIST} from
Maude's prelude.  This functional module provides a
sort \code{QidList} formed from elements of sort \code{Qid}
using the associative concatenation (empty syntax) operator 
\verb#__#, having \code{nil} (empty list) as its identity element.
Notice that the module \code{FT-COMM-CONF} provides\footnote{%
For a discussion on attributes not explained here, such as \texttt{gather},
please see \cite{maude-manual}.}
 definitions
for class names, the attribute definitions of the objects in these classes, 
and the messages. 

\begin{figure}[htb]
\begin{maude}
mod FT-COMM-CONF is 
  extending CONFIGURATION .
  protecting NAT + QID-LIST .

  ops Sender Receiver : -> Cid [ctor] .
  subsort Qid < Oid .

  op cnt:_ : Nat -> Attribute [ctor gather (&)] .
  op buff:_ : QidList -> Attribute [ctor gather (&)] .
  op snd:_ : Oid -> Attribute [ctor gather (&)] .
  op rec:_ : Oid -> Attribute [ctor gather (&)] .

  op to_from_val_cnt_ : Oid Oid Qid Nat -> Msg [ctor msg] .
  op to_from_ack_ : Oid Oid Nat -> Msg [ctor msg] .
endm
\end{maude}
\caption{\label{fig:FT-COMM-CONF}\code{FT-COMM-CONF} module}
\end{figure}

In a typical configuration for our desired fault-tolerant in-order 
communication protocol we will have senders and receivers, together with messages
`traveling' between them:

\begin{maude}
  < 'Alice : Sender | buff: 'a 'b 'c 'd, rec: 'Bob, cnt: 0 > 
  < 'Bob : Receiver | buff: 'a, snd: 'Alice, cnt: 1 >
  (to 'Alice from 'Bob ack 0)
\end{maude}

\noindent In this configuration, there is a sender object \code{'Alice} and 
a receiver object \code{'Bob} of respective classes \code{Sender} and 
\code{Receiver}. Both senders and receivers have a buffer attribute 
\code{buff:} whose value is a list of quoted identifiers, either remaining to be 
sent by the sender, or already received by the receiver. Both  also have 
a counter attribute \code{cnt:} which is used to ensure in-order communication.  
Initially, the value of the \code{cnt:} attribute is $0$ for both senders and 
receivers, which gets increased 
as sends are received and acknowledged. 
Furthermore, to 
establish the target of the communication, sender objects have a receiver 
attribute \code{rec:} with the name of the object to which values in the list 
should be sent. Likewise, receiver objects have a sender attribute \code{snd:} 
with the name of the sender from which data is expected. Sender objects like 
\code{'Alice} send values in messages such as 
\code{to 'Bob from 'Alice val 'a cnt 0}. 
This message means that it is the first value in the list being 
transmitted and its contents is \code{'a}. In the above configuration this 
first message was already received by \code{'Bob}, who now stores it in its 
buffer and is awaiting the second value, whose counter will be $1$. However, 
due to the asynchronous nature of the communication, sender \code{'Alice} is not 
yet aware that the first value has already been received and is still holding 
it in its send buffer in case it was lost and has to be re-sent. In the meantime, 
receiver \code{'Bob} did send a message 
\code{to 'Alice from 'Bob ack 0} acknowledging 
receipt of the first value \code{'a}. But this acknowledgment has not yet been 
received by sender \code{'Alice}.  

Of course, the functional module \code{FT-COMM-CONF} \emph{does not
do anything}. It describes, if you will, the \emph{statics}, i.e., just the 
distributed states of our system. \emph{Actions} themselves, the system's  
\emph{dynamics}, are defined in the system module \code{FT-COMM} in 
Figure~\ref{fig:FT-COMM}.
The rules in  \code{FT-COMM} are almost self-explanatory. Sender objects 
send the first value in their current list, plus a counter, to the receiver
with the \code{snd} rule. However, they still keep the sent value in
their buffer until an acknowledgment is received. If the expected
acknowledgment is received, the sent value can be cleared from the send
buffer and the counter is increased for the next value to be sent (rule 
\code{rec-ack1}).  Note that the case where the sender receives
an acknowledgment for counter value \code{M} with an empty buffer
will not happen for starting configurations with only objects.
The case of a duplicated acknowledgment message that \emph{was} already 
received before is handled by rule \code{rec-ack2}, where the acknowledgment
message is just discarded. Receiver objects perform two actions. Rule 
\code{rec1} describes the case where the ``expected'' value arrives, is put 
into the receive buffer, the counter is increased, and an acknowledgment 
message is sent to the sender. The case where the sent value \emph{was} 
already received is handled by the rule \code{rec2}, where the receiver's local 
state does not change, but an acknowledgment message is nevertheless sent, 
since a previous acknowledgment may have been lost.

\begin{figure}[htb]
\begin{maude}
mod FT-COMM is 
  including FT-COMM-CONF .
 
  var Q : Qid .   var L : QidList .   vars N M : Nat .   vars A B : Oid .
 
  rl [snd] : < A : Sender | buff: Q L, rec: B, cnt: M >
    => < A : Sender | buff: Q L, rec: B, cnt: M >
       (to B from A val Q cnt M)
    [print "[snd]: " A " sends " Q " to " B] .
    
  rl [rec1] :
    (to B from A val Q cnt M)
    < B : Receiver | buff: L, snd: A, cnt: M >
    => < B : Receiver | buff: L Q, snd: A, cnt: s M >
       (to A from B ack M)
    [print "[rec1]: " B " receives new " Q " from " A] .
    
  crl [rec2] :
    (to B from A val Q cnt N)
    < B : Receiver | buff: L, snd: A, cnt: M >
    => < B : Receiver | buff: L, snd: A, cnt: M >
       (to A from B ack N)
    if N < M
    [print "[rec2]: " B " receives old " Q " from " A] .
    
  rl [rec-ack1] :
    (to A from B ack M)
    < A : Sender | buff: Q L, rec: B, cnt: M >
    => < A : Sender | buff: L, rec: B, cnt: s M >
    [print "[rec-ack1]: " A " receives 1st ack " M " from " B] .
    
  crl [rec-ack2] :
    (to A from B ack N)
    < A : Sender | buff: L, rec: B, cnt: M >
    => < A : Sender | buff: L, rec: B, cnt: M >
    if N < M
    [print "[rec-ack2]: " A " receives old ack " N " from " B] .
endm
\end{maude}
\caption{\label{fig:FT-COMM}\code{FT-COMM} module}
\end{figure}

Although not necessary, for illustration purposes, to show the progress of the 
computation we have added \code{print} attributes
to each rule. In general, the print attribute allows one to specify information 
to be printed when a statement (equation, membership axiom, rule, or strategy) is 
executed, providing a minimized and flexible trace capability. 
If printing is turned on, when a statement with a print attribute is applied 
the pattern following \code{print} is instantiated using the corresponding 
matching substitution.

The \code{FT-COMM} module does indeed ensure in-order fault-tolerant
communication. Furthermore, if the sender was sending a list of length
$k$, counters in the sender and the receiver were originally $0$, and
the receiver's buffer was originally empty, there is a terminating
rewrite sequence in whose final state both the sender and the receiver
counters have the same value $k$, the sender's buffer is empty, and the original
list is now in the receiver's buffer. 

We use Maude's \code{frewrite} command to explore the behavior of \code{FT-COMM}.
As discussed in Section \ref{sec:prolog-system}, the \code{frewrite} command
implements a rule and position fair rewriting strategy. In the special case 
of object-message configurations, such as the \code{FT-COMM} configurations, 
\code{frewrite} implements  an object-message fair strategy\footnote{The precise definition of what an object-message is, the full 
specification of \code{frewrite}, and more examples can be found 
in~\cite{maude-book}.}. Roughly speaking, 
in each round, the strategy attempts to apply object-message rules to all 
existing object-message pairs and then attempts a single non-object-message 
rewrite of the resulting configuration using the remaining rules. %
%

We see in the output below that the protocol terminates as expected.

\begin{maude}
Maude> frew
  < 'Alice : Sender | cnt: 0, buff: 'a 'b 'c 'd, rec: 'Bob > 
  < 'Bob : Receiver | cnt: 0, buff: nil, snd: 'Alice > .
result Configuration: 
  < 'Alice : Sender | cnt: 4, buff: nil, rec: 'Bob > 
  < 'Bob : Receiver | cnt: 4, buff: 'a 'b 'c 'd, snd: 'Alice >
\end{maude}

\noindent 
To see how the protocol progresses, let us rewrite one step at
a time using Maude's \code{continue} (abbreviated \code{cont}) command.
We see that first \code{'Alice} sends \code{'a} to \code{'Bob} with count \code{0}. 

\begin{maude}
Maude> frew [1] < 'Alice : Sender | cnt: 0, buff: 'a 'b 'c 'd, rec: 'Bob > 
                < 'Bob : Receiver | cnt: 0, buff: nil, snd: 'Alice > .
result (sort not calculated): (
  < 'Alice : Sender | cnt: 0, buff: 'a 'b 'c 'd, rec: 'Bob > 
  to 'Bob from 'Alice val 'a cnt 0)
  < 'Bob : Receiver | cnt: 0, buff: nil, snd: 'Alice >
\end{maude}

\noindent
In the next step, \code{'Bob} receives \code{'a} and sends an acknowledgment 
message to \code{'Alice} with count \code{0}. 

\begin{maude}
Maude> cont 1 .
result Configuration: 
  < 'Alice : Sender | cnt: 0, buff: 'a 'b 'c 'd, rec: 'Bob > 
  < 'Bob : Receiver | cnt: 1, buff: 'a, snd: 'Alice > 
  to 'Alice from 'Bob ack 0
\end{maude}

\noindent
In the third step \code{'Alice} sends \code{'a} to \code{'Bob} with count 
\code{0} again, since it has not yet received an \code{ack} message. 

\begin{maude}
Maude> cont 1 .
result (sort not calculated): (
  < 'Alice : Sender | cnt: 0, buff: 'a 'b 'c 'd, rec: 'Bob > 
  to 'Bob from 'Alice val 'a cnt 0) 
  < 'Bob : Receiver | cnt: 1, buff: 'a, snd: 'Alice > 
  to 'Alice from 'Bob ack 0
\end{maude}

\noindent
In the fourth step, \code{'Alice} receives the count \code{0} acknowledgment, 
increments its counter, and removes \code{'a} from its list. \emph{Also}, 
\code{'Bob} receives the repeated \code{'a} with count \code{0} and sends another
\code{ack}.  This is two rewrites, although the command was
to continue 1 step.  This is because the \code{frewrite} strategy
attempts to deliver a message to each object in a given round.

\begin{maude}
Maude> cont 1 .
result Configuration: 
  < 'Alice : Sender | cnt: 1, buff: 'b 'c 'd, rec: 'Bob > 
  < 'Bob : Receiver | cnt: 1, buff: 'a, snd: 'Alice >
  to 'Alice from 'Bob ack 0
\end{maude}

\noindent
Continue again, \code{'Alice} sends \code{'b} to \code{'Bob} with count \code{1}.  

\begin{maude}
Maude> cont 1 .
result (sort not calculated): (
  < 'Alice : Sender | cnt: 1, buff: 'b 'c 'd, rec: 'Bob > 
  to 'Bob from 'Alice val 'b cnt 1 
  < 'Bob : Receiver | cnt: 1, buff: 'a, snd: 'Alice > 
  to 'Alice from 'Bob ack 0
\end{maude}

To see the difference between the strategies of the \code{rewrite} and 
\code{frewrite} commands, we use a configuration with two instances of the 
protocol, that is, two sender-receiver pairs. Using \code{frewrite} to execute 
the parallel protocol sessions, with the \code{print} attribute activated 
we see that activity of the two sessions is interleaved:

\begin{maude}
Maude> set print attribute on .
Maude> frew [24] 
  < 'Alice : Sender | cnt: 0, buff: ('a 'b 'c 'd), rec: 'Bob > 
  < 'Ada : Sender | cnt: 0, buff: ('a 'b 'c 'd), rec: 'Boris > 
  < 'Bob : Receiver | cnt: 0, buff: nil, snd: 'Alice > 
  < 'Boris : Receiver | cnt: 0, buff: nil, snd: 'Ada >  .
[snd]: 'Alice sends 'a to 'Bob
[snd]: 'Ada sends 'a to 'Boris
[rec1]: 'Bob receives new 'a from 'Alice
[rec1]: 'Boris receives new 'a from 'Ada
[snd]: 'Alice sends 'a to 'Bob
[snd]: 'Ada sends 'a to 'Boris
[rec-ack1]: 'Alice receives 1st ack 0 from 'Bob
[rec-ack1]: 'Ada receives 1st ack 0 from 'Boris
[rec2]: 'Bob receives old 'a from 'Alice
[rec2]: 'Boris receives old 'a from 'Ada
    ...
[snd]: 'Alice sends 'd to 'Bob
[snd]: 'Ada sends 'd to 'Boris
[rec-ack2]: 'Alice receives old ack 2 from 'Bob
[rec-ack2]: 'Ada receives old ack 2 from 'Boris
[rec1]: 'Bob receives new 'd from 'Alice
[rec1]: 'Boris receives new 'd from 'Ada
[snd]: 'Alice sends 'd to 'Bob
[snd]: 'Ada sends 'd to 'Boris
[rec-ack1]: 'Alice receives 1st ack 3 from 'Bob
[rec-ack1]: 'Ada receives 1st ack 3 from 'Boris
[rec2]: 'Bob receives old 'd from 'Alice
[rec2]: 'Boris receives old 'd from 'Ada
rewrites: 69 in 2ms cpu (3ms real) (23334 rewrites/second)
result Configuration: 
  < 'Alice : Sender | cnt: 4, buff: nil, rec: 'Bob > 
  < 'Ada : Sender | cnt: 4, buff: nil, rec: 'Boris > 
  < 'Bob : Receiver | cnt: 4, buff: ('a 'b 'c 'd), snd: 'Alice > 
  < 'Boris : Receiver | cnt: 4, buff: ('a 'b 'c 'd), snd: 'Ada >
  (to 'Alice from 'Bob ack 3) 
  (to 'Ada from 'Boris ack 3)
\end{maude}

\noindent
If instead we use \code{rewrite}, the rules are applied first to objects 
and messages in one session, and when that terminates, the rules are
applied to objects and messages of the other session.  

\begin{maude}
Maude> rew [24]  
  < 'Alice : Sender | cnt: 0, buff: ('a 'b 'c 'd), rec: 'Bob > 
  < 'Ada : Sender | cnt: 0, buff: ('a 'b 'c 'd), rec: 'Boris > 
  < 'Bob : Receiver | cnt: 0, buff: nil, snd: 'Alice > 
  < 'Boris : Receiver | cnt: 0, buff: nil, snd: 'Ada > .
[snd]: 'Alice sends 'a to 'Bob
[rec1]: 'Bob receives new 'a from 'Alice
[rec-ack1]: 'Alice receives 1st ack 0 from 'Bob
[snd]: 'Alice sends 'b to 'Bob
    ...
[snd]: 'Alice sends 'd to 'Bob
[rec1]: 'Bob receives new 'd from 'Alice
[rec-ack1]: 'Alice receives 1st ack 3 from 'Bob
[snd]: 'Ada sends 'a to 'Boris
[rec1]: 'Boris receives new 'a from 'Ada
[rec-ack1]: 'Ada receives 1st ack 0 from 'Boris
    ...
[snd]: 'Ada sends 'd to 'Boris
[rec1]: 'Boris receives new 'd from 'Ada
[rec-ack1]: 'Ada receives 1st ack 3 from 'Boris
rewrites: 25 in 1ms cpu (1ms real) (18037 rewrites/second)
result Configuration: 
  < 'Alice : Sender | cnt: 4, buff: nil, rec: 'Bob > 
  < 'Ada : Sender | cnt: 4, buff: nil, rec: 'Boris > 
  < 'Bob : Receiver | cnt: 4, buff: ('a 'b 'c 'd), snd: 'Alice > 
  < 'Boris : Receiver | cnt: 4, buff: ('a 'b 'c 'd), snd: 'Ada >
\end{maude}

\noindent
In the case of finite behaviors, in the end the result is the same, but 
if the protocol execution did not terminate, then using \code{rewrite} one of the 
sessions might never execute, while using \code{frewrite} both sessions 
make progress. 

Note that \code{FT-COMM} is a protocol that not only ensures in-order
communication, but is also \emph{fault-tolerant}. But can we also
\emph{model} a \emph{faulty} environment where messages can be lost?
Yes, we can do so in the system module \code{FT-COMM-IN-FAULTY-ENV}
in Figure~\ref{fig:FT-COMM-IN-FAULTY-ENV},
which adds such a faulty environment to \code{FT-COMM}.
What is remarkable about the communication protocol specified in 
\code{FT-COMM} is that it \emph{still works} in this faulty environment 
under suitable \emph{fairness assumptions}. Of course, if as soon as every 
message is sent it is immediately destroyed by the \code{loss1} or the
\code{loss2} rules, no communication will ever happen. But this is
clearly an unfair behavior which makes the protocol's rules hopeless. By
assuming fair executions and defining an \emph{equational abstraction}
\cite{equational-abstraction-tcs} that collapses a multiset of messages
into a set, the correctness of the \code{FT-COMM} protocol in such a
faulty environment can actually be model checked using Maude's LTLR
model checker \cite{DBLP:journals/scp/BaeM15}.
For more on Maude's LTLR model checker see Section~\ref{sec:tools-apps}.

\begin{figure}
\begin{maude}
mod FT-COMM-IN-FAULTY-ENV is
  including FT-COMM .

  var Q : Qid .   var M : Nat .   vars A B : Oid .

  rl [loss1] : (to B from A val Q cnt M) => none
    [print "[loss1]: lost val " Q " to " B] .
  rl [loss2] : (to A from B ack M) => none
    [print "[loss2]: lost ack " M " to " A] .
endm
\end{maude}
\caption{\label{fig:FT-COMM-IN-FAULTY-ENV}\code{FT-COMM-IN-FAULTY-ENV} module}
\end{figure}

If we repeat the above \code{frew} command in the 
\code{FT-COMM-IN-FAULTY-ENV} module, we see that the first few steps are the same 
as when done in the module \code{FT-COMM}. However, instead of \code{'Alice} 
receiving the \code{ack} from \code{'Bob} in the fourth step, the message 
is lost.

\begin{maude}
Maude> frew [1] < 'Alice : Sender | cnt: 0, buff: 'a 'b 'c 'd, rec: 'Bob > 
                < 'Bob : Receiver | cnt: 0, buff: nil, snd: 'Alice > .
[snd]: 'Alice sends 'a to 'Bob
[rec1]: 'Bob receives new 'a from 'Alice
[snd]: 'Alice sends 'a to 'Bob
[loss2]: lost ack 0 to 'Alice
[rec2]: 'Bob receives old 'a from 'Alice
...
\end{maude}

But is this all? Has the \code{FT-COMM} example illustrated all there
is to say about distributed objects in Maude? Not at all. It has
illustrated the \emph{most basic} possibilities, but many more remain.
Here are some: (1) The Full Maude extension supports an even more
expressive syntax for objects in which object classes can be structured
in \emph{multiple inheritance} hierarchies, rewrite rules can be
specified more succinctly and are automatically inherited by subclasses
\cite{ooconc,maude-book}. Furthermore, such class inheritance solves the
well-known \emph{inheritance anomaly} between subclassing and
concurrency \cite{anomaly}. (2) Configurations need not be \emph{flat}:
they can have a \emph{nested} structure ---what we call a \emph{Russian
dolls} structure. Furthermore, such a nested structure can provide very
useful mechanisms for \emph{meta-object-based reflection}
\cite{ecoop-02}. (3) Many distributed algorithms use time, and sometimes
physical space, in an essential way. Both time and space can be modeled
in an object-based manner in Maude using the Real-Time Maude tool
\cite{journ-rtm} (see, e.g., \cite{OlveczkyT09,liu-etal-MANETS-JLAMP}
for applications to sensor networks and to mobile ad-hoc networks). (4)
Not only time, but also \emph{randomness} in distributed object systems
can be modeled by \emph{probabilistic rewrite rules} \cite{qapl05}
(see, e.g., \cite{katelman-meseguer-hou,ACC-maude-techreport} for two
applications, respectively to sensor protocols and to cloud storage
systems).
Maude has been used for the specification and verification of many 
other distributed systems;
see, e.g., \cite{maude-book,20-years,maude-2-decades} for surveys on 
additional applications. 

\subsection{External Objects}
\label{sec:external-objects}

Maude objects should be able to interact by message-passing 
with  a variety of \emph{external
objects} that represent external entities with state, including the
user regarded as another external object.
Any configuration of Maude objects that wish to exchange messages with external objects
must include a special portal constructor, defined in module
\code{CONFIGURATION}:

\begin{maude}
  sort Portal .
  subsort Portal < Configuration .
  op <> : -> Portal [ctor] .
\end{maude}

From an implementation point of view, the main purpose of having
a portal subterm in a configuration is to avoid the degenerate case of a
configuration that consists just of an object waiting for a
message from outside of the configuration. This would be problematic
because the special behavior for object-message rewriting and exchanging
messages with external objects is attached to the configuration constructor:%
\footnote{While a single object or 
message has sort \code{Configuration} there is no configuration 
constructor for such a degenerate configuration.
Requiring a portal term ensures that there is a configuration
constructor for configurations which otherwise have only a single
object or message.
}

\begin{maude}
  op __ : Configuration Configuration -> Configuration 
    [ctor config assoc comm id: none] .
\end{maude}

Exchanging messages with external objects is enabled by the \code{erewrite} 
command. It performs fair rewriting and handles incoming and outgoing messages. 
It checks for messages in a configuration that are addressed to external objects, 
and checks for messages from external objects that are queued, waiting to enter 
a configuration containing a specific object.

Certain predefined external objects are available and some of them are
object managers that can create more ephemeral external objects that
represent entities such as files and sockets or, as we will see in
Section~\ref{sec:metaInterpreter}, virtual copies of the Maude interpreter 
itself.

\subsubsection{Standard Streams}
\label{sec:streams}

Each Unix process has three I/O channels, called standard streams:
standard input (\texttt{stdin}), standard output (\texttt{stdout}), and standard error (\texttt{stderr}).
In Maude, these are represented as three unique external objects, that are
defined in a predefined module \code{STD-STREAM}. 
Because some of the messages that are useful for streams are also useful for
file I/O, these messages are pulled out into a module
\code{COMMON-MESSAGES}. This module together with a fragment of
the module \code{STD-STREAM} are shown in Figure~\ref{fig:stdStream}.

\begin{figure}
\begin{maude}
mod COMMON-MESSAGES is
  protecting STRING .
  including  CONFIGURATION .

  op gotLine : Oid Oid String -> Msg [ctor msg ...] .
  op write : Oid Oid String -> Msg [ctor msg ...] .
  op wrote : Oid Oid -> Msg [ctor msg ...] .
endm
\end{maude}

\begin{maude}
mod STD-STREAM is
  including COMMON-MESSAGES .

  op getLine : Oid Oid String -> Msg [ctor msg ...] .
  op stdin : -> Oid [special (...)] .
  op stdout : -> Oid [special (...)] .
  op stderr : -> Oid [special (...)] .
endm
\end{maude}
\caption{\label{fig:stdStream}Fragments of modules \code{COMMON-MESSAGES} and  \code{STD-STREAM} (please, note the ellipses)}
\end{figure}

After more than 20 years you can now write a ``Hello World!'' program in Maude.
Module \code{HELLO} in Figure~\ref{fig:HELLO} shows a very simple program 
implementing an interaction with the user, which is asked to introduce 
his/her name to be properly greeted.
The equation for \code{run} produces a starting configuration, 
containing the portal, a user object to receive messages, and a message to 
\code{stdin} to read a line of text from the keyboard. When \code{stdin} 
has a line of text, it sends the text to the requesting object in a 
\code{gotLine} message. 

\begin{figure}
\begin{maude}
mod HELLO is
  including STD-STREAM .

  op myClass : -> Cid .
  op myObj : -> Oid .
  op run : -> Configuration .

  var O : Oid .
  var A : AttributeSet .
  var S : String .
  var C : Char .

  eq run
    = <>
      < myObj : myClass | none >
      getLine(stdin, myObj, "What is your name? ") .
  rl < myObj : myClass | A >
     gotLine(myObj, O, S) 
  => < myObj : myClass | A >
     if S =/= ""
     then write(stdout, myObj, "Hello " + S)
     else none
     fi .
endm
\end{maude}
\caption{\label{fig:HELLO}\code{HELLO} module}
\end{figure}

\begin{maude}
Maude> erew run .
What is your name? Joe
Hello Joe
result Configuration: <> wrote(myObj, stdout) < myObj : myClass | none >
\end{maude}

\subsubsection{File I/O}

Unlike standard streams, of which there are exactly three, a Unix process
may have many different files open at any one time. Thus,
in order to create new file handle objects as needed, we have a unique
external object called \code{fileManager}. To open a file, the 
\code{fileManager} is sent a message \code{openFile}. On success, an 
\code{openedFile} message is returned, with the name of an external object 
that is a handle on the open file as one of its arguments and to which 
messages to read and write the file can be directed. On failure, a 
\code{fileError} message is returned, with a text explanation of why the
file could not be opened as one of its arguments. These messages are defined 
in the module \code{FILE}, which is distributed as part of the Maude system.
A fragment of this predefined module is shown in Figure~\ref{fig:file}.

\begin{figure}
\begin{maude}
mod FILE is
  including COMMON-MESSAGES .
  protecting INT .
   ...
  op file : Nat -> Oid [ctor] .

  op openFile : Oid Oid String String -> Msg [ctor msg ...] .
  op openedFile : Oid Oid Oid -> Msg [ctor msg ...] .

  op getLine : Oid Oid -> Msg [ctor msg ...] .
  op getChars : Oid Oid Nat -> Msg [ctor msg ...] .
  op gotChars : Oid Oid String -> Msg [ctor msg ...] .
  op flush : Oid Oid -> Msg [ctor msg ...] .
  op flushed : Oid Oid -> Msg [ctor msg ...] .
   ...
  op closeFile : Oid Oid -> Msg [ctor msg ...] .
  op closedFile : Oid Oid -> Msg [ctor msg ...] .

  op fileError : Oid Oid String -> Msg [ctor msg ...] .
  op fileManager : -> Oid [special (...)] .
endm
\end{maude}
\caption{\label{fig:file}Fragment of the \code{FILE} module (notice the ellipses)}
\end{figure}

\begin{figure}
\begin{maude}
fmod MAYBE{X :: TRIV} is
  sort Maybe{X} .
  subsort X$Elt < Maybe{X} .
  op null : -> Maybe{X} .
endfm

view Oid from TRIV to CONFIGURATION is
  sort Elt to Oid .
endv

mod COPY-FILE is
  inc FILE .
  pr MAYBE{Oid} .

  op myClass : -> Cid .
  op myObj : -> Oid .
  ops in:_ out:_ : Maybe{Oid} -> Attribute .
  ops inFile:_ outFile:_ : String -> Attribute .

  op run : String String -> Configuration .
  vars Text Original Copy : String .
  vars FHIn FHOut : Oid .
  var  Attrs : AttributeSet .

  eq run(Original, Copy)
   = <>
     < myObj : myClass | in: null,  inFile: Original, 
                         out: null, outFile: Copy >
     openFile(fileManager, myObj, Original, "r") .
  rl < myObj : myClass | in: null, outFile: Copy, Attrs >
     openedFile(myObj, fileManager, FHIn)
  => < myObj : myClass | in: FHIn, outFile: Copy, Attrs >
     openFile(fileManager, myObj, Copy, "w") .
  rl < myObj : myClass | in: FHIn, out: null, Attrs >
     openedFile(myObj, fileManager, FHOut)
  => < myObj : myClass | in: FHIn, out: FHOut, Attrs >
     getLine(FHIn, myObj) .
  rl < myObj : myClass | in: FHIn, out: FHOut, Attrs >
     gotLine(myObj, FHIn, Text)
  => < myObj : myClass | in: FHIn, out: FHOut, Attrs >
     if Text == ""
     then closeFile(FHIn, myObj)
          closeFile(FHOut, myObj)
     else write(FHOut, myObj, Text)
     fi .
  rl < myObj : myClass | in: FHIn, out: FHOut, Attrs >
     wrote(myObj, FHOut)
  => < myObj : myClass | in: FHIn, out: FHOut, Attrs >
     getLine(FHIn, myObj) .
  rl < myObj : myClass | in: FHIn, out: FHOut, Attrs >
     closedFile(myObj, FHIn)
     closedFile(myObj, FHOut)
  => none .
endm
\end{maude}
\caption{\label{fig:ex-file-copy}File copy with external objects}
\end{figure}

The \code{COPY-FILE} module in Figure~\ref{fig:ex-file-copy} illustrates 
the basic use of files. It specifies a simple algorithm to copy files. 
In this case, the \code{run} operator takes two arguments: the name of the 
file to be copied and the name of the new file. As for the previous example, the 
equation for \code{run} produces a starting configuration, containing the 
portal, a user object to receive messages, and an initial message to open the 
original file. Once it is opened, the new file is created. Notice the 
\code{"w"} argument of the \code{openFile} message. Once both files are opened,
a loop in which a line is read from the original file and written in the 
copy file is initiated. This loop ends when the end of the file is reached. 
Both files are then closed. 

\subsubsection{Socket I/O}

Maude's support for sockets works in a similar way to that for
files. There is a unique object, \code{socketManager}, defined
in a module \code{SOCKET}, and messages
to this object can be used to create client or server TCP internet
sockets.  This feature is crucial to \emph{deploy} a Maude
concurrent object system as a \emph{distributed system}.
Consider, for example, a configuration containing 1,000 Maude objects.
They of course can be run on a single Maude interpreter, but then rewrites
corresponding to message sends and receives are necessarily
sequentialized by the interpreter.  That is, concurrency is only
\emph{simulated} that way as an interleaving of rewrite steps.
Using sockets we can easily distribute those 1,000 Maude objects
into, say, 10 machines, each running its own Maude interpreter and
holding a configuration of, say, 100 objects.  The number of objects is 
immaterial and is just  given for concreteness' sake; furthermore, new
objects can be created and destroyed, and Maude objects may
communicate not just with other Maude objects but also with various
external objects.  The three key conceptual points to keep in mind are:
(1) now the configuration or ``soup'' of 1,000 objects and messages
has been distributed into 10 such soups distributed over 10
machines and communicating through sockets; (2) message passing communication
between Maude objects belonging to one of those 10 sub-configurations
will happen as usual by rewriting performed by the Maude interpreter
for that configuration; (3) instead, a message generated in sub-configuration, say, number 2 but addressed to another object in
sub-configuration number 8 will be: (i) transformed into a string,
(ii) sent through a socket linking those two configurations, (iii)
transformed back into a message in sub-configuration 8, and (iv)
delivered to the addressee object there, that will then consume it
by an appropriate rewrite rule.

A fragment of this module is shown in Figure~\ref{fig:socket}.
Note that a number of details, such as DNS look-up, are hidden
by the \code{createClientTcpSocket}, which just takes a domain name
and a port number. 
For additional details on socket external objects in Maude 
see \cite{maude-manual,maude-book}. 

\begin{figure}
\begin{maude}
mod SOCKET is
  protecting STRING .
  including CONFIGURATION .

  op socket : Nat -> Oid [ctor] .

  op createClientTcpSocket : Oid Oid String Nat -> Msg [ctor msg ...] .
  op createServerTcpSocket : Oid Oid Nat Nat -> Msg [ctor msg ...] .
  op createdSocket : Oid Oid Oid -> Msg [ctor msg ...] .

  op acceptClient : Oid Oid -> Msg [ctor msg ...] .
  op acceptedClient : Oid Oid String Oid -> Msg [ctor msg ...] .
  op send : Oid Oid String -> Msg [ctor msg ...] .
  op sent : Oid Oid -> Msg [ctor msg ...] .
  op receive : Oid Oid -> Msg [ctor msg ...] .
  op received : Oid Oid String -> Msg [ctor msg ...] .
  op closeSocket : Oid Oid -> Msg [ctor msg ...] .
  op closedSocket : Oid Oid String -> Msg [ctor msg ...] .

  op socketError : Oid Oid String -> Msg [ctor msg ...] .
  op socketManager : -> Oid [special (...)] .
endm
\end{maude}
\caption{\label{fig:socket}Fragment of the \code{SOCKET} module (notice the ellipses)}
\end{figure}

\vspace{1ex}

\noindent {\bf Further Reading}.  The two most complete references for
the semantics of object-based systems in Maude are probably
\cite{ooconc,maude-book}.  How meta-objects that can control
other objects (or entire object sub-configurations) in ``Russian
dolls'' distributed architectures can easily be defined in Maude
is explained in \cite{ecoop-02}.  The specification of real-time concurrent
object systems in the Real-Time Maude extension is discussed in \cite{journ-rtm}.
An interesting application using 
sockets to specify and deploy  a mobile version of Maude called Mobile Maude is described in 
\cite{mobile-maude,mobile-maude-wrla06}.

\section{$B$-Unification, Variants, and $E\cup B$-unification}\label{sec:unif+variants}

Maude's predecessors envisioned the inclusion of several symbolic features
which were never included in Maude until quite recently: 
(i) Eqlog 
\cite{eqlog}
envisioned an integration of order-sorted equational logic with Horn logic, providing 
logical variables, constraint solving, and automated reasoning capabilities on top of 
order-sorted equational logic (see Section~\ref{sec:eqlog} for an
actual Eqlog interpreter);
and 
(ii) MaudeLog~\cite{volterra} envisioned an integration of order-sorted rewriting logic with queries including logical variables.
Among the many symbolic reasoning features that can be supported by
Maude,  in this paper we focus on order-sorted equational unification (this section)
and order-sorted narrowing-based symbolic reachability  analysis
(Section~\ref{sec:narrowing}).
For a broader discussion of other symbolic reasoning methods and tools in rewriting
logic and Maude,
see Section \ref{sec:tools-apps} and \cite{MeseguerWoLLIC18,GRT-coh-compl-symb-meth-JLAMP}.

At first sight, adding symbolic reasoning capabilities to Maude might
seem like an incremental improvement; but this is not at all the case.  
Let us focus, for the moment, on what it means to add equational unification.
Order-sorted unification modulo axioms first became available 
as a built-in feature in 2009 as part of the Maude~2.4 
release~\cite{maude-rta2009},
which supported any combination of order-sorted symbols declared to be
either free or associative-commutative (AC).
%
Unification was updated 
in 2011 as part of the Maude~2.6
release~\cite{maude-rta2011}.
Built-in equational unification was extended to allow any combination 
of symbols being either free, commutative (C), associative-commutative (AC), or associative-commutative 
with an identity symbol (ACU). 
The performance was dramatically improved, allowing further 
development of other techniques in Maude.
As we explain below, built-in order-sorted unification has been
further extended later to allow associativity-only (A) as well as identity
(U) axioms.  This all means that for axioms $B$ including 
combinations of these axioms, equational unification in
an order-sorted theory $(\Sigma,B)$ is supported by Maude.

The next natural but highly non-trivial step is supporting
equational unification in order-sorted theories
of the form $(\Sigma,E \cup B)$, where the equations $E$ oriented as
rules are convergent modulo such axioms $B$.  This is highly
non-trivial because, although it is well-known that narrowing with the equations $E$ modulo
axioms $B$ provides a complete $E\cup B$-unification \emph{semi-algorithm}
\cite{JKK83}, the prospects of obtaining a practical equational unification algorithm in
this general setting looked rather dim for the following reasons: (i)
without an efficient $E,B$-narrowing strategy the compounded
combinatorial explosion of $B$-unification and unrestricted
$E,B$-narrowing would make such a semi-algorithm hopeless;
(ii) almost nothing was known about $E,B$-narrowing strategies for $B
\not= \emptyset$; and (iii) almost nothing was known about
\emph{termination} results for complete $E,B$-narrowing strategies
for $B \not= \emptyset$
that would make the (in general undecidable) $E \cup B$-unification
semi-algorithm into a \emph{decidable unification algorithm}.
The key concepts making it possible to break through these daunting obstacles have been
those of  \emph{variant}~\cite{comon-delaune}, and of
\emph{folding variant narrowing} and \emph{variant unification} \cite{variant-JLAP}.
The introduction of these variant-based concepts in Maude (see
Sections~\ref{sec:variants}--\ref{sec:folding-narrowing} below) has
led to a drastic improvement in Maude's symbolic reasoning capabilities:
variant generation, variant-based $E \cup B$-unification, and symbolic reachability based on 
variant-based $E \cup B$-unification became all available for the first time.
Initially, 
all the variant-based features 
were only available in Full Maude,
and only for a restricted class of theories
called
\emph{strongly right irreducible}. However, all these variant-based
features are now efficiently supported in Core Maude as explained in
Sections \ref{sec:variants}--\ref{sec:folding-narrowing}.

%
Order-sorted unification modulo axioms $B$
 was extended again in 2016
as part of the Maude~2.7
release~\cite{maude-ijcar2016}.
First, the built-in unification algorithm allows any combination of symbols being free,
C, AC, ACU, CU (commutativity and identity), U (identity), Ul (left identity), and Ur (right identity). 
Second, variant generation and variant-based unification were implemented as built-in features in Maude.
This built-in implementation works for any convergent theory modulo the axioms described above,
both allowing 
very general 
equational theories (beyond the strongly right irreducible ones)
and
boosting the performance not only of these features but of their applications. 
%
$B$-unification has been recently further extended to the \emph{associative} case
as part of the Maude~2.7.1
release~\cite{DBLP:conf/wrla/DuranEEMMT18}.
This is a key contribution because associative unification is infinitary in general
and
the development of an efficient and \emph{effective in practice}
associative unification algorithm that furthermore supports
order-sorted typing and combination with any other symbols either
free or themselves combining some A and/or C and/or U axioms,
has been a highly non-trivial challenge.  A key concern in meeting this
challenge has been the identification of a fairly  broad class of unification
problems appearing in many practical applications for which our
algorithm is guaranteed to terminate with a \emph{finite} and \emph{complete}
set of unifiers.  
To deal with the possibility that a given unification problem may have an infinite minimal complete set of unifiers, or that the problem is outside the class for which the algorithm is known to be complete, the algorithm can return a finite set of unifiers with an \emph{explicit warning} that such a set may be incomplete.
In a good number of applications where we have used these new
associative symbolic features of Maude, 
unification problems falling outside the class supported by
our algorithm in a complete way often do not even arise\footnote{The
Maude-NPA protocol analyzer
has already been tested with various protocols using associative operators
without  encountering any  incompleteness warnings (see \cite{Gonzalez-Burgueno18}).} in practice.


\subsection{Order-Sorted Unification Modulo Axioms $B$}\label{sec:unification}

Maude currently provides an order-sorted $B$-unification algorithm for all order-sorted 
theories $(\Sigma,B)$ such that
the order-sorted signature $\Sigma$ is \emph{preregular} modulo $B$ 
(see \cite[Footnote 2]{crc-alp}) 
and
the axioms $B$ associated to function symbols 
can be any combination of:
(i) \texttt{iter} equational axioms, which can be declared for some unary symbols;%
\footnote{The iter, or iterated operator, theory is a built-in mechanism that allows the efficient input, output, and manipulation of very large stacks of a unary operator, see \cite{maude-manual}.}
(ii) \texttt{comm} (C) commutativity axioms; (iii) 
\texttt{assoc}  (A) associativity axioms; and (iv) \texttt{id:}
identity axioms (U) as well as the \texttt{left id:} left identity axioms (Ul)
and the \texttt{right id:}  right identity axioms (Ur), \emph{except}
for the following combinations not currently supported:
\texttt{assoc id}, \texttt{assoc left id}, and \texttt{assoc
  right id}.  However, these three remaining subcases are
easily supported by turning the respective identity axioms into
\emph{oriented equations} and then
using variant unification modulo the remaining axioms $B$ (see Section~\ref{sec:variants}).
Maude provides a $B$-unification command of the form
\begin{maude}[escapechar=^,basicstyle=\small\normalfont\ttfamily]
unify [^$n$^] in ^\nt{ModId}^ :
	^\nt{Term-1}^ =? ^\nt{Term'-1}^ /\ .../\ ^\nt{Term-k}^ =? ^\nt{Term'-k}^ .
\end{maude}
where  $k \geq 1$, $n$ is an optional argument providing a bound on the number of
unifiers requested, and 
\texttt{ModId} is the module where the command takes place.
The unification infrastructure now supports the notion of incomplete unification 
algorithms (e.g. for associative unification). 

Let us show some examples of unification with an associative attribute, which is the last feature available in Maude 2.7.1.
See \cite{maude-manual} for more examples of unification modulo axioms.
%

Consider a very simple module 
where the symbol    \verb+_._+ is associative:


\begin{maude}
fmod UNIFICATION-A is
  protecting NAT .
  sort NList .
  subsort Nat < NList .
  op _._ : NList NList -> NList [assoc] .
  vars X Y Z P Q : NList .
endfm
\end{maude}

\noindent
Even if associative unification is infinitary (we include concrete examples below)
there are many realistic unification problems that are still finitary.
The following unification problem returns five unifiers:

\begin{maude}
Maude> unify in UNIFICATION-A : X . Y . Z =? P . Q .

Solution 1                                      
X:NList --> #1:NList . #2:NList                 
Y:NList --> #3:NList                            
Z:NList --> #4:NList                            
P:NList --> #1:NList                            
Q:NList --> #2:NList . #3:NList . #4:NList      

Solution 2
X:NList --> #1:NList
Y:NList --> #2:NList . #3:NList
Z:NList --> #4:NList
P:NList --> #1:NList . #2:NList
Q:NList --> #3:NList . #4:NList

Solution 3                                      
X:NList --> #1:NList                            
Y:NList --> #2:NList                            
Z:NList --> #3:NList . #4:NList                 
P:NList --> #1:NList . #2:NList . #3:NList      
Q:NList --> #4:NList                            

Solution 4
X:NList --> #1:NList
Y:NList --> #2:NList
Z:NList --> #3:NList
P:NList --> #1:NList . #2:NList
Q:NList --> #3:NList

Solution 5
X:NList --> #1:NList
Y:NList --> #2:NList
Z:NList --> #3:NList
P:NList --> #1:NList
Q:NList --> #2:NList . #3:NList
\end{maude}

\noindent
The above output illustrates how fresh variables, not occurring in the original unification problem,
are introduced by Maude by using the notation \verb!#N:Sort!.

One possible condition for finitary associative unification 
(see \cite{DBLP:conf/wrla/DuranEEMMT18} for further details) 
is having linear  (i.e., unrepeated) \emph{list} variables, as in the example above.
On the other hand,
the unification problem may not be linear, but it may be easy to detect that there is no unifier,
e.g. it is impossible to unify a list \texttt{X} concatenated with itself with another list \texttt{Y}
concatenated also with itself but with a natural number, e.g. $1$, in between.

\begin{maude}
Maude> unify in UNIFICATION-A : X . X =? Y . 1 . Y .
No unifier.
\end{maude}

When nonlinear variables occur on both sides of an associative unification 
problem, Maude 
always ensures termination, but sometimes raises an incompleteness warning. 
Several cases are possible (see \cite{DBLP:conf/wrla/DuranEEMMT18} for further details):

\begin{enumerate}
\item 
One or more cycles are detected, but they do not give rise to unifiers.

\begin{maude}
Maude> unify in UNIFICATION-A : 0 . Q =? Q . 1 .
No unifier.
\end{maude}

%
%
%

\item 
There is at least one cycle that produces an infinite family of most general 
unifiers. In this case a warning will be issued and only the acyclic solutions are 
returned.

\begin{maude}
Maude> unify in UNIFICATION-A : 0 . X =? X . 0 .
Warning: Unification modulo the theory of operator _._ 
has encountered an instance for which it may not be complete.

Solution 1
X:NList --> 0
Warning: Some unifiers may have been missed due to incomplete 
unification algorithm(s).
\end{maude}

Note that the unification problem $\texttt{0\,.\,X} =^? \texttt{X\,.\,0}$ has an infinite family 
of most general unifiers $\{\texttt{X}\mapsto \texttt{0}^n\}$ for 
$\texttt{0}^n$ being a list of $n$ consecutive \texttt{0} elements.

\item 
There is at least one nonlinear variable with more than two occurrences and 
Maude will use a depth bound rather than cycle detection. If the search tree 
grows beyond the depth bound, the offending branches will be pruned, and a 
warning will be given.

\begin{maude}
Maude> unify in UNIFICATION-A : X . X . X =? Y . Y . Z . Y .
Warning: Unification modulo the theory of operator _._ 
has encountered an instance for which it may not be complete.

Solution 1
X:NList --> #1:NList . #1:NList . #1:NList . #1:NList
Y:NList --> #1:NList . #1:NList . #1:NList
Z:NList --> #1:NList . #1:NList . #1:NList

Solution 2
X:NList --> #1:NList . #1:NList . #1:NList
Y:NList --> #1:NList . #1:NList
Z:NList --> #1:NList . #1:NList . #1:NList

Solution 3
X:NList --> #1:NList . #1:NList
Y:NList --> #1:NList
Z:NList --> #1:NList . #1:NList . #1:NList
Warning: Some unifiers may have been missed due to incomplete unification algorithm(s).
\end{maude}

\end{enumerate}

\noindent
As many other Maude commands,
unification commands can also be
performed as metalevel operations
in the \texttt{META-LEVEL} module
described in Section~\ref{sec:reflection}.
See \cite{maude-manual} for details 
on the metalevel commands for unification,
which are extended with a new constant \texttt{noUnifierIncomplete},
and additional warnings generated during associative unification.


\subsection{Variants}\label{sec:variants}

Consider a term $t$ in a convergent order-sorted equational theory
$(\Sigma,E \cup B)$ where the equations $E$ are assumed unconditional.
Intuitively, a \emph{variant} of $t$ \cite{comon-delaune} is \emph{the
normal form} $u$ of an \emph{instance} $t \theta$ of $t$ by a
substitution $\theta$, which is computed by simplification with $E$
modulo $B$.  For example, for the unsorted signature $\Sigma =
\{0,s,+\}$ of addition in the Peano natural numbers, with
$E=\{x+0=x,x+s(y)=s(x+y)\}$ and $B = \emptyset$, the terms $x$ and
$s(x+y')$ are variants of the term $x+y$ for the respective
substitutions $\theta_{1} = \{y \mapsto 0\}$ and $\theta_{2} = \{y
\mapsto s(y')\}$.  Technically, it is useful to
tighten the notion of variant in two ways \cite{variant-JLAP}:
(i) by viewing a variant of
$t$ as a pair $(u,\theta)$ instead of just a normal form $u$ of an
instance term $t \theta$, and (ii) by requiring, without any real loss
of generality, that the substitution $\theta$ is in normal form, i.e.,
that for each variable $x$, the term $\theta(x)$ is in normal form.
Of course, \emph{some variants are more general than others}.  For
example, among the variants of $x+y$, $(x, \{y \mapsto 0\})$ is more
general than $(s(x'), \{x \mapsto s(x'),y \mapsto 0\})$, and
$(s(x+y'), \{y \mapsto s(y')\})$ is more general than $(s(s(x')+y'),
\{x \mapsto s(x'),y \mapsto s(y')\})$.  The general definition for
$(\Sigma,E \cup B)$ is that a variant $(u,\theta)$ of $t$ is
\emph{more general} than another variant $(v,\eta)$ of $t$ iff there
is a substitution $\gamma$ such that: (i) $u \gamma =_{B} v$, and (ii)
for each variable $z$ in $t$, $\gamma(\theta(z)) =_{B} \eta(z)$.

A convergent order-sorted theory $(\Sigma,E \cup
B)$ is said to have the \emph{finite variant property} (FVP) \cite{comon-delaune,variant-JLAP}
 iff each $\Sigma$-term $t$ has a \emph{finite} set of most general variants.
We can illustrate this property both by its absence and by its
presence.  For example,
$E=\{x+0=x,x+s(y)=s(x+y)\}$
is \emph{not} FVP, since $(x+y,\mathit{id})$, $(s(x+y_{1}),
\{y \mapsto s(y_{1})\})$, $(s(s(x+y_{2})),\{y \mapsto s(s(y_{2}))\})$, $\ldots$,
$(s^{n}(x+y_{n}),\{y \mapsto s^{n}(y_{n})\})$, $\ldots$, are all 
\emph{incomparable} variants of $x+y$.  Instead, the following
theory \emph{is} FVP:

\begin{example}\label{ex:xor}
Consider the equational theory for exclusive or in module 
	\linebreak
\code{EXCLUSIVE-OR} of Figure~\ref{fig:EXCLUSIVE-OR}.
The attribute \texttt{variant} specifies
that these equations will
be used for variant generation and variant-based unification.
The \texttt{owise} attribute  for equations should \emph{never} be used in variant equations.


\begin{figure}
\begin{maude}
fmod EXCLUSIVE-OR is 
  sorts Nat NatSet .  subsort Nat < NatSet .
  op 0 : -> Nat [ctor] .
  op s : Nat -> Nat [ctor] .
  op mt : -> NatSet [ctor] .
  op _*_ : NatSet NatSet -> NatSet [assoc comm] .
  vars X Y Z U V : [NatSet] .
  eq [idem] :     X * X = mt    [variant] .
  eq [idem-Coh] : X * X * Z = Z [variant] .
  eq [id] :       X * mt = X    [variant] .
endfm
\end{maude}
\caption{\label{fig:EXCLUSIVE-OR}\code{EXCLUSIVE-OR} module}
\end{figure}

Given the term \verb!X * Y!, 
we can construct several of its variants as follows:

\begin{enumerate}
\item The  pair 
$( {s(0) * s(0)}, \{{X} \mapsto {s(0)},  {Y} \mapsto   {s(0)} \} )$ is normalized into $( {mt}, \{{X} \mapsto {s(0)},  {Y} \mapsto   {s(0)}\} )$ 
since the term {s(0) * s(0)} is simplified into {mt};
\item The pair
$({s(0) * U * s(0)}, \{{X} \mapsto {s(0) * U},  {Y} \mapsto   {s(0)}\} )$ is normalized into  $({U}, \{{X} \mapsto {s(0) * U},  {Y} \mapsto   {s(0)}\} )$,
and; 
\item The pair
$({s(0) *  U * s(0)} * V, \{{X} \mapsto {s(0) * U},  {Y} \mapsto   {s(0) * V}\} )$ is normalized into $({U * V}, \{{X} \mapsto {s(0) * U},  {Y} \mapsto   {s(0) * V} \} )$.
\end{enumerate}

As claimed above, the module \texttt{EXCLUSIVE-OR} is FVP.  But how
can we \emph{know} this?  The answer is very simple.
Maude provides a variant generation command of the form: 
\begin{maude}[escapechar=^, basicstyle=\small\normalfont\ttfamily]
get variants [^$n$^] in ^\nt{ModId}^ : ^\nt{Term}^ .
\end{maude}
where $n$ is an optional argument providing a bound on the number of
variants requested, so that if the cardinality of the set of variants is greater than the 
specified bound, the variants beyond that bound are omitted; and 
\texttt{ModId} is the module where the command takes place.

Now, as proved in \cite{variants-of-variants}, a convergent theory $(\Sigma,E \cup
B)$ is FVP iff for each of its function symbols $f$, say, $f:
s_{1}\ldots s_{n} \rightarrow s$, the term $f(x_{1},\ldots x_{n})$
with $x_{i}$ of sort $s_{i}$, $1 \leq i \leq n$, has a \emph{finite}
number of variants.  Therefore,
we can check that the \texttt{EXCLUSIVE-OR} module of Figure~\ref{fig:EXCLUSIVE-OR} 
has the finite variant property by simply generating
the variants for the exclusive-or symbol.

%
%
%
%
%
%
%
%

\begin{maude}
Maude> get variants in EXCLUSIVE-OR : X * Y .
Variant 1                                        Variant 7
[NatSet]: #1:[NatSet] * #2:[NatSet]   .........  [NatSet]: %1:[NatSet]
X --> #1:[NatSet]                                X --> %1:[NatSet]
Y --> #2:[NatSet]                                Y --> mt
\end{maude}

\noindent  Note that all other symbols $f$ in this module, except the exclusive
or symbol, are constructors and therefore have the single, trivial variant
$(f(x_{1},\ldots x_{n}),\mathit{id})$, where $\mathit{id}$ denotes the
identity substitution.

The above output illustrates a difference between unifiers returned by 
the built-in unification modulo axioms and substitutions (or unifiers) returned by variant generation
or variant-based unification: there are two forms of fresh variables,
the former {\small\textit{\texttt{\#n:Sort}}} and the new {\small\textit{\texttt{\%n:Sort}}}.
Note that the two forms have different counters. 
\end{example}

The FVP property is extremely useful.  For example, we  show in
Section \ref{sec:folding-narrowing} below that if $(\Sigma,E \cup
B)$ is FVP and $B$ has a finitary $B$-unification algorithm, then
there is also a, variant-based, finitary $E \cup B$-unification
algorithm.  But how common is it for a convergent theory to be FVP?
Certainly not so common, but more common than one might think.
Roughly speaking, \emph{recursive equations} such as, for example, 
the addition equation $x+s(y)=s(x+y)$ push a theory outside the FVP
fold. This seems quite restrictive; but one can easily overlook the power of
equational simplification \emph{modulo} axioms such as associativity and commutativity (AC).  Specifying a function
with equations modulo AC can quite often make, what would typically
require a recursive function definition without using AC, into a
non-recursive one.  For example, we can extend the above example of
Peano addition with a new sort $\mathit{Bool}$  and a strict
order predicate $>$ defined by equations:
$\{0 > x = \mathit{false},s(x) > 0 = \mathit{true}, s(x) > s(y) = x > y\}$,
which are \emph{unavoidably recursive} in \emph{this} representation.
However,  $>$
and various other arithmetic functions are part of an FVP theory
when we define them modulo ACU by representing 
 the natural numbers with constants $0$ and $1$, and an ACU binary
constructor $+$, so that
$3$ is represented as $1 + 1 + 1$.  Then we can define, in a
\emph{non-recursive} way, arithmetic functions such as:
$p$ for the predecessor function,
$max$ (resp. $min$)
for
the biggest (resp. smallest) of two numbers,
\verb!\! for the ``monus'' function,
$d$ for the symmetric difference function, 
\verb!>!  for the strict order predicate,
and
\verb!~!  for the equality predicate, yielding the FVP theory in Figure~\ref{fig:NAT-FVP}.
We can \emph{check} that it is FVP by computing the variants of its
function symbols.  For example, the generation of variants for the following terms all stop with a finite number of variants:

\begin{figure}[htb]
\begin{maude}
fmod NAT-FVP is
  protecting TRUTH-VALUE .
  sorts Nat NzNat Zero .
  subsorts Zero NzNat < Nat .
  op 0 : -> Zero [ctor] .
  op 1 : -> NzNat [ctor] .
  op _+_ : Nat Nat -> Nat [ctor assoc comm id: 0 prec 33] .
  op _+_ : NzNat NzNat -> NzNat [ctor assoc comm id: 0 prec 33] .
  op p : NzNat -> Nat .                *** predecessor
  op max : Nat Nat -> Nat [comm] .
  op max : NzNat NzNat -> NzNat [comm] .
  op min : Nat Nat -> Nat [comm] .
  op min : NzNat NzNat -> NzNat [comm] .
  op d : Nat Nat -> Nat [comm] .       *** symmetric difference
  op _\_ : Nat Nat -> Nat .            *** monus
  op _~_ : Nat Nat -> Bool [comm] .    *** equality predicate
  op _>_ : Nat Nat -> Bool .

  vars N M : Nat .
  vars N' M' K' : NzNat .

  eq p(N + 1) = N [variant] .
  eq max(N + M, N) = N + M [variant] .
  eq min(N + M, N) = N [variant] .
  eq d(N + M, N) = M [variant] .
  eq (N + M) \ N = M [variant] .
  eq N \ (N + M) = 0 [variant] .
  eq N ~ N = true [variant] .
  eq (N + M') ~ N = false [variant] .
  eq M + N + 1 > N = true [variant] .
  eq N > N + M = false [variant] .
endfm
\end{maude}
\caption{\label{fig:NAT-FVP}\code{NAT-FVP} module}
\end{figure}

\begin{maude}
Maude> get variants in NAT-FVP : N \ M . --- 3 variants
Maude> get variants in NAT-FVP : N ~ M . --- 4 variants
Maude> get variants in NAT-FVP : N > M . --- 3 variants
\end{maude}

As shown in \cite{duran-meseguer-rocha-JLAMP}, this FVP example 
(borrowed from \cite{duran-meseguer-rocha-JLAMP})
can be
easily extended to an even richer FVP example \texttt{INT-FVP} where all
the above functions (except monus, which is superseded by actual integer
difference using unary minus and $+$) are extended to the integers,
and an absolute value function on integers is added.

Another interesting feature is that variant generation is \emph{incremental}. 
In this way we are able to support 
general convergent equational theories  $(\Sigma,E \cup
B)$ that need not be FVP, 
so that a term $t$ \emph{may} have an
infinite number of variants.
Let us consider the Maude specification \texttt{NAT-VARIANT}, given in Figure~\ref{fig:NAT-VARIANT},  of our previous
functional module for natural number addition in Peano notation
that  we already know does \emph{not} have the finite variant property.

\begin{figure}
\begin{maude} 
fmod NAT-VARIANT is
  sort Nat .
  op 0 : -> Nat [ctor] .
  op s : Nat -> Nat [ctor] .
  op _+_ : Nat Nat -> Nat .
  vars X Y : Nat .
  eq [base] : X + 0 = X [variant] .
  eq [ind] :  X + s(Y) = s(X + Y) [variant] .
endfm
\end{maude} 
\caption{\label{fig:NAT-VARIANT}\code{NAT-VARIANT} module}
\end{figure}

On the one hand, it is possible to have a term with a finite number of most general variants
although the theory 
is not FVP.
For instance, the term \texttt{X + s(0)} has the single variant \texttt{s(X)}.

\begin{maude}
Maude> get variants in NAT-VARIANT : X + s(0) .
Variant 1
Nat: s(#1:Nat)
X --> #1:Nat
\end{maude}

\noindent
On the other hand, we can incrementally generate the variants
of a term that we suspect does not have a finite number of most general variants.
For instance, the term \texttt{s(0) + X} has an infinite number of most general variants.
In such a case, Maude can either output all the variants to the screen (and the user can stop the process whenever she wants), 
or generate the first $n$ variants by including a
bound $n$ in the command.

\begin{maude}
Maude> get variants [10] in NAT-VARIANT : s(0) + X .
Variant 1                                           Variant 10
Nat: s(0) + #1:Nat    .........................     Nat: s(s(s(s(s(0)))))
X --> #1:Nat                                        X --> s(s(s(s(0))))
\end{maude}

\noindent
A third approach, particularly when we are \emph{not} sure
whether a term has a finite number of variants, is to incrementally increase the bound and, if we obtain a 
number of variants smaller than the bound, then we know for sure that it had a finite 
number of most general variants. 

\subsection{Equational Narrowing, Folding Variant Narrowing, and $E\cup B$-unification}\label{sec:folding-narrowing}

Variant generation relies on \emph{folding variant narrowing} \cite{variant-JLAP}, which we
informally describe in this section.
Since folding variant narrowing is a narrowing
\emph{strategy}  (more on this below), we begin by
explaining\footnote{To explain this notion, we recall the
standard notation for a position $p$
in a term $t$  (a string of numbers characterizing the path to that position
term $t$ seen as a tree), and of subterm
$t|_{p}$ of $t$ at position $p$ (see, e.g.,
\cite{dershowitz-jouannaud}).  For
example, the subterm of $0+s(s(0))$
at position $2.1$ is $s(0)$.}
 \emph{unrestricted equational 
narrowing} in a convergent order-sorted equational theory $(\Sigma,E \cup
B)$. Recall from Section \ref{rewriterel-sect}, that
in ordinary rewriting with the oriented equations $\vec{E}$
modulo $B$,  to perform a rewrite $t\rightarrow_{\vec{E},B}t'$
one must choose a subterm $t|_p$ of the subject term $t$,
a rule $l \to r$ in $\vec{E}$ and a  substitution $\sigma$
such that: (i) $t|_p$ \emph{matches} $l$ modulo $B$ with $\sigma$,
i.e., $t|_p =_{B} l \sigma$, and (ii) $t'=t[r \sigma]_{p}$.
Instead, in narrowing a term $t$ with the oriented equations $\vec{E}$
modulo $B$,
one must choose a subterm $t|_p$ of the subject term $t$, 
a rule $l \to r$ in $\vec{E}$, and a \emph{unifier} $\sigma$ 
\emph{modulo} $B$ of the equation $t|_p = l$, i.e., a
substitution $\sigma$ such that $\sigma(t|_{p}) =_{B} \sigma(l)$.
Then, the analogous of the rewrite relation
$t\rightarrow_{\vec{E},B}t'$
is the \emph{narrowing} with $\vec{E}$ \emph{modulo} $B$
 relation
$t\leadsto_{\vec{E},B}t'$, where $t'=(t[r]_{p})\sigma$.
%
%

Consider the functional module \texttt{NAT-VARIANT} of Figure~\ref{fig:NAT-VARIANT}, 
%
%
%
the term \verb!s(0) + X! and the two equations \texttt{base} and \texttt{ind}.
Narrowing will instantiate\footnote{New variables in Maude are introduced as \texttt{\#1:Nat} 
or \texttt{\%1:Nat} instead of \texttt{X'}.} variable \texttt{X} with \texttt{0} and \texttt{s(X')},
respectively.
The following two narrowing steps are generated:

{\small
$$
\begin{array}{@{}l@{\hspace{15ex}}r@{}}
\mathtt{s(0) + X}
\; \leadsto_{\{\mathtt{X} \mapsto \mathtt{0}\},\mathtt{base}} \;
\mathtt{s(0)}
\\[2ex]
\mathtt{s(0) + X}
\; \leadsto_{\{\mathtt{X} \mapsto \mathtt{s(\#1{:}Nat)}\},\mathtt{ind}} \;
     \mathtt{s(s(0) + \#1{:}Nat)} 
\end{array}
$$
}

\noindent
Note that, for simplicity,  we show only the bindings of the unifier that affect the input term.
There are infinitely many narrowing derivations starting at the input
expression  \verb!s(0) + X!
(at each step the reduced subterm is underlined):

{\small
\begin{enumerate}
\item 
$\ul{\mathtt{s(0) + X}} 
\; \leadsto_{\{\mathtt{X} \mapsto \mathtt{0}\},\mathtt{base}} \;
\mathtt{s(0)}$

\item
$\ul{\mathtt{s(0) + X}} \;\;
\begin{array}[t]{@{}l@{}}
\leadsto_{\{\mathtt{X} \mapsto \mathtt{s(\#1{:}Nat)}\},\mathtt{ind}} \;
     \mathtt{s(\ul{s(0) + \#1{:}Nat})} 
\leadsto_{\{\mathtt{\#1{:}Nat} \mapsto \mathtt{0}\},\mathtt{base}} \;
     \mathtt{s(s(0))}
\end{array}$

\item
$\ul{\mathtt{s(0) + X}} \;\;
\begin{array}[t]{@{}l@{}}
\leadsto_{\{\mathtt{X} \mapsto \mathtt{s(\#1{:}Nat)}\},\mathtt{ind}} \;
     \mathtt{s(\ul{s(0) + \#1{:}Nat})} 
\\
\leadsto_{\{\mathtt{\#1{:}Nat} \mapsto \mathtt{s(\#2{:}Nat)}\},\mathtt{ind}} \;
     \mathtt{s(s(\ul{s(0) + \#2{:}Nat}))} 
\leadsto_{\{\mathtt{\#2{:}Nat} \mapsto \mathtt{0}\},\mathtt{base}} \;
     \mathtt{s(s(s(0)))} 
\end{array}$
\end{enumerate}
}

\noindent
And some of those, infinitely many, narrowing derivations are infinite in length,
e.g.
by applying
rule \texttt{ind} infinitely many times:
$$
\ul{\mathtt{s(0) + X}} \;\;
\begin{array}[t]{@{}l@{}}
\leadsto_{\{\mathtt{X} \mapsto \mathtt{s(\#1{:}Nat)}\},\mathtt{ind}} \;
     \mathtt{s(\ul{s(0) + \#1{:}Nat})} 
\\
\leadsto_{\{\mathtt{\#1{:}Nat} \mapsto \mathtt{s(\#2{:}Nat)}\},\mathtt{ind}} \;
     \mathtt{s(s(\ul{s(0) + \#2{:}Nat}))} 
\\
\leadsto_{\{\mathtt{\#2{:}Nat} \mapsto \mathtt{s(\#3{:}Nat)}\},\mathtt{ind}} \;
     \mathtt{s(s(s(\ul{s(0) + \#3{:}Nat})))} 
\\
\ldots     
\end{array}
$$

A \emph{narrowing path} $u_{0} \leadsto_{\theta_{1}} \! u_{1} \ldots u_{n-1}
\leadsto_{\theta_{n}}\!  u_{n}$, $n \geq 0$,
is denoted $u_{0} \leadsto^{*}_{\theta} \! u_{n}$, where $\theta =
\theta_{1} \ldots \theta_{n}$ is the so-called \emph{accumulated
substitution} obtained by composing the substitutions $\theta_{1},
\ldots, \theta_{n}$ for each step.

For a convergent order-sorted equational theory $(\Sigma,E \cup
B)$  any $E \cup B$-unification problem
can be reduced to a
narrowing problem as follows:
\begin{enumerate}
\item 
we add a fresh new sort \texttt{Truth} to $\Sigma$ with a constant \texttt{tt};
\item 
for each top sort of each connected component of sorts we add a binary predicate 
\texttt{eq} of sort \texttt{Truth} and add to $E$ the equation \texttt{eq(x,x) = tt}, 
where \texttt{x} has such a top sort;
\item 
we then reduce an $E \cup B$-unification problem $t =^? t'$ to the narrowing 
reachability problem
\[ \texttt{eq}(t,t') \leadsto ^{*} \texttt{tt} \]
modulo $B$ in the theory extending $(\Sigma,E \cup B)$ with these new sorts, operators, 
and equations, where $E$ and the new equations are used as rewrite rules.  
\end{enumerate}
That is, we search for all narrowing paths modulo $B$ from $\texttt{eq}(t,t')$ to $\texttt{tt}$.  
The accumulated substitution $\theta$ associated to each
such path then gives us an $E \cup B$-\emph{unifier} of the equation $t =^? t'$. 
%
However, as already pointed out, \emph{unrestricted} equational  narrowing with an
equational theory $(\Sigma,E \cup B)$ can be very wasteful because of the
compounded combinatorial explosion of potentially many $B$-unifiers at
each step and the existence of many, often redundant, narrowing paths.
Furthermore, only if all such narrowing paths terminate can
we be sure to have a \emph{finite}, complete set  of $E\cup B$-unifiers for a
given unification problem.  But, as
pointed out before, modulo axioms $B$ such as AC, unrestricted
narrowing almost never terminates, so we are in practice condemned to 
an $E \cup B$-unification \emph{semi-algorithm}.
The upshot is that unrestricted
narrowing modulo $B$ is actually hopeless in practice: without a
suitable narrowing strategy drastically restricting the narrowing
paths and able to detect when narrowing paths can be stopped there is
no hope for a practical $E \cup B$-unification semi-algorithm, and
even less hope for a, narrowing based,  $E \cup B$-unification algorithm.

\vspace{1ex}

\noindent {\bf Folding Variant Narrowing}.  The \emph{folding variant
narrowing strategy} proposed in \cite{variant-JLAP} solves all these
problems in one blow.   It  furthermore provides a method to compute a
complete set of variants for \emph{any} convergent equational theory
$(\Sigma,E \cup B)$ such that $B$ has a $B$-unification algorithm.  We
briefly explain this strategy and how it is used by Maude to compute a
complete set of variants of a term, and a complete set of $E \cup
B$-unifiers for \emph{any} convergent $(\Sigma,E \cup B)$
having a $B$-unification algorithm.

Roughly speaking, given a convergent theory $(\Sigma,E \cup B)$,
 the folding variant narrowing strategy defines a
\emph{subset} of narrowing paths, so that only those in such subset are
computed.  To begin with, only narrowing paths of the form
$u_{0} \leadsto^{*}_{\theta} \! u_{n}$ with $u_{n}$ and $\theta$
 \emph{normalized} are considered.  This exactly means that $(u_{n},\theta)$
 is a \emph{variant} of $u_{0}$.  In fact, it is shown in
 \cite{variant-JLAP} that such sequences compute a \emph{complete set of
most general variants} of $u_{0}$.  But folding variant narrowing goes further in
 two ways: (i) it furthermore discards \emph{redundant} narrowing
 paths $u_{0} \leadsto^{*}_{\theta} \! u_{n}$, where such a path is
 redundant if there is another path $u_{0} \leadsto^{*}_{\theta'} \! u'_{m}$
such that the variant $(u'_{m},\theta')$ is \emph{more general} than
 the variant $(u_{n},\theta)$.  We can then ``fold,'' i.e., subsume,
 the less general path into the more general one; and (ii) the folding
 variant  narrowing strategy can \emph{safely stop} when all new paths
 thus computed in a breadth-first manner can be folded into previously
 computed paths.  The following are the most remarkable properties
 about folding variant narrowing for a convergent  equational theory $(\Sigma,E \cup B)$:

\begin{enumerate}
\item It computes a complete set of most general variants for any term
$t$.  In general  this set may be infinite and is computed incrementally by Maude.

\item 
When the theory $(\Sigma,E \cup B)$ is FVP, i.e., 
the complete set of most general variants of any term is finite, the narrowing strategy \emph{terminates} for any input term $t$. 

\item Extending $(\Sigma,E \cup B)$ with equality operators and the
constant \texttt{tt} as explained above, a \emph{complete} set of most
general $E\cup B$-unifiers of $t =^? t'$ is obtained as the set of all substitutions $\theta$
such that $(\texttt{tt},\theta)$ belongs to the complete set of most
general variants of the term $\texttt{eq}(t,t')$.  In general this set
is infinite and is computed by Maude incrementally, so we only have a \emph{semi-algorithm}.

\item This variant-based $E\cup B$-unification semi-algorithm becomes a
\emph{finitary} $E\cup B$-unification algorithm, therefore terminating
for any unification problem $t =^? t'$,  iff $(\Sigma,E \cup B)$ is FVP.
\end{enumerate}

Consider for instance a variant unification problem between terms 
$X * Y$ and $U * V$ in the \texttt{EXCLUSIVE-OR} module in Figure~\ref{fig:EXCLUSIVE-OR},
which is FVP.

\begin{maude}
Maude> variant unify in EXCLUSIVE-OR : X * Y =? U * V  .
--- 57 unifiers are reported
\end{maude}

\noindent
Similarly, we can call variant unification between terms $X + Y$ and $0$,
which has only one possible solution, in
the \texttt{NAT-VARIANT} module in Figure~\ref{fig:NAT-VARIANT},
which is not FVP.
The variant unify command terminates
if we limit the number of solutions to $1$.

\begin{maude}
Maude> variant unify [1] in NAT-VARIANT : X + Y =? 0  .

Unifier #1
rewrites: 4 in 0ms cpu (0ms real) (12903 rewrites/second)
X --> 0
Y --> 0
\end{maude}

\noindent
However, it does not terminate if we limit the number of solutions to $2$,
since it keeps trying to generate more and more solutions without being able
to realize that there is only one.
\\

\noindent {\bf Further Reading}.   For order-sorted unification modulo
axioms $B$ see
\cite{DBLP:journals/jsc/MeseguerG89,hendrix-meseguer-rule08,DBLP:conf/birthday/Eker11}.
For Maude's  order-sorted associative unification
algorithm and its combination with other axioms $B$ see
\cite{DBLP:conf/wrla/DuranEEMMT18}.  The original paper on variants is
\cite{comon-delaune}.  The correctness of the method for checking that 
a theory is FVP as well as several formulations of the variant notion
can be found in \cite{variants-of-variants}.  Folding variant
narrowing and variant unification are studied in \cite{variant-JLAP}.
Note that if an equational theory $(\Sigma,E \cup B)$ is FVP,
with a $B$-unification algorithm, $E \cup B$-unifiability of a conjunction
of equalities is decidable.  Assuming
non-empty sorts, satisfiability
in the initial algebra $T_{\Sigma/E \cup B}$ of any positive 
quantifier-free (QF) $\Sigma$-formula is then
\emph{decidable}.  But one can go further. Under mild assumptions
about a constructor subspecification for $(\Sigma,E \cup B)$ 
a theory-generic satisfiability algorithm for \emph{all} QF
$\Sigma$-formulas can be given (see
\cite{var-sat-scp,gutierrez-meseguer-var-pred-LOPSTR}, and for 
detailed algorithms and an implementation in Maude \cite{skeirik-meseguer-var-sat-JLAMP}).

\section{Narrowing with Rules and Narrowing Search}\label{sec:narrowing}

 When formally analyzing 
the properties of a rewrite theory $(\Sigma,E \cup B,R,\phi)$, an important problem is ascertaining for specific patterns 
(i.e., terms with variables) $t$ and $t'$ 
the following \emph{symbolic reachability problem:}
\index{reachability problem}%
\[\exists X\; t \longrightarrow^{*} t'\]
with $X$ the set of variables appearing in $t$ and $t'$, which for this discussion we may 
assume are a disjoint union of those in $t$ and those in $t'$. That is, $t$ and $t'$
symbolically describe sets of concurrent states $[\![t]\!]$ and $[\![t']\!]$ (namely, all the 
ground substitution instances of $t$, resp.\ $t'$, or, more precisely, the 
$E \cup B$-equivalence classes associated to such ground instances).  And we are asking: 
is there a state in $[\![t]\!]$ from which we can \emph{reach} a state in $[\![t']\!]$ after 
a finite number of rewriting steps?  Solving this problem means
\emph{searching} for a symbolic solution to it in a hopefully
\emph{complete} way (so that if a solution exists it will be found).
This has the flavor of the \texttt{search} command, and therefore of
some kind of model checking: for example, $t'$ could be a pattern
describing the \emph{violation} of an invariant, and $t$ a pattern
describing a set of initial states.  The main difference (and the
advantage in this case) is that with
the  \texttt{search} command we explore the \emph{concrete} states
reachable from some given \emph{concrete} initial state.  Instead,
here we explore \emph{symbolically}  reachability between possibly
infinite \emph{sets} of states such as $[\![t]\!]$ and $[\![t']\!]$.
The way we do so is by \emph{narrowing} $t$ with the rewrite rules
$R$ in the given system module $(\Sigma,E \cup B,R,\phi)$,
\emph{modulo} the equations $E \cup B$.

Provided the rewrite theory
$(\Sigma,E \cup B,R,\phi)$ is \emph{topmost} (that is, all rewrites take place 
at the root of a term), or, as in the case of AC-rewriting of object-oriented systems, 
$\mathcal{R}$ is ``essentially topmost,'' and the rules $R$ are coherent with $E$ modulo 
$B$, narrowing with the rules $R$ modulo the equations $E \cup B$ 
(recall the definition of 
narrowing in Section \ref{sec:folding-narrowing})
\index{narrowing!with rules}%
gives a constructive, 
sound, and complete method to solve reachability problems of the form 
$\exists X\; t \longrightarrow^{*} t'$.
That is, such a problem has an affirmative answer 
if and only if we can find a finite narrowing sequence 
with the rules $R$ modulo $E \cup B$  of the form 
$t \leadsto_{R,E \cup B}^{*} u$ such that $u$ and $t'$ are unifiable modulo $E \cup B$
\cite{narrowing-hosc}.  The method is \emph{constructive}, because 
instantiating $t$ with the composition of the unifiers for each step
in the narrowing sequence, plus a $E \cup B$-unifier for $u = t'$,
gives us a concrete rewrite sequence witnessing the existential formula.

Of course, narrowing with $R$ modulo $E \cup B$ requires performing $E \cup B$-unification 
at each narrowing step.  As explained in Section~\ref{sec:variants}, $E \cup B$-unification 
can itself be performed by folding variant narrowing with the equations $E$ modulo $B$, provided $E$ is 
convergent  modulo $B$.  Therefore, in performing symbolic 
reachability analysis in a rewrite theory $(\Sigma,E \cup B,R)$ there are 
\emph{two levels of narrowing} involved: (i)  narrowing with rules $R$ 
modulo $E \cup B$ for reachability, and (ii) folding variant narrowing with equations $E$ modulo $B$
to compute the $E \cup B$-unifiers needed for narrowing with $R$ 
modulo $E \cup B$.

Maude provides a \texttt{vu-narrow} command similar to the
\texttt{search} command for rewriting.
Specifically, \texttt{vu-narrow} searches in a breadth-first manner
for a substitution instance of the given
goal pattern that can be reached by rewriting from a substitution instance 
of the given pattern for initial states.
The general form of the command is:
\texttt{vu-narrow} $t$ \texttt{=>}$\diamond$ $t'$ \texttt{.} 
where $t$ is the pattern of initial states and $t'$ is the goal pattern.
The  $\diamond$ symbol is a place holder for the 
options: $\diamond=$ \texttt{1} (exactly one rewrite step), $\diamond=$
\texttt{+} (one or more steps), $\diamond=$
\texttt{*} (zero or more steps), and $\diamond=$
\texttt{!} (terminating states).
Since the narrowing search may either never terminate
and/or find an infinite number of solutions, two
\emph{bounds} can be added to a \texttt{vu-narrow}
command: one bounding the number of solutions
requested, and another bounding the depth of the rewrite
steps from the initial term $t$ (see below).

Let us illustrate the power of performing narrowing-based reachability analysis modulo variant equations and axioms, including associativity.
	Consider the specification of a generic grammar interpreter in Maude, based on \cite{ACEM-LOPSTR16}, given in Figure~\ref{fig:GRAMMAR}.
	We define a symbol \verb!_@_! to represent the interpreter configurations,
	where the first underscore represents the  current  string (of terminal and non-terminal symbols), 
	and the second underscore stands for the considered grammar. 
	For simplicity, we provide four non-terminal symbols \texttt{S}, \texttt{A}, \texttt{B}, and \texttt{C} for sort \texttt{NSymbol}
	and 
	four terminal symbols $0$, $1$, $2$, and the finalizing mark \texttt{eps} (the empty string) for sort \texttt{TSymbol},
	but a parametric specification would have been more appropriate.
	
\begin{figure}	
\begin{maude}
mod GRAMMAR is 
  sorts Symbol NSymbol TSymbol String Production Grammar Conf . 
  subsorts TSymbol NSymbol < Symbol < String .
  subsort Production < Grammar .
  ops 0 1 2 eps : -> TSymbol .
  ops S A B C : -> NSymbol .
  op _@_ : String Grammar -> Conf .
  op _->_ : String String -> Production .
  op __ : String String -> String [assoc id: eps] .
  op mt : -> Grammar .
  op _;_ : Grammar Grammar -> Grammar [assoc comm id: mt] .
  vars  L1 L2 U V : String .
  var G : Grammar .
  var N : NSymbol .
  var T : TSymbol .
  rl ( L1 U L2 @ (U -> V) ; G) => ( L1 V L2 @ (U -> V) ; G) [narrowing] .
endm
\end{maude}
\caption{\texttt{GRAMMAR} module}%
\label{fig:GRAMMAR}%
\end{figure}	

Only those rules including the \texttt{narrowing} attribute 
are used for narrowing-based reachability analysis.
Note the important fact that the string concatenation symbol \verb!__! is not
just \texttt{assoc}, but has also \texttt{eps} as its \emph{identity}
element. This means that in each narrowing step with the interpreter's
rule equational unification must be performed modulo AU
and not just modulo A.  
This is not directly supported by the order-sorted $B$-unification of
Section~\ref{sec:unification}, but \emph{is} supported by the variant-based
$E \cup B$-unification of Section~\ref{sec:folding-narrowing}.
That is, AU-unification is achieved by transforming 
the identity property into the FVP variant equations: 

\begin{maude}
  eq eps U = U [variant] .       
  eq U eps V = U V [variant] .      
  eq V eps = V [variant] .
\end{maude}	

	The interpreter can be used in two ways thanks to narrowing: to generate words of the given grammar, 
	but also to parse a given string
	(see \cite{Caballero-FLOPS99} for further references on this topic).
	Generating the words of a given grammar is defined by rewriting the configuration
	(\texttt{S}~@~\texttt{$\Gamma$}) into (\textit{st}~@~\texttt{$\Gamma$}) where 
	\textit{st} is a string of terminal symbols
	using the rules of the grammar $\Gamma$.
	For example, we have the following search query associated to a context-free grammar defining the language $0^n1^n$:

\begin{maude}
Maude> vu-narrow [4] S @ (S -> eps) ; (S -> 0 S 1) 
                 =>! U @ (S -> eps) ; (S -> 0 S 1) .

Solution 1       Solution 2       Solution 3          Solution 4
U --> eps        U --> 0 1        U --> 0 0 1 1       U --> 0 0 0 1 1 1
\end{maude}

	Parsing a   string \textit{st} according to a given   grammar $\Gamma$ is defined by 
	narrowing the configuration
	(\textit{N}~@~\texttt{$\Gamma$}) into (\textit{st}~@~\texttt{$\Gamma$}) 
	where 
	\emph{N} is a logical variable denoting a non-terminal symbol.
	For example, we have the following search query:

\begin{maude}
Maude> vu-narrow [1] N       @ (S -> eps) ; (S -> 0 S 1)  
                 =>* 0 0 1 1 @ (S -> eps) ; (S -> 0 S 1) .
Solution 1
N --> S
\end{maude}

Moreover, we can use narrowing to answer a more complex question: 
{\em What is the missing production so that the string ``0 0 1" is parsed into the non-terminal symbol \texttt{S}?}

\begin{maude}
Maude> vu-narrow [1] S     @ (N -> T) ; (S -> eps) ; (S -> 0 S 1) 
                 =>* 0 0 1 @ (N -> T) ; (S -> eps) ; (S -> 0 S 1) .
Solution 1
N --> S ;
T --> 0
\end{maude}

And we can use any grammar, e.g. a Type-0 grammar defining the language $0^n1^n2^n$.

\begin{maude}
Maude> vu-narrow [1] N @ (S -> eps) ; (S -> 0 S B C) ; (C B -> B C) ; 
                         (0 B -> 0 1) ; (1 B -> 1 1) ; (1 C -> 1 2) ; 
                         (2 C -> 2 2) 
       =>* 0 0 1 1 2 2 @ (S -> eps) ; (S -> 0 S B C) ; (C B -> B C) ;
                         (0 B -> 0 1) ; (1 B -> 1 1) ; (1 C -> 1 2) ; 
                         (2 C -> 2 2) .
Solution 1
N --> S
\end{maude}

Note that we must restrict the search in the previous narrowing-based search commands, 
because narrowing does not 
terminate for these reachability problems. 
However, it is extremely important that no warning about
A-unification incompleteness
 is shown, ensuring that the symbolic analysis is complete modulo AU,
despite associative unification being infinite for some uncommon cases.
The key reason is that string variables (\texttt{L1},
\texttt{L2}, and \texttt{U}) in 
the transition rule are \emph{linear} (\texttt{L1} and \texttt{L2}) or under order-sorted restrictions (\texttt{U}).

%


\subsection{Logic Programming as Symbolic Reachability}\label{sec:prolog-system-narrowing}

In this section we show how narrowing-based symbolic reachability
analysis can be used to provide a very simple alternative implementation
of logic programming.  The key idea is that there is a simple
theory transformation:
\[R[\_] : \mathit{HornLogicTheories} \rightarrow \mathit{RewriteTheories}\]
so that given a logic program $H$ we obtain an associated rewrite
theory $R[H]$ such that any query for $H$ can be solved by a corresponding 
\texttt{vu-narrow} search command for $R[H]$.  We explain and
illustrate below this theory transformation.


All theories $R[H]$ for any logic program $H$
extend the following module \texttt{LP-SEMANTICS},
that imports the \texttt{LP-SYNTAX} module, by adding to it the rules
of $R[H]$.
We no longer need any auxiliary unification or renaming machinery,
since narrowing performs those automatically. 

\begin{maude}
mod LP-SEMANTICS is
  protecting LP-SYNTAX .

  sort PredicateList .
  op nil : -> PredicateList .
  op _,_ : Predicate PredicateList -> PredicateList .

  var PL : PredicateList .
  vars X Y Z : Term .

  sort Configuration .
  op <_> : PredicateList -> Configuration .
\end{maude}

\noindent
For each Horn theory $H$, $R[H]$ just adds to the above signature
the rewrite rules into which the Horn clauses of $H$ are transformed.
Specifically, each
logic clause $P\ \verb!:-!\ P_1,\ldots,P_n$ is transformed into the
rewrite rule $\langle P,\textit{PL} \rangle \rightarrow \langle P_1,\ldots,P_n,\textit{PL}\rangle$,
where $\textit{PL}$ is a new variable of sort \texttt{PredicateList}
and where the leftmost predicate $P$ is replaced by $P_1,\ldots,P_n$.

Let us illustrate how this transformation is
used by means of our running logic programming example.

\begin{example}[Symbolic Search
  LP-evaluation]\label{ex:relatives-system-narrowing}
For $H$ the logic program of Example
\ref{ex:relatives-system}, $R[H]$ adds to \emph{\texttt{LP-SEMANTICS}} the following rewrite rules:



\begin{maude}
  rl < 'mother('jane, 'mike),PL >  => < PL > [narrowing] .
  rl < 'mother('sally, 'john),PL > => < PL > [narrowing] .
  rl < 'father('tom, 'sally),PL >  => < PL > [narrowing] .
  rl < 'father('tom, 'erica),PL >  => < PL > [narrowing] .
  rl < 'father('mike, 'john),PL >  => < PL > [narrowing] .
  rl < 'parent(X,Y),PL > => < 'father(X,Y), PL > [narrowing] .
  rl < 'parent(X,Y),PL > => < 'mother(X,Y), PL > [narrowing] .
  rl < 'sibling(X,Y),PL > => < 'parent(Z,X), 'parent(Z,Y),PL > 
    [narrowing nonexec] .
  rl < 'relative(X,Y),PL > => < 'parent(X,Z), 'parent(Z,Y),PL > 
    [narrowing nonexec] .
  rl < 'relative(X,Y),PL > => < 'sibling(X,Z), 'relative(Z,Y),PL > 
    [narrowing nonexec] .
\end{maude}

\noindent
Note that Maude requires that rules with extra variables in the righthand side
must be labeled with the \texttt{nonexec} keyword, 
even though the narrowing machinery uses them to perform narrowing
steps without any problem.

We can now evaluate different queries for our example logic program $H$
by giving the corresponding \texttt{vu-narrow} search commands for $R[H]$
with goal \linebreak \emph{\texttt{<\ nil\ >}}. 

First, whether Sally and Erica are sisters; the associated reachability graph is finite and no bound is necessary.

\begin{maude}
Maude> vu-narrow < 'sibling('sally,'erica),nil > =>* < nil > .
Solution 1
\end{maude}

\noindent
Who 
are the siblings of Erica? Sally and herself.

\begin{maude}
Maude> vu-narrow < 'sibling(X,'erica),nil > =>* < nil > .
Solution 1
X --> 'sally

Solution 2
X --> 'erica
\end{maude}

\noindent
How many pairs of possible siblings are there? Sally and Sally, Sally and Erica, Erica and Sally, Erica and Erica, John and John, and Mike and Mike.

\begin{maude}
Maude> vu-narrow < 'sibling(X,Y),nil > =>* < nil > .
Solution 1
X --> 'sally
Y --> 'sally

Solution 2
X --> 'sally
Y --> 'erica

Solution 3
X --> 'erica
Y --> 'sally

Solution 4
X --> 'erica
Y --> 'erica

Solution 5
X --> 'john
Y --> 'john

Solution 6
X --> 'mike
Y --> 'mike

Solution 7
X --> 'john
Y --> 'john
\end{maude}

\noindent
Are Jane and John relatives? Yes

\begin{maude}
Maude> vu-narrow < 'relative('jane,'john),nil > =>* < nil > .
Solution 1
\end{maude}

\noindent
Who 
are the relatives of John? Tom and Jane.

\begin{maude}
Maude> vu-narrow [2] < 'relative(X,'john),nil > =>* < nil > .
Solution 1
X --> 'tom

Solution 2
X --> 'jane
\end{maude}

\noindent
As explained in Section~\ref{sec:prolog-system},
this last call produces an infinite narrowing search,
so we must restrict the search, asking for two solutions only.
\end{example}

In retrospect, the deep connection between logic programming and
narrowing-based reachability analysis is not surprising at all: both
are based on \emph{unification}, and Horn clauses can easily be
transformed into rules so that solving logic programming queries
just \emph{becomes} narrowing search for the \texttt{< nil >} list of
atomic predicates.  But this leaves two pending questions: (1) how can we
\emph{mechanize} the $H \mapsto R[H]$ transformation; and (2) how can
we obtain a \emph{programming environment} for logic programming in Maude
based on narrowing?  Both questions can be easily answered by
a very powerful Maude feature, namely, \emph{reflection}.
In fact, Sections \ref{eqlog-transf-sect} and \ref{sec:eqlog}
will, respectively, answer questions (1) and (2) not just for logic
programming, but for the much more general functional-logic
programming language Eqlog \cite{eqlog-jlp}.

\vspace{1ex}

\noindent {\bf Further Reading}.  Narrowing-based symbolic
reachability analysis of concurrent systems was first studied
and proved complete in \cite{narrowing-hosc}.  To ensure 
that the narrowing tree is finitely branching, and for 
performance reasons, we have here assumed that in the topmost rewrite theory
$\mathcal{R}=(\Sigma,E \cup B,R,\phi)$,  (i) the equations $E \cup B$ are FVP, and
(ii) the rules $R$ are unconditional.  This of course restricts substantially
the class of concurrent systems that can be symbolically model checked
by narrowing.  As explained in \cite{GRT-coh-compl-symb-meth-JLAMP},
using a semantics-preserving theory
transformation and the concept of \emph{constrained narrowing},
restrictions (i)--(ii) can be dropped and a much wider class of
systems can be symbolically model checked.
Under the same just-mentioned assumptions (i)--(ii) on $\mathcal{R}$
it is possible to symbolically model check not only invariants
using \texttt{vu-narrow}, but arbitrary LTL formulas using
Maude's LTL logical model checker \cite{bae-narrow-check}. 



\section{Reflection, \texttt{META-LEVEL}, and Meta-Interpreters} 

\label{sec:reflection}

Rewriting logic is reflective~\cite{clavel-meseguer-tcs,clavel-meseguer-palomino-tcs}, 
in the sense that important aspects of its metatheory can be represented 
at the object level in a consistent way. That is, the object-level 
representation correctly simulates the relevant metatheoretic aspects,
just as a universal Turing machine correctly simulates any
other Turing machine, including itself.
This fact is systematically used in the design and implementation of the 
Maude language, making the metatheory of rewriting logic accessible to the 
user in a clear, principled, and efficient way. 

Rewriting logic being reflective  means that there is a finitely 
presented rewrite theory $\mathcal{U}$ in which we can represent any finitely 
presented rewrite theory $\mathcal{R}$ (including $\mathcal{U}$ itself) as a 
term $\overline{\mathcal{R}}$, any terms $t,t'$ in $\mathcal{R}$ as terms 
$\overline{t}, \overline{t'}$, and any pair $(\mathcal{R}, t)$ as a term 
$\langle\overline{\mathcal{R}}, \overline{t}\rangle$, in such a way
that we have the following equivalence:%

\noindent
\begin{align*}\label{eqn:reflection}
\mathcal{R} \vdash t \longrightarrow^*
t' \;\;\Leftrightarrow\;\; \mathcal{U}\vdash \langle\overline{\mathcal{R}},
\overline{t}\rangle \longrightarrow^* \langle\overline{\mathcal{R}},
\overline{t'}\rangle
\end{align*}

\noindent 
where $\mathcal{R} \vdash t \longrightarrow^* t'$ 
denotes that $t$ rewrites into $t'$ using the rewrite theory $\mathcal{R}$.
Since $\mathcal{U}$ is representable in itself, we can 
have an arbitrary number of levels of reflection, giving place to what 
is known as a ``reflective tower'':

\begin{center}
$\mathcal{R}\vdash t \rightarrow^* t' \;\Leftrightarrow\;
\mathcal{U}\vdash \langle\overline{\mathcal{R}}, \overline{t}\rangle
\rightarrow^* \langle\overline{\mathcal{R}}, \overline{t'}\rangle
\;\Leftrightarrow\; \mathcal{U}\vdash \langle\overline{\mathcal{U}},
\overline{\langle\overline{\mathcal{R}}, \overline{t}\rangle}\rangle
\rightarrow^* \langle\overline{\mathcal{U}},
\overline{\langle\overline{\mathcal{R}}, \overline{t'}\rangle}\rangle
\ldots$
\end{center}

This section explains how this is achieved in Maude through its predefined 
\code{META-LEVEL} and \code{META-INTERPRETER} modules. While the 
\code{META-LEVEL} module provides a purely functional
access to key functionality of
the universal theory $\mathcal{U}$,
the \code{META-INTERPRETER} module can also handle reflective object-oriented 
computations that interact with the outside world. Indeed, the meta-interpreter
manager and the created meta-interpreters are external objects like internet 
sockets, files, or standard \textsc{I/O} 
(see Section~\ref{sec:external-objects}). 

In a naive implementation of reflection, each step up the above reflective tower comes at
considerable computational cost, because simulating a single step of
rewriting at one level involves many rewriting steps one level up.  It
is therefore important to have systematic ways of lowering the levels
of reflective computations as much as possible, so that a rewriting
subcomputation happens at a higher level in the tower only when this
is strictly necessary.  In Maude, key functionality of the universal theory $\mathcal{U}$
has been efficiently implemented in the functional module \code{META-LEVEL}.
This module includes definitions of sorts and operations for representing 
every element in a structured specification. For example,
 terms are metarepresented 
as elements of a data type \code{Term} of terms; modules are metarepresented 
as terms in a data type \code{Module} of modules; and views are metarepresented 
as terms in a data type \code{View} of views. \code{META-LEVEL} also contains
so-called \emph{descent functions} that use the equivalences
in the reflective tower from right to left to lower as much as
possible the level of reflective computation for boosting performance.
In many cases, the performance cost is just a simple, linear change of data
representation before and after the given ``descended'' computation.
In fact, virtually all Maude commands plus many metalevel
operations such as unification, matching, rule application, rewriting, search, and so on,
are supported one level up the hierarchy as descent functions.  For example,
\code{metaReduce}, \code{metaRewrite}, \code{metaApply}, 
and \code{metaMatch} are some of these descent functions in \code{META-LEVEL}.
Furthermore, reflective operations like \code{upModule}, 
\code{upTerm}, \code{downTerm}, and other similar ones allow moving 
various kinds of entities one level up or down in the reflective
hierarchy.  

Giving a \emph{full} account of the \code{META-LEVEL} module
is beyond the scope of this paper.  Full details can be found in \cite{maude-manual,maude-book}.
However, we  give here a taste of how reflection is supported in Maude  by:
(1) explaining how a term $t$ 
is meta-represented 
as a meta-term $\overline{t}$ of sort \code{Term}, (2)
explaining how a rewrite theory  $\mathcal{R}$ (resp. an equational
theory $\mathcal{E}$) is meta-represented as a term
$\overline{\mathcal{R}}$ (resp. 
$\overline{\mathcal{E}}$) of sort \code{Module}, and (3) illustrating 
in Section \ref{eqlog-transf-sect} how easy it is to define
\emph{program transformations} by reflection by means of an
example transformation that  mechanizes within Maude the Eqlog
functional-logic language \cite{eqlog-jlp}.

\subsection{The \code{META-TERM} module}
\label{sec:metalevel}

In the \code{META-TERM} submodule of \code{META-LEVEL},
sorts and kinds are metarepresented as
terms in subsorts \code{Sort} and \code{Kind} of the sort \code{Qid} of quoted 
identifiers. Since characters, parentheses, brackets, and commas break 
identifiers in Maude, they must be \emph{escaped} with back quotes. For example, 
\verb!'NzNat!, 
\verb~'Map`{Int`,String`}~, 
and \verb~'Map`{Int`,Tuple`{String`,Set`{Rat`}`}`}~ are terms of sort 
\code{Sort}. Similarly, \verb!'`[Bool`]!, \verb!'`[List`{Int`}`]! and 
	\linebreak
\verb!'`[NzNat`,Zero`,Nat`]! are valid elements of the sort
\code{Kind}. 

A term $t$ is meta-represented as a so-called meta-term $\overline{t}$
 of the data type \code{Term} of terms. 
The base cases in the metarepresentation of terms are given by subsorts 
\code{Constant} and \code{Variable} of the sort \code{Qid}.
Constants are quoted identifiers that contain the constant's name and its type
separated by a `\code{.}', e.g., \code{'0.Nat}. Similarly, variables
contain their name and type separated by a `\code{:}', e.g., \code{'N:Nat}.
Appropriate selectors then extract their names and types.

A (non-constant) \emph{function symbol} is meta-represented as a
quoted identifier of sort \code{Qid}.
A \emph{term} different from a constant or a variable is meta-represented
by applying an operator symbol to a nonempty list of meta-terms using the
constructor 

\begin{maude}
  op _[_] : Qid NeTermList -> Term [ctor] .
\end{maude}

\noindent For example, the natural number term 
 \verb!s(N) + M! is meta-represented as the meta-term \verb!'_+_['s['N:Nat], 'M:Nat]!.

\subsection{The \code{META-MODULE} module}
\label{sec:meta-module}

In the submodule \code{META-MODULE} of \code{META-LEVEL},
which imports \code{META-TERM},
functional and system modules, as well as functional and system theories,
are metarepresented in a syntax very similar to their original user syntax.
Given meta-representations of sorts, operations, equations, membership
axioms, and rules, modules and theories are meta-represented as terms
of sort \code{Module} (and corresponding subsorts, like \code{FModule}
for functional modules and \code{SModule} for system modules). 
For example, a system module is meta-represented using the following constructor:

\begin{maude}
  op mod_is_sorts_._____endm : Header ImportList SortSet 
       SubsortDeclSet OpDeclSet MembAxSet EquationSet RuleSet 
       -> SModule [ctor gather (& & & & & & & &)] .
\end{maude}

Sort \code{Header} can take as values just an identifier
in the case of non-parameterized modules or an identifier together with
a list of parameter declarations in the case of a parameterized module.
Let us get a taste for how each of the different elements in modules
and theories are meta-represented by looking at how equations are meta-represented.

\begin{maude}
  sorts Equation EquationSet .
  subsort Equation < EquationSet .
  op eq_=_[_]. : Term Term AttrSet -> Equation [ctor] .
  op none : -> EquationSet [ctor] .
  op __ : EquationSet EquationSet -> EquationSet [ctor assoc comm id: none] .
\end{maude}

Similar definitions allow us to represent the rest of the components 
of modules. To get a feeling about the similarity between the
object-level and meta-level notations, 
let us consider the  metarepresentation of the module on the left as the 
term (called a \emph{meta-module})  displayed on the right:

\begin{maude}
  fmod NAT is                        fmod 'NAT is
    pr BOOL .                          protecting 'BOOL .
    sorts Zero Nat .                   sorts 'Zero ; 'Nat .
    subsort Zero < Nat .               subsort 'Zero < 'Nat .
    op 0 : -> Zero [ctor] .            op '0 : nil -> 'Zero [ctor] .
    op s : Nat -> Nat [ctor] .         op 's : 'Nat -> 'Nat [ctor] .
    op _+_ : Nat Nat -> Nat [comm] .   op '_+_ : 'Nat 'Nat -> 'Nat [comm] .
    vars N M : Nat .                   --- no variable declarations
    --- no mbs                         none
    eq 0 + N = N .                     eq '_+_['0.Nat, 'N:Nat] = 'N:Nat .
    eq s(N) + M = s(N + M) .           eq '_+_['s['N:Nat], 'M:Nat]
                                         = 's['_+_['N:Nat, 'M:Nat]] .
  endfm                              endfm
\end{maude}

To prepare the ground for our program transformation example
in Section~\ref{eqlog-transf-sect}, just think for a moment 
about what a program transformation is in its simplest possible form,
and why reflection should provide a powerful way of
\emph{meta-programming} such transformations.
In Maude, a program is a rewrite theory $\mathcal{R}$.
Therefore, the simplest kind of program transformation we can think of
is some kind of function, say, $\mathit{Tr}$, that maps
any rewrite theory $\mathcal{R}$ to its transformed theory
 $\mathit{Tr}(\mathcal{R})$.  But \emph{where} does this function exist?
In a stratosphere called the \emph{metalevel} of rewriting logic.
What reflection does is to bring such a stratosphere down to
earth, namely, down to the \texttt{META-LEVEL} module.
Of course, for the program transformation $\mathit{Tr}$ to be of any
use at all, it should be effective, that is, it should be a
\emph{computable} function.  But we know by the meta-theorem
of Bergstra and Tucker \cite{bt80} that \emph{any} computable
function can be defined by a \emph{convergent}, finite set of
equations.  Since by reflection we already have an algebraic data type
of rewrite theories, namely, the data type defined
by the \code{META-MODULE} functional module,
this all means that we can \emph{meta-program}
any program transformation $\mathit{Tr}$ of our choice
as an \emph{equationally-defined function}

\begin{maude}
  op Tr : Module -> Module .
\end{maude}

\noindent in a functional module extending \code{META-MODULE}.
Let us see all this for Eqlog!

\subsection{A Program Transformation for Eqlog} \label{eqlog-transf-sect}
Program transformation is one of the applications of meta-programming. 
The following example illustrates the power of program transformations
in a way that generalizes the program transformation
$H \mapsto R[H]$ from Horn clause theories to rewrite theories
 defined in Section \ref{sec:prolog-system-narrowing} and
illustrated in Example \ref{ex:relatives-system-narrowing}.

The generalization has to do with considering a much more general
class of Horn theories, namely, \emph{order-sorted Horn theories with
  equality}, which are the theories on which the Eqlog \cite{eqlog-jlp}
functional-logic language is based.  Such theories have the form
$((\Sigma,\Pi),E \cup B \cup H)$, where
$(\Sigma,E \cup B)$ is a convergent order-sorted equational
theory that, to make sure $E \cup B$-unification terminates, we will
assume FVP, and $H$ is a collection of \emph{Horn clauses} defined on the
signature $\Pi$ of \emph{predicate symbols}.  We
could easily  define in Maude a data type whose terms are
exactly (reflective versions of) such order-sorted Horn theories with equality, and then
we could define by reflection the transformation mapping
any such theory to a corresponding meta-module term in Maude.
But there is a shortcut that we will take to ease the presentation.

The shortcut has to do with the fact that each order-sorted Horn
theory with equality $((\Sigma,\Pi),E \cup B \cup H)$ can be
transformed into a \emph{semantically equivalent} order-sorted
\emph{equational theory} of the form 
$(\Sigma \cup \Pi,E \cup B \cup E_{H})$,
where the predicates $\Pi$ have been transformed into
additional \emph{function symbols}, and the Horn clauses $H$
into additional \emph{conditional equations} $E_{H}$ by:
(i) adding a fresh new sort $\mathit{Pred}$ of predicates to $\Sigma$
having a constant $\top$ denoting ``truth,''
(ii) turning each predicate $p$ in $\Pi$ having argument  sorts
$s_{1} \ldots s_{n}$ into a function symbol
$p: s_{1} \ldots s_{n} \rightarrow \mathit{Pred}$, and (iii)
transforming each Horn clause $p(\overline{u}) \Leftarrow 
p_{1}(\overline{u_{1}}) \wedge \ldots \wedge p_{n}(\overline{u_{n}})$
into the conditional equation
$p(\overline{u}) = \top \Leftarrow 
p_{1}(\overline{u_{1}})= \top \wedge \ldots \wedge
p_{n}(\overline{u_{n}})= \top$.
For simplicity we will assume that each Eqlog theory $T$ has been
specified as an equational theory of the form $T=(\Sigma \cup \Pi,E \cup
B \cup E_{H})$.  This has the advantage of allowing us to express $T$
inside Maude as a functional module, so that our desired
transformation $T \mapsto R[T]$ turning $T$ into a rewrite theory
can be defined as a metalevel function

{\footnotesize
\begin{verbatim}
  op meta-R[_] : FModule -> SModule .
\end{verbatim}
}

To illustrate these ideas, let us consider an Eqlog program that
extends that of Example \ref{ex:relatives-system} by adding age
information for the relatives in the example and an order predicate
to compare ages.  In its functional
version such an Eqlog program can be specified as the 
functional module in Figure \ref{ex:ext-relatives}.

\begin{figure}
\begin{maude}
fmod EXAMPLE is
  protecting TRUTH-VALUE .
  sorts Person Nat Pred .

  ops jane tom sally mike john erica : -> Person [ctor] .

  op T : -> Pred  [ctor] .    *** true
  op mother : Person Person -> Pred [ctor] .
  op father : Person Person -> Pred [ctor] .
  op sibling : Person Person -> Pred [ctor] .
  op parent : Person Person -> Pred [ctor] .
  op relative : Person Person -> Pred [ctor] .

  vars X1 X2 X3 : Person .

  *** Horn Clauses as conditional equations:
  eq mother(jane, mike) = T .
  eq mother(sally, john) = T .
  eq father(tom, sally) = T .
  eq father(tom, erica) = T .
  eq father(mike, john) = T .
  ceq sibling(X1, X2) = T 
    if parent(X3, X1) = T /\ parent(X3, X2) = T [nonexec] .
  ceq parent(X1, X2) = T if father(X1, X2) = T .
  ceq parent(X1, X2) = T if mother(X1, X2) = T .
  ceq relative(X1, X2) = T 
    if parent(X1, X3) = T /\ parent(X3, X2) = T [nonexec] .
  ceq relative(X1, X2) = T 
    if sibling(X1, X3) = T /\ relative(X3, X2) = T [nonexec] .

  ops 0 1 : -> Nat [ctor] .
  op _+_ : Nat Nat -> Nat [ctor assoc comm id: 0] .
  op _>_ : Nat Nat -> Bool .
  op _>=_ : Nat Nat -> Bool .

  vars n m k : Nat .

  eq n + m + 1 > n = true [variant] .
  eq n > n + m = false [variant] .

  eq n + m >= n = true [variant] .
  eq n >= n + m + 1 = false [variant] .

  op age : Person -> Nat .
  eq age(tom)
    = 1 + 1 + ... + 1 [variant] . *** 50
  eq age(sally)
    = 1 + 1 + ... + 1 [variant] . *** 30
  eq age(john)
    = 1 + 1 + ... + 1 [variant] . *** 10
  eq age(jane)
    = 1 + 1 + ... + 1 [variant] . *** 52
  eq age(mike)
    = 1 + 1 + ... + 1 [variant] . *** 32
  eq age(erica)
  = 1 + 1 + 1 ... + 1 [variant] . *** 28
endfm
\end{maude}
\caption{\label{ex:ext-relatives}Eqlog program extending the relatives program in Example~\ref{ex:relatives} (notice the ellipses)}
\end{figure}

The transformation $T \mapsto R[T]$ is in essence very
simple.  It has the form
$R : (\Sigma \cup \Pi,E \cup
B \cup E_{H}) \mapsto
(\Sigma \cup \Pi \cup \Omega,E \cup
B, R[H] \cup R_{\mathit{eq}})$, where
$\Omega$ adds new sorts \code{PredList} and \code{Configuration} 
and operator declarations 
      \code{<\_>} of sort \code{Configuration} 
      and 
      \code{nil} and \code{\_,\_} of sort \code{PredList} 
just as we did in the $H \mapsto R[H]$ transformation of 
Section \ref{sec:prolog-system-narrowing},
and the Horn clauses $H$ (here expressed as conditional
equations $ E_{H}$ but this is immaterial) are transformed into
rewrite rules exactly as in the $H \mapsto R[H]$ transformation.
Furthermore, 
 for each connected component, \code{[s]}, other than that of \code{Pred}, 
      a binary equality predicate \code{\_==\_ : [s] [s] -> Pred} is
      added to $\Omega$,
and a rule 
      defining this predicate for equalities  of that kind: 
      \code{rl < X:[s] == X:[s],PL > => < PL >} is added
to the set of rules (these are the rules denoted $R_{\mathit{eq}}$).

Given self-explanatory auxiliary functions \code{addOps}, \code{setName}, 
\code{setEqs}, \code{setRls}, \code{getSorts}, \code{getEqs}, 
\code{getRls}, and \code{getName}, the following equation implements the 
$T \mapsto R[T]$ transformation:

\begin{maude}
  op meta-R[_] : FModule -> SModule .
  eq meta-R[M]
    = addSorts('PredList ; 'Configuration,
        addOps(
          op 'nil : nil -> 'PredList [none] .
          op '_`,_ : 'Pred 'PredList -> 'PredList [none] .
          op '<_> : 'PredList -> 'Configuration [none] .
          mkEqOps(getSorts(M)),
          transformEqs(
            getEqs(M),
            setEqs(
              setRls(setName(M, qid("R[" + string(getName(M)) + "]")),
                mkEqRls(getSorts(M))),
              none),
            M))) .
\end{maude}

\noindent
Auxiliary functions \code{mkEqOps} and \code{mkEqRls} create, given a set of 
sorts, the operator declarations and equations for \code{\_==\_} as explained 
above. 

\begin{maude}
  op mkEqOps : SortSet -> OpDeclSet .
  eq mkEqOps(S ; SS)
    = if S == 'Pred
      then none
      else op '_==_ : kind(S) kind(S) -> 'Pred [none] .
      fi
      mkEqOps(SS) .
  eq mkEqOps(none) = none .

  op mkEqRls : SortSet -> RuleSet .
  eq mkEqRls(S ; SS)
    = if S == 'Pred
      then none
      else rl '<_>['_`,_['_==_[qid("X:" + string(kind(S))), 
                               qid("X:" + string(kind(S)))], 'PL:PredList]]
             => '<_>['PL:PredList] [narrowing] .
      fi
      mkEqRls(SS) .
  eq mkEqRls(none) = none .
\end{maude}

\noindent
The operation \code{transformEqs} transforms equations of sort \code{Pred} into the corresponding rules:

\begin{maude}
  op transformEqs : EquationSet Module Module -> Module .
  op transformCd : EqCondition -> Term .
  ceq transformEqs(eq T = 'T.Pred [none] . Eqs, M, M')
    = transformEqs(Eqs,
        addRls(rl '<_>['_`,_[T, 'PL:PredList]] 
                 => '<_>['PL:PredList] [narrowing] ., M),
        M')
    if leastSort(M', T) = 'Pred .
  ceq transformEqs(ceq T = 'T.Pred if Cd [Atts] . Eqs, M, M')
    = transformEqs(
           Eqs,
           addRls(rl '<_>['_`,_[T, 'PL:PredList]]
                    => '<_>[transformCd(Cd)] [Atts narrowing] ., M),
           M')
    if leastSort(M', T) = 'Pred .
  eq transformEqs(Eqs, M, M') = addEqs(Eqs, M) [owise] .

  ceq transformCd(T = 'T.Pred /\ Cd)
    = '_`,_[T, transformCd(Cd)]
    if Cd =/= nil .
  eq transformCd(T = 'T.Pred) = '_`,_[T, 'PL:PredList]  .
\end{maude}

\noindent
Notice that rules are added to the module as they are generated. The
second module does not change, it is used just for checking sorts.

\begin{example}[Eqlog Example as Narrowing Search]\label{ex:eqlog-example-narrowing}
Consider the functional module \code{EXAMPLE} in Figure~\ref{ex:ext-relatives}, 
which is the already-discussed Eqlog program extending
the relatives logic program of Example~\ref{ex:relatives} with an \code{age} operation and order predicates 
to compare natural numbers. 
Its transformed system meta-module \verb!meta-R[!$\overline{\tt EXAMPLE}$\verb!]!
obtained using the \verb!meta-R[_]!
metalevel function
has (meta-represented)
\emph{unconditional} rewrite rules such as the following ones:


\begin{maude}
   rl < mother(jane, mike), PL:PredList > 
     => < PL:PredList > .
   rl < sibling(X1,X2), PL:PredList > 
     => < PL:PredList, parent(X3, X1), parent(X3, X2) > [nonexec] .
\end{maude}

\noindent 
The system meta-module \verb!meta-R[!$\overline{\tt EXAMPLE}$\verb!]!
can then be used at the 
metalevel to perform Eqlog-based symbolic computation for this example
using narrowing search.\footnote{An alternative way of representing 
 Eqlog programs can be found
in \cite{DBLP:conf/birthday/Escobar14} using system modules and narrowing search,
and
also in \cite{DBLP:conf/wrla/Escobar18} using functional modules and folding variant narrowing.}
For example, we can then use the \code{metaNarrowingSearch} operation 
to find persons with a father and a mother in the transformed module:

\begin{maude}
red metaNarrowingSearch(
      meta-R[upModule('EXAMPLE, true)],
      '<_>['_`,_['father['X:Person, 'Y:Person],
                 '_`,_['mother['Z:Person, 'Y:Person], 'nil.PredList]]],
      '<_>['nil.PredList],
      '*,
      unbounded,
      'none,     ---- vu-narrow folding strategy
      0) .
result NarrowingSearchResult: {
  '<_>['nil.PredList],
  'Configuration,
  ('X:Person <- 'mike.Person ;
   'Y:Person <- 'john.Person ;
   'Z:Person <- 'sally.Person),
  '%,
  (none).Substitution,
  '#
}
\end{maude}

\noindent Or to find out that there are no fathers younger than their children:

\begin{maude}
red metaNarrowingSearch(
      meta-R[upModule('EXAMPLE, true)],
      '<_>['_`,_['father['X:Person, 'Y:Person],
                 '_`,_['_==_['_>_['age['Y:Person], 'age['X:Person]], 'true.Bool],
                       'nil.PredList]]],
      '<_>['nil.PredList],
      '*,
      unbounded,
      'none,     ---- vu-narrow folding strategy
      0) .
result NarrowingSearchResult?: (failure).NarrowingSearchResult?
\end{maude}
\end{example}

\subsection{An Eqlog Execution Environment}
\label{sec:eqlog}
\label{sec:env-mod-expr}

Maude provides meta-programming facilities for the generation of execution
environments for a wide variety of languages and logics. We explain here how 
these facilities may be used to develop a user-friendly notation for the 
introduction of  Eqlog programs. We have extended Full Maude with a new 
module expression to be able to use Eqlog  programs as functional modules 
as in Example~\ref{ex:relatives} and corresponding commands to execute 
queries on them.


Full Maude \cite{maude-book} is an extension of Maude written in Maude itself using its 
reflective capabilities. It was developed as a place in which to experiment
with new features, and provide facilities not yet available in the core 
implementation. Indeed, many of the features now available in Core Maude, 
like strategies, unification, variants, narrowing, parameterized modules, views, and module expressions 
like summation, renaming and instantiation, were available in Full Maude long 
before they were available in Maude 
(see, e.g.,~\cite{duran-meseguer-wrla98,Duran00}). This same setting represents
a perfect place to add new features with which to experiment or develop new 
prototypes. 

The interested reader may find at \url{http://maude.cs.illinois.edu}
a module extending Full Maude that provides a new module expression
\code{R[\_]} to transform an Eqlog program $T$ already entered into Maude
as a functional module into the rewrite theory $R[T]$ defined
in Section \ref{eqlog-transf-sect},
and a command \code{solve[\_]\_.} to get the first $n$ solutions to a
query for such an Eqlog program. 
The extension has been performed as in many previous cases; see 
guidelines in \cite{DuranO09} or \cite{maude-book,maude-manual}.

Once the module expression is available, it can be used as any other module
expression in Maude, in importation declarations in other modules or in 
commands. For example, given the \code{EXAMPLE} module one can select the 
generated module with
\begin{maude}
(select R[EXAMPLE] .)
\end{maude}
And then, at the object level, one can write commands of the form:
\begin{maude}[mathescape]
(solve [$n$] $A1$,...,$An$ .)
\end{maude}
to ask for the first $n$ solutions to the query \code{\(A1\),...,\(An\)} 
where the $Ai$ are predicate atoms, including equality predicates 
of the form \code{\(t\) == \(t'\)}.

We can now solve the queries in Example~\ref{ex:eqlog-example-narrowing} in a more 
user-friendly syntax:

\begin{maude}
(solve father(X:Person, Y:Person), mother(Z:Person, Y:Person), nil .)

Solution 1
state: < nil >
accumulated substitution:
X:Person --> mike ;
Y:Person --> john ;
Z:Person --> sally
variant unifier: empty substitution

No more solutions.
\end{maude}

\begin{maude}
(solve father(X:Person, Y:Person), 
       age(Y:Person) > age(X:Person) == true, 
       nil .)

No more solutions.
\end{maude}

\subsection{Meta-interpreters}
\label{sec:metaInterpreter}


The \code{META-LEVEL} module is \emph{purely functional}.  This is because
all its \emph{descent} functions are deterministic, even though they
may manipulate intrinsically nondeterministic entities such as
rewrite theories.  For example, the \code{metaSearch} descent function
with a bound of, say, 3, is entirely deterministic, since given the
meta-representations $\overline{\mathcal{R}}$ of the desired
 system module and $\overline{t}$ of the initial term plus the bound
 3, the result yielded by \code{search} for $\mathcal{R}$, $t$ and 3 at the object level, and
therefore by \code{metaSearch} at the metalevel, is uniquely determined.

 Although \code{META-LEVEL} is very powerful, its
purely functional nature means that it has \emph{no notion of state}.
Therefore reflective applications where \emph{user
  interaction} in a state-changing manner is essential require using \code{META-LEVEL}
in the context of additional features supporting such interaction.   Until recently, all such reflective
interactions were mediated by the built-in \code{LOOP-MODE} module \cite{maude-book}:
a simple read-eval-print loop where a Maude user can interact from the
terminal with a Maude module $M$ already stored in Maude through an object
(the \emph{state} of \code{LOOP-MODE}) consisting of
 a 3-tuple that holds a Maude term $t$ in module $M$ ---thought of as the current
 ``internal state'' of the loop--- together with input and output
 buffers.  The user interactions do change the state of that 3-tuple
by consuming user input, producing output, and possibly changing the
internal state $t$ to a new state $t'$ 
according to the user-given rewrite rules defining the desired
interaction.  For example, Full Maude, the Eqlog  extension of
it presented in Section \ref{sec:eqlog}, and many other interactive Maude
tools use suitable extensions of  \code{META-LEVEL} and \code{LOOP-MODE}
to support user interaction.  This is adequate for many purposes, but
limits the type of interactions to simple real-eval-print ones.  Much
more flexible kinds of reflective interactions are possible by means
of Maude's new \emph{meta-interpreters} feature, in which Maude
interpreters are encapsulated as external objects and can reflectively
interact with both other Maude interpreters and with various other
external objects, including the user.

Conceptually a meta-interpreter is an external object that is an independent 
Maude interpreter, complete with module and view databases, which sends and 
receives messages.
The module \code{META-INTERPRETER} in Maude's standard prelude
contains command and reply messages
that cover almost the entirety of the Maude interpreter. 
For example, it can be instructed to insert or show modules and
views, or carry out computations in a named module. As response, the 
meta-interpreter replies with messages acknowledging operations carried out 
or containing results. Meta-interpreters can be created and destroyed
as needed, and because a meta-interpreter is a complete Maude interpreter,
it can host meta-interpreters itself and so on in a tower of reflection.
Furthermore, the original \code{META-LEVEL}
 functional module  can itself be used from
inside a meta-interpreter.


The meta-representation of terms, modules, and views is shared with
the \code{META-LEVEL} functional module. 
The API to 
meta-interpreters defined in the 
	\linebreak
\code{META-INTERPRETER}  module
includes several sorts and constructors, a built-in object identifier 
\code{interpreterManager} and a large collection of command and response 
messages. 
%
%
%
%
%
%
%
The \code{interpreterManager} object identifier refers to a special external 
object that is responsible for creating new meta-interpreters in the current 
execution context. Such meta-interpreters have object identifiers of the form 
\code{interpreter(n)} for natural number $n$. 

\begin{example}
Let us illustrate the flexibility and generality of meta-interpreters with a 
short example. The example, which we call \code{RUSSIAN-DOLLS} after the 
Russian nesting dolls, is shown in Figure~\ref{fig:russianDolls}. It performs 
a computation in a meta-interpreter that itself exists in a tower of 
meta-interpreters nested to a user-definable depth and requires only two equations and two rules. 

\begin{figure}
\begin{maude}
mod RUSSIAN-DOLLS is
  extending META-INTERPRETER .

  op me : -> Oid .
  op User : -> Cid .
  op depth:_ : Nat -> Attribute .
  op computation:_ : Term -> Attribute .

  vars X Y Z : Oid .
  var AS : AttributeSet .
  var N : Nat .
  var T : Term .

  op newMetaState : Nat Term -> Term .
  eq newMetaState(0, T) = T .
  eq newMetaState(s N, T) 
    = upTerm(
        <> 
        < me : User | depth: N, computation: T >
        createInterpreter(interpreterManager, me, none)) .

  rl < X : User | AS > 
     createdInterpreter(X, Y, Z) 
  => < X : User | AS > 
     insertModule(Z, X, upModule('RUSSIAN-DOLLS, true)) .
  rl < X : User | depth: N, computation: T, AS > 
     insertedModule(X, Y) 
  => < X : User | AS > 
     erewriteTerm(Y, X, unbounded, 1, 'RUSSIAN-DOLLS, newMetaState(N, T)) .
endm
\end{maude}
\caption{\label{fig:russianDolls}Nested meta-interpreter example}
\end{figure}

The visible state of the computation resides in a Maude object of identifier 
\code{me} and class \code{User}. The object holds two values in respective 
attributes: the depth of the meta-interpreter, which is recorded as a \code{Nat}, 
with \code{0} as the top level, and the computation to perform, which is recorded 
as a \code{Term}. 

The operator \code{newMetaState} takes a depth and a meta-term to evaluate. If 
the depth is zero, then it simply returns the meta-term as the new meta-state. 
Otherwise a new configuration is created, consisting of a portal (needed for 
rewriting with external objects to locate where messages exchanged with external 
objects leave and enter the configuration), the user-visible object holding the 
decremented depth and computation, and a message directed at the 
\code{interpreterManager} external object, requesting the creation of a new 
meta-interpreter, and this configuration is lifted to the metalevel using the 
built-in \code{upTerm} operator imported from the functional metalevel.

The first rule of the module handles the \code{createdInterpreter} message from  
\code{interpreterManager}, which carries the object identifier of the newly 
created meta-interpreter. It uses \code{upModule} to lift its own module, 
\code{RUSSIAN-DOLLS}, to the metalevel and sends a request to insert this 
meta-module into the new meta-interpreter. The second rule handles the 
\code{insertedModule} message from the new meta-interpreter. It calls the 
\code{newMetaState} operator to create a new meta-state and then sends a request 
to the new meta-interpreter to perform an unbounded number of rewrites, with
external object support and one rewrite per location per traversal in the 
metalevel copy of the \code{RUSSIAN-DOLLS} module that was just inserted.

We start the computation with an \code{erewrite} command on a configuration 
that consists of a portal, a user object, and a \code{createInterpreter} message:

\begin{maude}
erewrite <>
         < me : User | depth: 0, 
           computation: ('_+_['s_^2['0.Zero], 's_^2['0.Zero]]) >
         createInterpreter(interpreterManager, me, none) .
result Configuration: 
  <> 
  < me : User | none > 
  erewroteTerm(me, interpreter(0), 1, 's_^4['0.Zero], 'NzNat)
\end{maude}

\noindent With depth 0, this results in the evaluation of the meta-representation
of $2 + 2$ directly in a meta-interpreter, with no nesting.
Passing a depth of 1 results in the evaluation instead being done in a nested
meta-interpreter.

\begin{maude}
erewrite <>
         < me : User | depth: 1, 
           computation: ('_+_['s_^2['0.Zero],'s_^2['0.Zero]]) > 
         createInterpreter(interpreterManager, me, none) .
result Configuration: 
  <> 
  < me : User | none > 
  erewroteTerm(me, interpreter(0), 5, 
    '__['<>.Portal,  
        '<_:_|_>['me.Oid,'User.Cid,'none.AttributeSet],
        'erewroteTerm['me.Oid,'interpreter['0.Zero],'s_['0.Zero],
          '_`[_`][''s_^4.Sort,''0.Zero.Constant],''NzNat.Sort]], 'Configuration)
\end{maude}
Notice here that the top level reply message \code{erewroteTerm(...)} contains
a result that is a meta-configuration, which contains the reply
\code{'erewroteTerm[...]} meta-message from the inner meta-interpreter.
\end{example}

\vspace{1ex}

\noindent {\bf Further Reading}.  As already mentioned, full details
can be found in \cite{maude-manual,maude-book}.  The most complete
treatment of reflection in both rewriting logic and membership
equational logic, including proofs of correctness of the
meta-representations in the reflective tower, can be found in \cite{clavel-meseguer-palomino-tcs}.
About program transformations we only scratched the surface.
Inside Maude they are generalized to \emph{module
  operations} that make Maude \emph{user-extensible} with new module
composition features~\cite{DBLP:journals/scp/DuranM07}.  Outside
Maude ---or transforming programs in other logics to programs in Maude,
or conversely---  ``program transformations'' are  \emph{maps between logics}
in the sense of \cite{general} that can be implemented inside
Maude when we use it as a \emph{meta-logical framework} (see
\cite{20-years} and references there).
Such metalevel mappings are very useful to use Maude as
a \emph{formal meta-tool} to build formal tools for many other logics
(see again \cite{20-years}).

\section{Tools and Applications}
\label{sec:tools-apps}

As its title suggests, this paper has a twofold purpose.  On the one hand, it tries to give a gentle introduction to Maude's \emph{declarative programming style} without assuming prior familiarity with the language.  On the other hand, it provides, for the first time, a unified account of the most recent Maude features supporting \emph{symbolic computation} as well as other important new features.  
To keep the paper within reasonable size bounds, other important topics already well covered in the Maude book \cite{maude-book} had to be omitted or be mentioned only in passing.  In particular, two important topics have not been fully explained: \emph{model checking} has only been treated in the form of search-based (with either the standard \texttt{search} command or with narrowing-based symbolic search) reachability analysis; and Full Maude has only made a cameo appearance through its extension into an Eqlog  interpreter in Section \ref{sec:eqlog}.
For more details on Full Maude, 
including its advanced features for object-based programming already mentioned in Section
\ref{sec:oo}, we refer the reader to the detailed account in \cite{maude-book}, and for
how to build a wide range of formal tools as Full Maude extensions (as we did in this paper for
Eqlog) to the quite useful methodological paper \cite{DuranO09}.

For model checking, the first important distinction to be made is between \emph{explicit-state} model checking, where the search space of all concrete states of a system are explored, and \emph{symbolic} model checking, where sets of states, as opposed to individual concrete states, are represented and explored symbolically.  Maude supports \emph{both} kinds of model checking by model checkers directly supported by 
Maude and by additional model checkers built on top of Maude.  We first discuss Maude's support for explicit-state model checking.  Discussion of symbolic model checking is postponed until we discuss symbolic computation 
in Section \ref{symb-comp-tools-and-apps}.

\vspace{1ex}

\noindent {\bf Explicit-State Model Checking in Maude}.
The most basic form of explicit-state model checking has already been illustrated in this paper, since it is supported by the \texttt{search} command.  Note that, as further explained and illustrated with examples in \cite{maude-book}, \texttt{search} can be used to both verify \emph{invariants} or to find violations of invariants in the following sense.  Suppose that an invariant has been specified as a Boolean-valued predicate, say \texttt{p}, on states of sort \texttt{State}, and we wish to verify that \texttt{p} holds in every state reachable from an initial state \texttt{init}.  Then we can search for a violation of \texttt{p} by giving the search command:
\begin{maude}
search init =>* S:State s.t. p(S:State) =/= true .
\end{maude}
If the invariant \texttt{p} fails to hold, it will do so for some finite sequence
of transitions from \texttt{init}, and this will be uncovered by the above \texttt{search}
command since all reachable sates are explored in a breadth-first manner.  If, instead,
the invariant \texttt{p} does hold, two things can happen: (i) if the set of states reachable
from \texttt{init} is \emph{finite}, the \texttt{search} command will report failure to
find a violation of  \texttt{p} and therefore  \texttt{p} holds; but (ii) if there is an infinite
number of states reachable from \texttt{init}, \texttt{search} will never terminate.
Two options are then available: (ii)-(a) we can instead perform \emph{bounded model checking}
of \texttt{p} by specifying a depth bound for the \texttt{search} command; or (ii)-(b), as
explained in \cite{maude-book}, it may be possible to define an \emph{equational abstraction}
\cite{equational-abstraction-tcs}
of the given system module by identifying states by additional equations, so that
the system becomes finite-state and the invariant \texttt{p} can be verified.

Under the assumption that the set of states reachable from an initial state \texttt{init}  is
finite, Core Maude also supports explicit-state model checking verification of
any properties in linear time temporal logic (LTL) through its \emph{LTL model
checker}.  We refer to \cite{maude-book} for a detailed account of LTL model checking
in Maude, including the use of equational abstractions \cite{equational-abstraction-tcs} 
to abstract an infinite-state system into a finite-state one that can actually be
model checked for LTL properties.  But this is not all.  Some important system properties
go beyond LTL ones.  We did mention in passing properties of this kind when discussing
the fault-tolerant communication protocol of Section \ref{sec:oo}, namely, that only under
suitable object and message fairness assumptions could successful termination of
the protocol be guaranteed.  The most satisfactory way to express these advance properties
and effectively model check them is by specifying them in the \emph{linear time temporal
logic of rewriting} (LTLR) \cite{tlr-ugo} and verifying them using Maude's
\emph{LTLR model checker} \cite{DBLP:journals/scp/BaeM15}. 

\subsection{Symbolic Reasoning: Tools and Applications} \label{symb-comp-tools-and-apps}

This paper has placed special emphasis on Maude's novel features supporting symbolic
computation, including: (i) $B$-unification and $E \cup B$-unification; (ii) variants
and equational narrowing with the folding variant narrowing strategy; and
(iii) narrowing-based symbolic reachability analysis for topmost rewrite theories
of the form $\mathcal{R}=(\Sigma,E \cup B,R)$, where: (a) $(\Sigma,E \cup B)$ is FVP, and
(b) the rules $R$ are unconditional.  The best way to understand features (i)--(iii) 
is to see them as basic building blocks on top of which a wide range of symbolic reasoning
tools and applications can be built.  What follows is an attempt to provide an overview
of the tools and applications that support symbolic reasoning on top of features (i)--(iii).
More detailed accounts can be found in \cite{MeseguerWoLLIC18,GRT-coh-compl-symb-meth-JLAMP}.

\vspace{1ex}

\noindent {\bf Symbolic Model Checking}.  In complete analogy with the explicit-state case, the simplest kind of symbolic model checking supported by Maude is the narrowing-based symbolic reachability analysis provided by feature (iii) above.  As in the explicit-state case, such symbolic reachability analysis can be used to verify \emph{invariants}.  The simplest (but not the only: see below)
way to specify invariants is by providing a finite set $\{u_{1},\ldots, u_{n}\}$ of constructor patterns, so that the invariant's \emph{complement} is the set of ground instances of any of those patterns.  As in the explicit-state case, \emph{if} an invariant fails to hold, narrowing-based symbolic reachability analysis is guaranteed to detect the invariant's violation at some finite depth.  If, instead, the invariant does hold two things can happen: (i) if the narrowing-based search terminates without finding a violation, the invariant holds; otherwise, several possibilities remain open: (ii)-(a) perform bounded symbolic model checking by fixing a depth bound; (ii)-(b) use \emph{state space reduction} techniques to hopefully make the number of reachable symbolic states \emph{finite}%
; and (ii)-(c) use an equational abstraction, where the underlying equational theory remains FVP, in conjunction with (ii)-(b) to make the space of symbolic reachable states finite.  In cases (ii)-(b) and (ii)-(c) full verification of the given invariant can be achieved.  

An important domain-specific symbolic model checker also based on 
	\linebreak
narrowing-based symbolic reachability analysis is the \emph{Maude-NPA} tool for symbolic verification of cryptographic protocols \cite{EscobarMM-fosad}.  The point is that a cryptographic protocol $\mathcal{P}$ can be specified in Maude as a topmost rewrite theory $\mathcal{P}=(\Sigma,E \cup B,R)$ whose FVP equational part $(\Sigma,E \cup B)$ specifies the algebraic properties of the protocol's cryptographic functions. As before, security violations (invariant failures) can be specified by constructor patterns $\{u_{1},\ldots, u_{n}\}$ here called \emph{attack states}. The strongest points of the Maude-NPA tool are perhaps that: (1) it has very advanced state space reduction techniques \cite{DBLP:journals/iandc/EscobarMMS14}, so that a finite symbolic state space is actually reached in many cases, thus achieving full verification; and (iii) because of its support for reasoning modulo an FVP theory $(\Sigma,E \cup B)$, Maude-NPA is arguably the most general tool currently available
 for verifying cryptographic protocols modulo their algebraic properties.

In complete analogy with the explicit-state model checking case, the above narrowing-based symbolic model checking techniques extend to similar \emph{narrowing-based symbolic LTL model-checking} techniques \cite{bae-narrow-check} supported by Maude's logical LTL model checker available in the Maude web page.  This symbolic technique has been further
 extended to \emph{narrowing-based symbolic LTLR model checking} in \cite{bae-narrow-ltlr-check}.  Furthermore, symbolic methods can also be used to define \emph{predicate abstractions} that can effectively model check LTL properties \cite{DBLP:conf/rta/BaeM14}.

\vspace{1ex}

\noindent {\bf Term Pattern Predicates}.  If we take to heart the above-mentioned
idea of describing a possibly infinite set of states by a finite set 
$\{u_{1},\ldots, u_{n}\}$ of constructor patterns, what we can arrive at is
a series of increasingly more expressive languages for defining
\emph{state predicates} based on patterns.  In such languages,
logical operations  can be
 effectively computed by  symbolic techniques
in a way completely similar to how operations on finite automata
can effectively perform Boolean operations on their associated
regular languages.  In fact, if we have a constructor subspecification
$(\Omega,E_{\Omega} \cup B_{\Omega}) \subseteq (\Sigma,E \cup B)$
such that $(\Omega,E_{\Omega} \cup B_{\Omega})$ is FVP, then pattern
conjunction can be effectively computed by variable-disjoint $E_{\Omega} \cup B_{\Omega}$-variant
unification, and disjunction is just union of patterns.  The good properties
for the free case $E_{\Omega} \cup B_{\Omega}= \emptyset$, including
also negation for order-sorted linear patterns, have been investigated in
\cite{meseguer-skeirik-patterns-FAC}.  But we can go further by considering more
expressive \emph{constrained patterns} of the form $u \mid \varphi$, where
$u$ is an $\Omega$-term and $\varphi$ is a QF $\Sigma$-formula, so that $u \mid \varphi$
specifies the ground instances of $u$ for which $\varphi$ holds.
State predicates having such constrained patterns $u \mid \varphi$ as atomic predicates
and closed under conjunction and disjunction in an effectively, symbolically computable manner
have been studied in \cite{DBLP:conf/lopstr/SkeirikSM17,GRT-coh-compl-symb-meth-JLAMP}.
Such a language of pattern predicates is very useful to specify sets of states
both in reachability logic (see below), and in the constrained style of narrowing-based
reachability analysis defined in \cite{GRT-coh-compl-symb-meth-JLAMP}.

Another technique where pattern predicates are extremely useful is
in \emph{rewriting modulo SMT} \cite{rw-SMT-JLAMP}, where sets of states are represented by
pattern predicates $u \mid \varphi$ where satisfiability of $\varphi$ is decidable
by an SMT solver.  Roughly speaking, rewriting modulo SMT is  a symbolic reachability analysis
technique closely related to narrowing-based reachability analysis and
even more so to the narrowing-based constrained reachability analysis proposed in
\cite{GRT-coh-compl-symb-meth-JLAMP}.  The main differences with these two other approaches are: (i) instead of narrowing
modulo an FVP theory $E \cup B$, we perform $B$-matching;  (ii) the rules $R$ can
be conditional, but their conditions are SMT-solvable formulas; and (iii) after rewriting a symbolic
state $u \mid \varphi$
we accumulate SMT solvable constraints coming from a rule's condition
in the new symbolic state and
check for their satisfiability of the new constraint.

\vspace{1ex}

\noindent {\bf Variant Satisfiability}.  As already pointed out at the end of Section \ref{sec:variants}, under mild conditions on the constructors of an FVP theory $(\Sigma,E \cup B)$, satisfiability of QF formulas in the initial algebra $T_{\Sigma/E \cup B}$ is \emph{decidable} 
by theory-generic variant satisfiability algorithms \cite{var-sat-scp,gutierrez-meseguer-var-pred-LOPSTR}.
This is important, since the initial algebra $T_{\Sigma/E \cup B}$ is the initial model
of the functional module specified by the theory $(\Sigma,E \cup B)$, so that satisfiability of
QF formulas in  many \emph{user-defined} algebraic data types can be
decided this way.  For example, satisfiability of QF formulas in the initial algebras of
 the \texttt{NAT-FVP}  and \texttt{INT-FVP} examples
discussed in Section \ref{sec:variants} is decidable by this method, and many
more examples, including parameterized data types, are given in \cite{var-sat-scp,gutierrez-meseguer-var-pred-LOPSTR}.
For details on the algorithms and implementation see \cite{skeirik-meseguer-var-sat-JLAMP}.
One tool where these satisfiability results and algorithms are routinely used is in 
the reachability logic theorem prover (more on this below).

\vspace{1ex}

\noindent {\bf Generalization, Homeomorphic Embedding, and Partial Evaluation}.
 Generalization is the dual of unification.  When $B$-unifying
  terms $t$ and $t'$ we look for a term $u$ and substitution $\sigma$
  such that $t \sigma =_{B} u =_{B} t'\sigma$.  Instead, when we
  want to $B$-\emph{generalize} two term patterns  $t$ and $t'$ we look for
  a term $g$ and substitutions $\sigma,\tau$ of which they are instances
  up to $B$-equality, i.e., such that $g\sigma =_{B} t$ and $g \tau =_{B} t'$.
  In unification we look for most general unifiers (mgu's).  Instead, in generalization
  we look for \emph{least general generalizers} (lgg's).
 The relevance of  \cite{DBLP:journals/iandc/AlpuenteEEM14,DBLP:conf/jelia/AlpuenteBCEM19} and its associated ACUOS$^2$ tool
as a symbolic technique is that it supports
reasoning about generalization in a setting that is both
order-sorted and modulo axioms $B$, and does so in a modular way.
Specifically, the work in \cite{DBLP:journals/iandc/AlpuenteEEM14} 
and its Maude implementation provide a modular order-sorted equational
generalization algorithm modulo $B$, where 
$B$ can be any combination of associativity and/or commutativity and/or identity axioms.

The \emph{homeomorphic embedding} relation $u \lhd v$, where, roughly speaking, $u$ can be
obtained from $v$ by dropping some of $v$'s function symbols, gives a general method for stopping
any sequence of terms $t_{0}, t_{1},\ldots, t_{n},\ldots$ as soon as we can find $i < j$
such that $ t_{i} \lhd t_{j}$.  This important relation has been studied for untyped terms;
but in the context of Maude we often need to use the homeomorphic embedding relation $u \lhd v$ 
when $u$ and $v$ are order-sorted terms and, furthermore,  we need to reason not syntactically but
\emph{modulo} axioms $B$ such as associativity and/or commutativity, that is, with a relation
$u \lhd_{B} v$.  
In \cite{ACEM-LOPSTR18}, the relation $\lhd_{B}$ is studied
and 
efficient algorithms for computing it are designed for Maude.
They have been implemented in the tool HEMS.

Both order-sorted generalization modulo $B$ and homeomorphic embedding modulo $B$ 
are crucial components of a \emph{partial evaluator} for Maude functional modules.  Partial evaluation of
equational specifications had never been considered before in the order-sorted and modulo $B$ level of
generality  needed for Maude  equational programs with convergent theories of the
form $(\Sigma,E \cup B)$.  Partial evaluation methods that can work in this very general
setting (note that the usual ``vanilla flavored'' case where
 $\Sigma$ is unsorted and $B = \emptyset$ is indeed a very special subcase) have been
developed in \cite{ACEM-LOPSTR16} and have been implemented
in Maude in the Victoria tool \cite{VictoriaPEMaude}.

\vspace{1ex}

\noindent {\bf Theorem Provers}. Using rewriting logic's nice properties
as a logical framework (see the survey \cite{20-years}),
the symbolic techniques currently supported by Maude can
be applied to a wide range of theorem provers not just for Maude and rewriting
logic but also for many other logics. We will focus here on theorem-proving
tools more closely related to Maude.  To begin with, let us discuss tools for 
\emph{reachability logic}.  This logic was originally proposed in 
\cite{DBLP:conf/fm/RosuS12,DBLP:conf/oopsla/RosuS12,DBLP:conf/rta/StefanescuCMMSR14,DBLP:conf/oopsla/StefanescuPYLR16}
as a language-generic approach to program verification parametric on
the operational semantics of a programming language.  Both Hoare logic
and separation logic can be naturally mapped into reachability
logic~\cite{DBLP:conf/fm/RosuS12,DBLP:conf/oopsla/RosuS12}.  The work
in \cite{DBLP:conf/lopstr/SkeirikSM17}
extends  reachability logic from a
programming-language-generic logic of programs to a
rewrite-theory-generic logic to reason about \emph{both} distributed system
designs and  programs, based on their rewriting logic
semantics.  This extension is non-trivial and requires a number
of new concepts and results (see~\cite{DBLP:conf/lopstr/SkeirikSM17}).
In particular, concepts such as: (i) constructor pattern predicates, (ii) narrowing with conditional
rules, and (iii) variant satisfiability, go a long way in making the \emph{constructor-based}
version of reachability logic proposed in \cite{DBLP:conf/lopstr/SkeirikSM17} much more
easily mechanizable by exploiting the recent symbolic features of Maude
and the Maude-based variant satisfiability algorithms in \cite{skeirik-meseguer-var-sat-JLAMP}.
Indeed, the work in~\cite{DBLP:conf/lopstr/SkeirikSM17} has been implemented in Maude.
It was originally inspired by the also Maude-based work in
\cite{DBLP:conf/birthday/LucanuRAN15}, but it adds to that work a
substantial number of new results and methods.

The most recent Maude-based work on reachability logic provers
 closest  to the work in  \cite{DBLP:conf/lopstr/SkeirikSM17} is that
 in \cite{DBLP:journals/jsc/LucanuRA17}
and, even more so, in \cite{DBLP:conf/cade/CiobacaL18}.
The approach in~\cite{DBLP:journals/jsc/LucanuRA17} adopts a semantic
framework for models similar to the already-discussed work in
\cite{DBLP:conf/rta/StefanescuCMMSR14,DBLP:conf/oopsla/StefanescuPYLR16}, i.e.,
 state properties are specified using matching logic
and assume a given first-order logic model.  Therefore, the
semantic framework is different from the one in  \cite{DBLP:conf/lopstr/SkeirikSM17}.
An important contribution
of the work in \cite{DBLP:journals/jsc/LucanuRA17}
is its  coinductive semantics and justification for circular
co-inductive reasoning. 
 Perhaps the recent work closest to  \cite{DBLP:conf/lopstr/SkeirikSM17}
in the coinductive approach is that of  Ciob\^{a}c\u{a} and Lucanu
in \cite{DBLP:conf/cade/CiobacaL18}.  In summary, for verification of reachability properties of rewrite theories
---including Hoare logic properties as a special case---  the reachability logic
theorem provers in \cite{DBLP:conf/lopstr/SkeirikSM17}, \cite{DBLP:journals/jsc/LucanuRA17}
and \cite{DBLP:conf/cade/CiobacaL18} seem to be the most advanced and most promising,
and all do make use of the Maude symbolic techniques described in this paper.

Last, but not least, let us mention two other theorem-proving tools.
The Tamarin theorem-proving tool~\cite{MeierSCB13} for verification of cryptographic protocols
 uses Maude's  variant-generation algorithm, 
initially only for the Diffie-Hellman theory, but recently extended to finite variant theories in Maude \cite{DBLP:conf/post/DreierDKS17}.
Finally,  several decision procedures 
for formula satisfiability modulo equational theories
have been provided based on 
narrowing in the tool  \cite{TushkanovaGRK15}.

\vspace{1ex}

\noindent {\bf Further Reading for a Broader Perspective}.  This entire section can be misleading, since we have said nothing at all about many other application areas such as, for example: (i) specification and verification of programming languages based on their rewriting logic definitions; (ii) real-time and cyber-physical systems; (iii) probabilisitic systems; (iv) logical framework applications; and (v) bioinformatics applications, to mention just a few areas.  There is no space here for discussing tools and applications on all those and other areas, or just for discussing many other Maude-based tools.  Fortunately, the survey paper \cite{20-years} gives a quite complete account of this broader perspective and, in spite of being a few years old, is still a good starting point to obtain a broad overview of the many applications made possible by Maude.

\section{Conclusions and Future Work} \label{sec:concl}

In this paper we have both tried to give an introduction to
Maude that does not assume prior
acquaintance with the language, and to describe important new features
that have been added to the language since 2007, when the Maude
book \cite{maude-book} appeared.  Our intention has been  to provide
a journal-level entry point to the language as it currently exists,
both for readers new to Maude and for readers familiar with Maude
who would like to have a comprehensive explanation of these important new features.

In particular, we have described those features enabling Maude's very
general support for \emph{symbolic computation}, including 
\emph{order-sorted unification algorithms}: (i) modulo axioms $B$ like
associativity and/or
commutativity and/or identity, and (ii) modulo equations $E \cup B$
where the equations $E$ are convergent modulo $B$.
Of particular importance is the existence of
an infinite class of theories $E \cup B$ (namely those having the 
finite variant property) for which Maude's $E \cup B$-unification 
algorithm always terminates with a complete set of most general
solutions for any unification problem.
Furthermore, we have also described Maude's support for
\emph{narrowing-based symbolic reachability analysis} (that
builds on the $E \cup B$-unification capability). This functionality allows the
user to leverage the power of symbolic computation to carry out
symbolic model-checking analyses of systems that would otherwise be
unfeasible due to the need to explore  infinite or very large state
spaces.  As we have explained in Section \ref{symb-comp-tools-and-apps},
these symbolic  features make possible a wide range of formal tools built
using them and many formal analysis applications.

We also discussed Maude's \emph{strategy language}, which provides
a declarative and modular way to carve out subsets of a system's behavior
without in any way changing the rules specifying the system.  

Finally, we introduced new external objects that allow Maude
specifications to interact with the external world: input/output
objects---the three standard IO objects and file objects (plus the
prior socket objects); and a powerful new kind of external objects
called \emph{meta-interpreters}.  A meta-interpreter encapsulates a Maude
interpreter as an object, and can interact with other stateful objects
both internal and external (including other meta-interpreters).

Regarding future work, perhaps the most important symbolic computation
 topic missing in the present paper is \emph{SMT solving}.  We have
explained in Section \ref{symb-comp-tools-and-apps} that
\emph{variant-based satisfiability} of quantifier-free formulas
for  algebraic data types specified by functional modules
having the finite variant property and satisfying mild additional
assumptions is already available in an extension of the
\texttt{META-LEVEL} module and is used in reachability logic theorem
proving.  But there is the additional fact that in recent years
experimental versions of Maude supporting access to the CVC4 \cite{cvc4-reference}
and Yices \cite{Yices-reference} SMT solvers have been available
and have been used in various applications.  The main reason for not
including SMT solving in this paper is that we are still experimenting
with SMT solving features and it seems preferable to leave this topic
for a future publication.  

Without trying to be exhaustive, three future
directions seem both clear and strategically important:

\begin{enumerate}
\item \emph{Symbolic computation}.  Besides further advancing Maude's
  support for SMT solving, important new advances are needed
in narrowing-based symbolic model checking and in many
theorem-proving applications.

\item \emph{Distributed programming}.  The present, much more flexible
  support for interaction with external objects opens up as never
  before the possibility of a seamless and correct-by-construction passage from Maude
  specifications of concurrent object systems to their deployment as
  distributed systems.  This can have important advantages for
  developing highly reliable distributed systems and for doing so in a
  fully declarative way.

\item \emph{Strategies}.  Now that strategies are available and
  efficiently supported at the Core Maude level, many applications
  seem ripe, including, for example, the following: (i) strategy-based model-checking algorithms; 
  (ii) support for strategies in Maude-based
  theorem-proving tools; and (iii) further advances of the rewriting
  logic semantics project
  \cite{meseguer-rosu-tcs,DBLP:journals/iandc/MeseguerR13}
  made possible by using strategies in the semantic definition of languages.
\end{enumerate}

\paragraph{Acknowledgements}
Dur\'an has been partially supported by MINECO/FEDER project TIN2014-52034-R.
Escobar has been 
partially supported by the EU (FEDER) and the Spanish
MCIU under grant RTI2018-094403-B-C32,
by the Spanish Generalitat Valenciana under grant PROMETEO/2019/098,
and by the US Air Force Office of Scientific Research 
under award number FA9550-17-1-0286.
Mart\'{\i}-Oliet and Rubio
have been partially supported by
MCIU Spanish project TRACES (TIN2015-67522-C3-3-R).
Rubio has also been partially supported by a MCIU grant FPU17/02319.
Meseguer and Talcott
 have been partially supported by NRL Grant N00173-17-1-G002.
Talcott has also been
 partially supported by ONR Grant N00014-15-1-2202.

\phantomsection
\addcontentsline{toc}{section}{References}

 \bibliographystyle{elsarticle-harv} 
 \bibliography{maude-jlamp}

\end{document}